\documentclass{optica-article}

\journal{opticajournal} 

\articletype{Research Article}

\usepackage{graphicx}
\usepackage{amsmath}
\usepackage{booktabs}
\usepackage{xcolor}
\usepackage{multirow}
\usepackage{subcaption}
\usepackage{pifont}
\usepackage{lineno}
\usepackage{appendix}
\usepackage{svg}

\usepackage{tikz}
\usetikzlibrary{spy}
\usetikzlibrary{arrows.meta}

\linenumbers 

\begin{document}
\nolinenumbers
\title{MetaTele: Compact Refractive Metasurface Computational Telephoto Camera}

\author{Harshana Weligampola\authormark{1,\textdagger}, Yuanrui Chen\authormark{1,\textdagger}, Abhiram Gnanasambandam\authormark{2}, Dilshan Godaliyadda\authormark{2}, Hamid R. Sheikh\authormark{2}, Stanley H. Chan\authormark{1}, and Qi Guo\authormark{1,*}}

\address{\authormark{1}Elmore Family School of Electrical and Computer Engineering, Purdue University\\
\authormark{2}Samsung Research America}

\address{\authormark{\textdagger}Co-first authors with equal contribution}
\address{\authormark{*}\email{qiguo@purdue.edu}}

\begin{abstract*}
Smartphone cameras face fundamental form-factor constraints that limit their optical magnification, primarily due to the difficulty of reducing a lens assembly's telephoto ratio, the ratio between total track length (TTL) and effective focal length (EFL). Currently, conventional refractive optics struggle to achieve a telephoto ratio below 0.5 without requiring multiple bulky elements to correct optical aberrations. In this paper, we introduce MetaTele, a novel optics-algorithm co-design that breaks this bottleneck. MetaTele explicitly decouples the acquisition of scene structure and color information. First, it utilizes a compact refractive-metasurface optical assembly to capture a fine-detail structure image under a narrow wavelength band, inherently avoiding severe chromatic aberrations. Second, it captures a broadband color cue using the same optics; although this cue is heavily corrupted by chromatic aberrations, it retains sufficient spectral information to guide post-processing. We then employ a custom one-step diffusion model to computationally fuse these two raw measurements, successfully colorizing the structure image while correcting for system aberrations. We demonstrate a MetaTele prototype, achieving an unprecedented telephoto ratio of 0.44 with a TTL of just 13 mm for RGB imaging, paving the way for DSLR-level telephoto capabilities within smartphone form factors.
\end{abstract*}

\section{Introduction}\label{sec:intro}

\begin{figure}
\centering
\includegraphics[width=\linewidth]{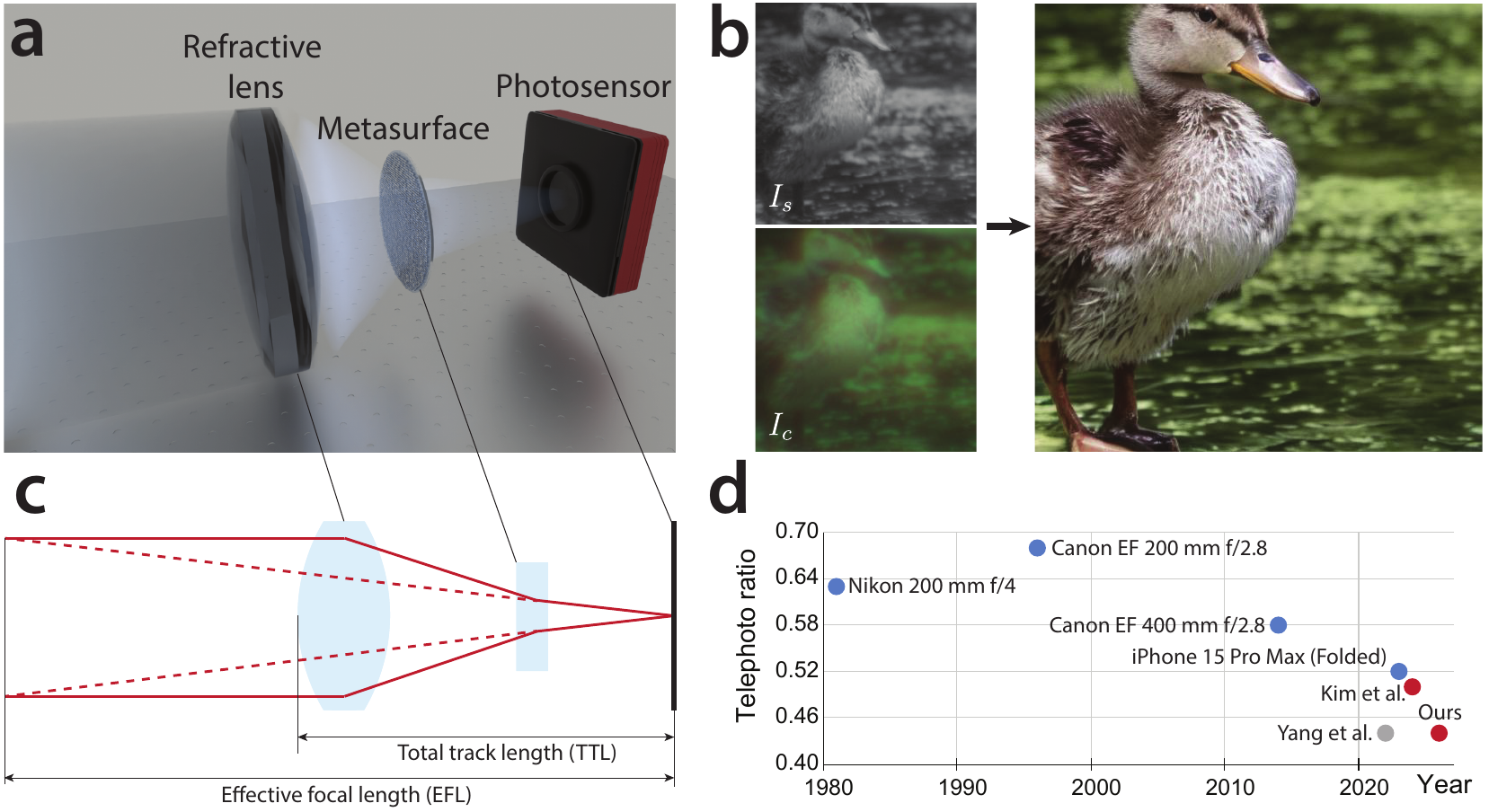}
\caption{Overview.
(a) The proposed MetaTele imaging system consists of a hybrid refractive–metasurface assembly, forming a compact telephoto architecture. (b) The hardware sequentially captures (i) a structure image $I_s$ with fine details under a narrow spectral bandwidth by inserting a spectral filter into the optical path, and (ii) a color cue $I_c$ over the full visible spectrum without the filter, where strong aberrations are present. In future implementations, these two measurements can be acquired simultaneously using a dedicated spectral filter array. The captured measurements are then computationally fused to reconstruct a high-quality RGB telephoto image. (c) The \textit{telephoto ratio}, defined as the ratio between the total track length (TTL) and the effective focal length (EFL), quantifies telephoto compactness; smaller values indicate stronger telephoto capability. (d) MetaTele achieves, to our knowledge, the lowest reported telephoto ratio. Blue dots denote commercially available lenses. Gray and red dots represent research prototypes, where gray indicates monochrome-only demonstrations and red indicates full RGB imaging capability. }
\label{fig:teaser}
\end{figure}

Telephoto lenses employ assemblies of optical elements to achieve an effective focal length (EFL) that exceeds the total track length (TTL) of the imaging system. Such lenses are widely used in photography, scientific imaging, and national defense applications. However, conventional refractive telephoto lenses are bulky, as they require multiple rigid, curved elements to correct optical aberrations~\cite{sawant2021aberration}. Consequently, the telephoto ratio—defined as the ratio between the TTL and the EFL—is typically limited to values no lower than approximately 0.5 (Fig.~\ref{fig:teaser}). This form-factor constraint fundamentally limits the integration of high-resolution cameras into compact platforms such as smartphones, micro-robots, and mixed-reality headsets. 

We propose \textit{MetaTele}, a novel RGB telephoto camera that pushes the limit of the telephoto ratio through a co-designed optical assembly and computational post-processing. MetaTele is built upon two core ideas. First, it explicitly decouples the acquisition of scene structure and color information. The optical system is optimized to achieve high optical zoom and imaging fidelity within a narrow spectral band, where chromatic aberrations are inherently minimal. By avoiding the need to optically correct chromatic aberrations across the full visible spectrum, the lens design problem is substantially relaxed, enabling a telephoto ratio lower than that achievable with conventional achromatic telephoto lenses.

As illustrated in Fig.~\ref{fig:teaser}, MetaTele captures a high-optical-zoom, fine-detail \textit{structure image} within the designed spectral band, followed by a \textit{color cue} acquired using the same optics over the full visible spectrum. Although the color cue suffers from severe chromatic aberrations, it retains sufficient color information to guide the colorization of the structure image during post-processing. To this end, we develop a custom one-step diffusion model that fuses the two raw measurements and reconstructs high-quality RGB telephoto images.

Second, enabled by the relaxed achromaticity requirement, we demonstrate that the telephoto ratio can be further reduced by replacing bulky refractive optics with a metasurface~\cite{khorasaninejad2016metalenses}. Moreover, our analysis shows that, at small aperture sizes, metasurfaces exhibit higher tolerance to fabrication and assembly nonidealities compared to conventional refractive optics.

In this paper, we present a MetaTele prototype composed of two optical elements: an off-the-shelf refractive lens serving as the objective and a custom-fabricated metasurface functioning as the eyepiece (Fig.~\ref{fig:teaser}). To support the development and evaluation of post-processing algorithms, we collect a large-scale dataset comprising 2,650 paired raw measurements captured by the real-world MetaTele prototype, including both the structure image and the color cue, along with their corresponding ground-truth images. This dataset is used to systematically analyze the reconstruction performance of various learning-based post-processing methods. The proposed MetaTele prototype achieves a total track length (TTL) of only 13~mm and a telephoto ratio of $0.44$, exceeding the performance of conventional refractive telephoto lenses.

The contributions of this paper are summarized as follows:
\begin{enumerate}
    \item We introduce a two-shot computational imaging framework for capturing high-quality RGB telephoto images.
    \item We present a large-scale, real-world metasurface imaging dataset to facilitate the development and benchmarking of image restoration algorithms.
    \item We demonstrate a compact RGB camera prototype that achieves a telephoto ratio of 0.44, which, to the best of our knowledge, represents the lowest reported telephoto ratio.
\end{enumerate}

\section{Related works}

\paragraph{Metasurface Computational Imaging.}
Metasurfaces' compactness and versatile modulation in terms of amplitude, phase, and polarization towards incident light make them an emerging technology for computational imaging~\cite{metasurface_review}. In recent years, people have demonstrated metasurface-based computational imagers with unprecedented form factors, latency, or accuracy for achromatic~\cite{colburn,rgbeyepiece, Tseng2021NeuralNanoOptics},~HDR~\cite{Brookshire:24,mandal2026enabling},~depth~\cite{jumping},~hyperspectral~\cite{hyperspectral},~full-Stokes polarization~\cite{Stokes}, superresolution imaging~\cite{subdiffraction}, etc. 

To streamline the design of metasurfaces given specific imaging applications, people have also developed computational frameworks that model the metasurface and enable gradient-based optimizations over the metasurface shape parameters~\cite{pestourie2018inverse, jiang2019global, dflat, hammond2022high}. However, such simulators are computationally expensive due to the excessive memory required to store the metasurface parameters or the sophisticated computation to numerically solve Maxwell's equations. To bypass this, Pinilla \emph{et al.} explored directly optimizing the optical design using hardware-in-the-loop (HIL), which led to similar performance as a simulation design but more than 100$\times$ lower computational cost~\cite{Pinilla2023}. 

\paragraph{Metasurface Zoom and Telephoto Cameras.} People have demonstrated metasurface-based zoom optics, which vary their EFLs by mechanically adjusting the relative angle~\cite{Rotation-zoom} or distance~\cite{Metaoptical_zoom} between multiple metasurface elements. These systems primarily aim to achieve smoothly-varying optical magnification. In contrast, metasurface-based \textit{telephoto} cameras, i.e., telephoto ratio smaller than 1, remain largely unexplored. Yang \emph{et al.} report a parfocal zoom metasurface camera that incidentally attains a telephoto ratio of 0.44~\cite{parfocal}. Kim \emph{et al.} employ folded metasurfaces to realize an effective telephoto ratio of 0.5 within an ultra-slim system thickness of 0.7~mm~\cite{Kim2024}. However, both systems are limited to monochrome imaging. To the best of our knowledge, a metasurface-based telephoto camera capable of producing full-color RGB photographs has not yet been demonstrated.

\paragraph{Image Colorization.}
The concept of independently measuring scene structure and color/spectral information has been extensively studied in hyperspectral and multispectral imaging. Existing approaches typically fuse a high-resolution monochrome image with a low-resolution or spatially sparse spectral cue, using either learning-based methods~\cite{masi2016pansharpening, meng2023pandiff} or model-based, non-learning approaches~\cite{palsson2013new, loncan2015hyperspectral, wang2018high, hypersectral}. Beyond simple fusion, prior work has also investigated the joint optimization of sparse spectral sampling patterns on the photosensor and the corresponding post-processing algorithms to further enhance reconstruction quality~\cite{Zickler, sparsesensor}. In contrast to these studies, we extend the image colorization paradigm to telephoto imaging and investigate a previously unexplored regime in which the color cue is intentionally corrupted by severe optical aberrations.

\paragraph{Diffusion-based computational imaging.} Recently, people have demonstrated diffusion models as powerful generative priors to solve inverse problems in computational imaging~\cite{kawar2022denoising, fei2023generative, garber2024image, liu2024residual, luo2025visual}. These works demonstrate the synthesis of high-quality photographs from degraded sensor measurement, sometimes even from unconventional sensors~\cite{purohit2024generative}. However, most diffusion-based computational imaging models are ``physics-agnostic", lack the explicit physical modeling required to decouple spatially-varying, significant aberrations from the underlying scene content in novel imaging systems like ours. To address the compact telephoto problem, a model must function not only as a semantic synthesizer, but also as a physically grounded solver capable of correcting for such aberrations while maintaining fidelity. 

\begin{table}[ht]
    \centering
    \renewcommand{\arraystretch}{1.2}
    \setlength{\tabcolsep}{8pt} 
    \resizebox{\linewidth}{!}{  
    \begin{tabular}{l c c c c c c c}
        \toprule
         & Method & Telephoto ratio & f/\#  & TTL (mm) & Inputs & Output & Postprocessing  \\
        \midrule
         \multirow{2}{*}{Telephoto} & Ours, 2024 & 0.44 & 6 & 13 & 2 & Color & Diffusion\\
         & Yang \emph{et al.}, 2022\cite{parfocal} & 0.44 & 6.8 & 10.8 & 1 & Monochrome & N/A\\
        \midrule
         Folded & Kim et al, 2024 \cite{Kim2024} & 0.5 & 4 & 0.7 & 1 & Monochrome & N/A\\
        \midrule
         \multirow{2}{*}{Zoom} & Wei \emph{et al.}, 2020 \cite{Rotation-zoom} & - & 27.5 & 12 & 1 & Monochrome & N/A\\
         & Zhang \emph{et al.}, 2024 \cite{Metaoptical_zoom} & - & 4.5 & 7.5 & 1 & Monochrome & N/A\\
        \midrule
         \multirow{5}{*}{Others} & Heide \emph{et al.}, 2016~\cite{Heide2016} & - & 12.5 & 100 & 1 & Color & Non-learning \\
         & Fröch \emph{et al.}, 2025~\cite{Froch2025} & - & 2 & 20 & 1 & Color & Diffusion \\
         & Tseng \emph{et al.}, 2021~\cite{Tseng2021NeuralNanoOptics} & - & 2 & 1 & 1 & Color & U-net \\
         & Liu \emph{et al.}, 2024~\cite{Liu2024} & - & 3 & 1.57 & 1 & Color & Attention \\
         & Pinilla \emph{et al.} 2023~\cite{Pinilla2023} & - & 1 & 10.5 & 1 & Color & DRU-net \\
        \bottomrule
    \end{tabular}
    }
    \noindent \scriptsize{- The telephoto ratio is greater than 1, or not reported}\\
    \caption{Comparison of specifications of recent metasurface-based imaging systems. Ours achieves the smallest telephoto ratio for color imaging.}
    \label{tab:related_works}
\end{table}
\vspace{-0.1in}

\section{System}

\subsection{Measurement model}
\begin{figure}
    \centering
    \includegraphics[width=0.8\linewidth]{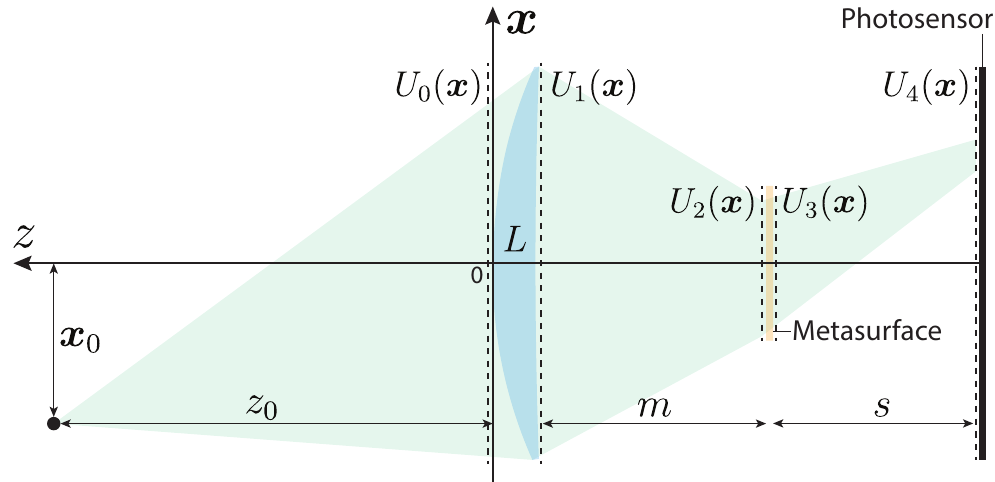}
    \caption{Optical Model. MetaTele consists of a refractive objective and a metasurface eyepiece. The assembly magnifies the incident angle of the incoming light waves. }
    \label{fig:optical-model}
\end{figure}

As illustrated in Fig.~\ref{fig:optical-model}, consider the MetaTele system, comprising an achromatic spherical lens $L$ and a metasurface $M$, imaging a point source emitting wavelength $\lambda$ located at position $(\mathbf{x}_0, z_0)$. We assume thin optics and paraxial approximation, and utilize the Fresnel propagator:
\begin{align}
    \text{Fresnel}_z(U)(\mathbf{x}) = \frac{e^{jkz}}{j\lambda z} \int U(\mathbf{s}) \exp\left(j\frac{k}{2z}\lVert\mathbf{x} -\mathbf{s}\rVert^2\right) d\mathbf{s}, 
\end{align}
when the wavefront is propagated in free space for the axial distance $z$. 

\paragraph{Wave propagation.} The wavefront immediately before entering the system is:
\begin{align}
    U_0(\mathbf{x}) \propto \exp\left(j\frac{k(\lambda)}{2z_0}\lVert\mathbf{x}_0 - \mathbf{x}\rVert^2\right).
\end{align}
The spherical lens $L$ exerts an equivalent optical modulation:
\begin{align}
    L(\mathbf{x}) &= \exp\left(-jk(\lambda)(n-1)\left(R - \sqrt{R^2 - \lVert\mathbf{x}\rVert^2}\right)\right),
\end{align}
where $n$ is the index of refraction of the lens.
By applying the second-order approximation to the phase profile of $L$, the wavefront after the spherical lens $U_1$ is:
\begin{align}
    U_1(\mathbf{x}) = U_0(\mathbf{x}) L(\mathbf{x}) \propto  \exp\left(j\frac{k(\lambda)}{2}\left[\left(\frac{1}{z_0} - \frac{1}{f_1}\right)\lVert\mathbf{x}\rVert^2 - \frac{2}{z_0}\mathbf{x}_0\cdot \mathbf{x} \right]\right),
\end{align}
where $f_1 = \frac{R}{n-1}$ is the focal length of $L$. The wavefront $U_1$ propagates axially by $m$ and becomes $U_2$ before entering the metasurface, which is, according to the Gaussian integral:
\begin{equation}
\begin{aligned}
    U_2(\mathbf{x})  = \text{Fresnel}_m(U_1)(\mathbf{x})  \propto
\exp\left(
j\frac{k(\lambda)}{2}\,A_2\,\|\mathbf{x}\|^2
\right)
\exp\left(
-ik(\lambda)\,B_2\,\mathbf{x}_0 \cdot \mathbf{x}
\right),
\end{aligned}
\end{equation}
where
\begin{align*}
    A_2
=
\frac{1}{m}
-
\frac{1}{m^2}
\Bigg/
\left(
\frac{1}{m}
+
\frac{1}{z_0}
-
\frac{1}{f_1}
\right),
\qquad
B_2
=
\frac{1}{z_0m}
\Bigg/
\left(
\frac{1}{m}
+
\frac{1}{z_0}
-
\frac{1}{f_1}
\right).
\end{align*}
This conclusion is derived under the assumption that aperture diameter of refractive lens is sufficiently large. Consider that the metasurface $M$ is designed to exert a quadratic phase delay profile with focal length $f_2$ to the incident wavefront at the design wavelength $\lambda_0$:
\begin{align}
    M(\mathbf{x};\lambda_0) = P(\mathbf{x})\exp\left(-j\frac{k_0}{2f_2}\lVert\mathbf{x}\rVert^2\right), 
    \label{eq:M}
\end{align}
where $k_0 = \frac{2\pi}{\lambda_0}$ and $P(\mathbf{x})$ is the transmittance profile of the metasurface. According to previous studies, we can safely assume the metasurface's modulation at other visible wavelengths $\lambda$ to be constant: $M(\mathbf{x}; \lambda) = M(\mathbf{x}; \lambda_0)$~\cite{liu2025metah2}. The wavefront after the metasurface is: $U_3(\mathbf{x}) = U_2(\mathbf{x})M(\mathbf{x};\lambda)$. Therefore, the wavefront at the photosensor, $U_4$, is:
\begin{equation}
\begin{aligned}
        U_4(\mathbf{x}) &= \text{Fresnel}_s(U_3)(\mathbf{x}) \\ &\propto e^{j\frac{k(\lambda)}{2s}\|\mathbf{x}\|^2}\int P(\mathbf{s})\exp\left(j\frac{k(\lambda)}{2}\Delta(\lambda)\|\mathbf{s}\|^2-jk(\lambda)\left(\frac{\mathbf{x}}{s}+B_2\mathbf{x}_0\right)\cdot\mathbf{s}\right)d\mathbf{s},
    \label{eq:U4}
\end{aligned}
\end{equation}
where the residual defocus coefficient $\Delta(\lambda)$ is:
\begin{align*}
    \Delta(\lambda)=\frac{1}{s}+A_2-\frac{k_0}{k(\lambda)}\frac{1}{f_2}.
\end{align*}

\paragraph{Point spread functions (PSFs).} Define the PSF of the MetaTele system at distance $z$ and wavelength $\lambda$ as:
\begin{align}
    h(\mathbf{x};z, \lambda) = \left|\int P(\mathbf{s})\exp\left(j\frac{k(\lambda)}{2}\Delta(\lambda)\|\mathbf{s}\|^2-j\frac{k(\lambda)}{s}\mathbf{x}\cdot\mathbf{s}\right)d\mathbf{s} \right|^2.
    \label{eq:psf}
\end{align}
Eq.~\ref{eq:psf} becomes a focused PSF when the residual defocus coefficient $\Delta(\lambda) = 0$, which provides the solution of focal plane distance $z_f$ of the proposed system:
\begin{equation}
z_f(\lambda)=\left(\frac{1}{m^2 R(\lambda)}-\frac{1}{m}+\frac{1}{f_1}\right)^{-1},\text{ where } R(\lambda) = \frac{1}{s}+\frac{1}{m}-\frac{k_0}{k(\lambda)}\frac{1}{f_2}.
\end{equation}
This suggests that the focal plane of MetaTele varies according to the wavelength. When the point source is out of the focal plane $Z_f(\lambda)$, the PSF expands according to Eq.~\ref{eq:psf}. 

The image of the point source $(\mathbf{x}_0, z_0)$ on the photosensor is the translation of the PSF:
\begin{align}
    \left|U_4(\mathbf{x})\right|^2 = h\left(\mathbf{x} + \gamma\mathbf{x}_0; z_0, \lambda\right),
    \label{eq:E}
\end{align}
where $\gamma = -s B_2$ is the magnification. The EFL of the system can be calculated as:
\begin{align}
    \text{EFL} = \lim_{z_0\rightarrow \infty}\left|\gamma z_f(\lambda)\right| = \frac{sf_1}{f_1 - m}. 
\end{align}

\paragraph{Image formation model.}
We model the target scene as a collection of point sources $\{(\mathbf{x}_i, z_i)\}_{i=1}^M$. MetaTele captures two measurements: a \emph{structure image} $I_s$ and a \emph{color cue} $I_c$. The structure image $I_s$ is acquired with a bandpass filter that restricts the spectrum to a narrow bandwidth centered at the design wavelength $\lambda_0$, whereas the color cue $I_c$ is captured over the full visible spectrum. Both measurements are described by the following image formation model:
\begin{align}
    I_j(\mathbf{x}) = G \cdot \mathrm{Poisson}\!\left(
    \eta\, t \sum_{i} \int_{\lambda} S_j(\lambda)\, E_i(\lambda)\,
    h(\mathbf{x}+\gamma \mathbf{x}_i; z_i, \lambda)\, d\lambda
    \right)
    + \mathcal{N}(0,\sigma^2),~j\in\{s,c\}
\end{align}
where $E_i(\lambda)$ denotes the spectral irradiance of the $i$th point source and $S_j(\lambda)$ is the effective spectral response of the imaging system for the structure image or the color cue. The parameters $t$, $\eta$, and $G$ denote the exposure time, quantum efficiency, and electronic gain, respectively, while the additive Gaussian term models read noise with variance $\sigma^2$. This image formation model follows that of Brookshire \emph{et al.}~\cite{Brookshire:24}. 

While this is derived for a static system, we show that the architecture supports autofocus and continuous zoom (from 20 to 50 mm), as detailed in Supplement 1. 

\subsection{Computational model}
\label{secsec:comp}

The goal of post-processing is to learn a mapping function
$G_{\boldsymbol{\mathbf{\theta}}}(I_s, I_c) \rightarrow \hat{I}$ that synthesizes a high-quality telephotograph $\hat{I}$ by fusing the high spatial fidelity of the structure image $I_s$ with the chromatic information provided by the color cue $I_c$, while compensating for optical aberrations introduced by the imaging system. We use $\mathbf{\theta}$ to denote the parameters of the generator.

\begin{figure*}
    \centering
    \includegraphics[width=\linewidth]{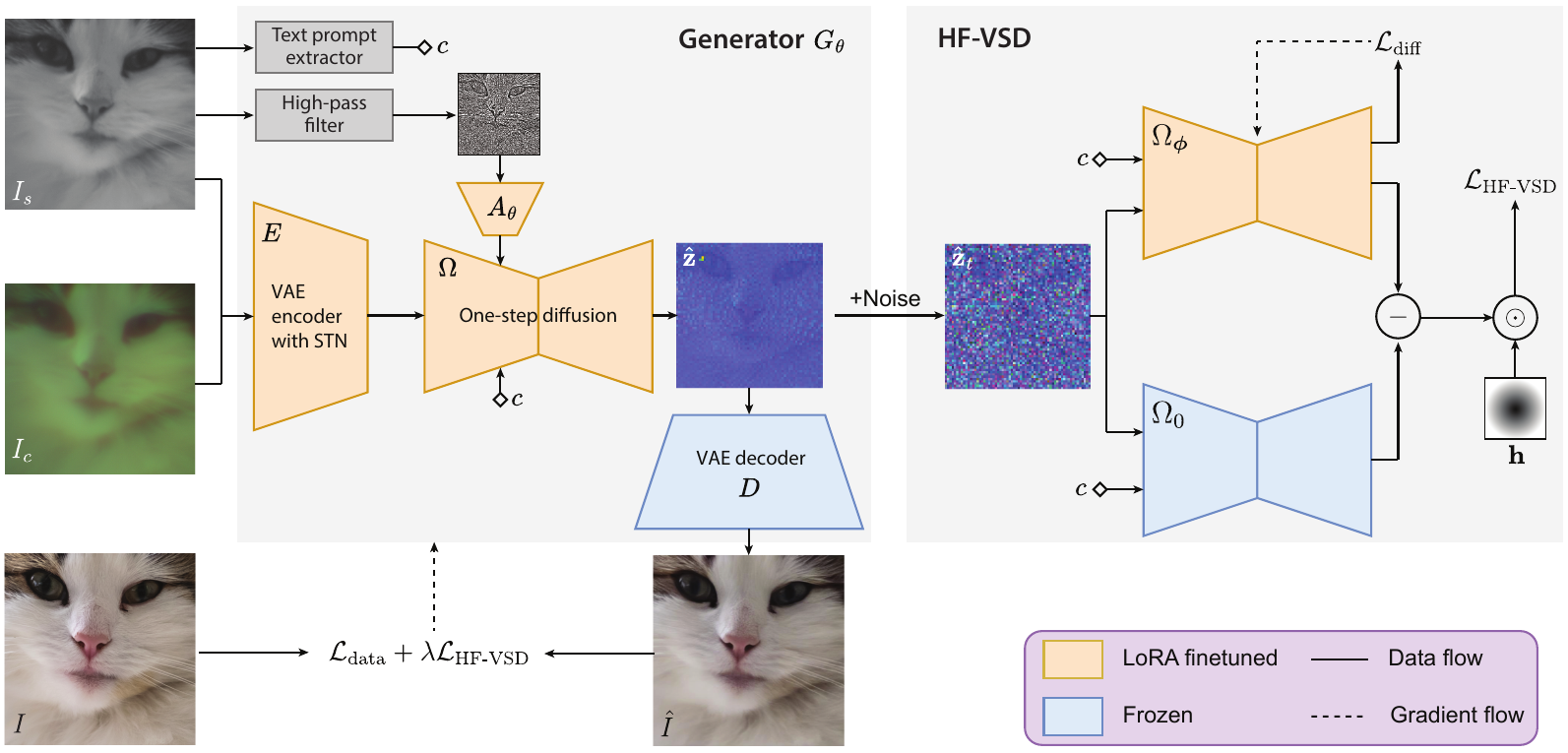}
    \caption{Computational model and training framework. MetaTele utilizes a generator $G_{\mathbf{\theta}}$ built upon a one-step diffusion model $\Omega$. The model is fine-tuned using the combination of the data fidelity loss $\mathcal{L}_{\text{data}}$ and the high-frequency varitional score distillation (HF-VSD) loss $\mathcal{L}_{\text{HF-VSD}}$ modified from the standard VSD~\cite{wang2023prolificdreamer}. }
    \label{fig:framework}
\end{figure*}

\paragraph{Network architecture.}
To realize this fusion and aberration-correction task, we propose a one-step generative neural network. As illustrated in Fig.~\ref{fig:framework}, the framework adopts a variational encoder--decoder architecture comprising an encoder $E$ and a decoder $D$, with a one-step diffusion module $\Omega$ embedded between them. The diffusion module is conditioned on two complementary sources of information: (i) text prompts $\mathbf{c}$ extracted from the structure image $I_s$, and (ii) learned feature embeddings generated by an adaptor network $A$ that operates on the high spatial-frequency components of $I_s$. The former is a standard condition of diffusion models and the latter guides the reconstruction process to enhance the high-frequency texture information. We reduce the diffusion process to one step to keep relatively low computational cost and latency of postprocessing, compared to the classic Stable Diffusion~\cite{rombach2021highresolution}. 

\paragraph{Training.}
We initialize the encoder $E$, decoder $D$, and diffusion module $\Omega$ using pre-trained weights from Stable Diffusion~\cite{rombach2021highresolution}, and introduce trainable Low-Rank Adaptation (LoRA) modules~\cite{hu2022lora} to selected layers and fine-tune only these parameters. The resulting optimization problem is formulated as
\begin{equation}
    \label{eq:problem}
    \boldsymbol{\theta}^* = \arg\min_{\boldsymbol{\theta}} 
    \mathbb{E}_{I_s, I_c, I}
    \big[
    \mathcal{L}_{\text{data}}(G_{\boldsymbol{\theta}}(I_s, I_c), I)
    + \lambda \mathcal{L}_{\text{HF-VSD}}(G_{\boldsymbol{\theta}}(I_s, I_c))
    \big],
\end{equation}
where $\mathcal{L}_{\text{data}}$ enforces reconstruction fidelity and $\mathcal{L}_{\text{HF-VSD}}$ serves as a regularization term that promotes high-frequency detail synthesis.

Given a supervised dataset containing paired structure images $I_s$, color cues $I_c$, and ground-truth telephotographs $I$, the data loss penalizes deviations between the reconstructed image $\hat{I}$ and the ground truth $I$ using a weighted combination of pixel-wise and perceptual metrics:
\begin{equation}
\label{eq:loss_data}
\mathcal{L}_{\text{data}}(\hat{I}, I)
= \mathrm{MSE}(\hat{I}, I)
+ \lambda_1 \,\mathrm{LPIPS}(\hat{I}, I),
\end{equation}
where LPIPS~\cite{zhang2018lpips} encourages perceptual similarity.

\paragraph{High-frequency variational score distillation.}
To further enhance fine-detail synthesis, we introduce a High-Frequency Variational Score Distillation (HF-VSD) loss, denoted $\mathcal{L}_{\text{HF-VSD}}$. Compared to the original VSD loss~\cite{wang2023prolificdreamer}, HF-VSD explicitly emphasizes high spatial-frequency components guided by the monochrome structure image $I_s$, while preserving low-frequency chromatic consistency from the color cue $I_c$.

The HF-VSD loss is defined as
\begin{align}
    \mathcal{L}_{\text{HF-VSD}}(\hat{I})
    =
    \mathbb{E}_{t,\boldsymbol{\varepsilon}}
    \left[
    \mathcal{L}_{\text{MSE}}
    \left(
    \boldsymbol{\omega}(t)\,
    \mathcal{F}^{-1}
    \left[
    \mathbf{h}(u,v)\odot
    \mathcal{F}
    \left[
    \Omega_0(\hat{\mathbf{z}}_t; t, \mathbf{c})
    - \Omega_{\mathbf{\phi}}(\hat{\mathbf{z}}_t; t, \mathbf{c})
    \right]
    \right]
    \right)
    \right].
\end{align}
Here, $\Omega_0$ and $\Omega_{\mathbf{\phi}}$ denote frozen and trainable Stable Diffusion models, respectively, where $\mathbf{\phi}$ indicates trainable parameters (Fig.~\ref{fig:framework}). The variable $\hat{\mathbf{z}}_t$ represents the noisy latent variable re-corrupted from the generator output $\hat{\mathbf{z}}$, $t$ is the diffusion timestep, and $\boldsymbol{\omega}(t)$ is a timestep-dependent weighting function.

The frequency reweighting is controlled by a 2D high-pass filter $\mathbf{h}(u,v)$ defined as
\begin{equation*}
\mathbf{h}(u,v)
=
\mathrm{clip}
\left[
\left(
\left(\frac{\alpha u}{R}\right)^2
+
\left(\frac{\alpha v}{R}\right)^2
\right)^{\gamma}
+
\beta,
\,
0,
\,
1
\right],
\end{equation*}
where $(u,v)$ denote spatial frequency coordinates, $R$ is the half-maximum frequency, $\alpha$ and $\gamma$ control the frequency scaling, and $\beta$ is a bias term. This design amplifies high-frequency components in the latent space, encouraging the one-step diffusion model $\Omega$ to distill fine-detail generation capabilities from the pre-trained diffusion prior.

\paragraph{Optimization strategy.}
Following the standard VSD framework, the generator's parameters $\boldsymbol{\theta}$ and the HF-VSD's trainable parameters $\mathbf{\phi}$ are updated alternatively via Eq.~\ref{eq:problem} and by minimizing the following difference loss, respectively:
\begin{align}
    \mathcal{L}_{\text{diff}}(\hat{I})
    =
    \mathbb{E}_{t,\boldsymbol{\varepsilon}, \hat{\mathbf{z}}_t}
    \left[
    \mathcal{L}_{\text{MSE}}
    \big(
    \Omega_{\boldsymbol{\phi}}(\alpha_t \hat{\mathbf{z}}_t + \beta_t \boldsymbol{\varepsilon}; t, \mathbf{c}),
    \boldsymbol{\varepsilon}
    \big)
    \right],
\end{align}
where $\alpha_t$ and $\beta_t$ are noise scheduling coefficients and $\boldsymbol{\varepsilon}$ is Gaussian noise.

\section{Experimental results}\label{sec:results}
\subsection{Optical design and fabrication}\label{sec:lens_optimize}
We construct the MetaTele prototype following the optical schematics in Fig.~\ref{fig:optical-model}, using an off-the-shelf refractive objective lens with a custom-designed metasurface eyepiece. The target total track length (TTL) and effective focal length (EFL) are around 14~mm and 30~mm, respectively.

To meet these specifications, we select a Thorlabs achromatic doublet (AC050-008-A) with a diameter of 5 mm and a focal length of 7.5 mm as the objective lens. Among commercially available options with comparable focal lengths, it provides the largest entrance pupil, enabling a sufficiently low f-number. We select the achromatic doublet, instead of the singlet, to suppress dispersion in the hybrid refractive–metasurface system. 

\paragraph{Metasurface design.} Given the objective lens, we explore using optimization to design the metasurface eyepiece. The design problem can be formulated as:
\begin{equation}
\begin{aligned}
        \arg\min_\phi\; &l(\phi(\mathbf{x};\lambda_0),s, m), \\
        &\text{s.t.}\quad
         \text{strehl}(\lambda_0, \theta) > C, \
        f_c(\lambda_0, \theta) \ge f_N, \ s + m \le s_M, \forall \theta \in [0, \theta_{\text{max}}]
    \label{eq:opt}
\end{aligned}
\end{equation}
which optimizes the telephoto ratio $l$ while satsifying the minimal Strehl ratio $C$ and cutoff frequency $f_N$ for all incident angles $\theta\in [0, \theta_{\text{max}}]$, and the separation, $s$ and $m$ as indicated in Fig.~\ref{fig:optical-model}, satisfy the spatial constraints. The optimization variables include the metasurface phase profile at the design wavelength $\lambda_0$, $\phi(\mathbf{x};\lambda_0)$, and the separations $s$ and $m$.

We perform the optimization at $\lambda_0=532~\text{nm}$, $C = 0.13$, $f_N = 250~\text{lp/mm}$, and $\theta_{\text{max}} = 3^\circ$. The minimal Strehl ratio $C$ is set relatively low, as post-processing can partially compensate for image blur. The optimization variables include the position of the objective and the metasurface phase parameters. The metasurface phase profile is parameterized using radially symmetric even-order polynomials up to the fourteenth order:
\begin{equation}
\label{eq:quad_phase_profile}
    \phi(\mathbf{x},\lambda_0) = \frac{2\pi}{\lambda_0}\sum_{i=1}^{7} c_i \|\mathbf{x}\|^{2i}.
\end{equation}
We utilize Code V to carry out the optimization. Interestingly, the converged metasurface phase profile $\Tilde{\phi}(\mathbf{x},\lambda_0)$ closely resembles a quadratic function, corresponding to a diverging lens:
\begin{align}
    \Tilde{\phi}(\mathbf{x},\lambda_0) \approx -\frac{2\pi}{\lambda_0}\frac{\|\mathbf{x}\|^2}{2f}, 
    \label{eq:quad_phase}
\end{align}
with a focal length $f=-2~\text{mm}$. The exact converged coefficients $\{c_i\}_{i=1}^7$ and the benefits of the converged quadratic phase profile w.r.t. other phase profiles are provided in Supplement 1. Consequently, we adopt the quadratic phase profile shown in Eq.~\ref{eq:quad_phase} for the metasurface design, which yields nearly identical performance in terms of the modulation transfer function (MTF) according to our simulation. 

\paragraph{Fabrication.}
The metasurface is modeled as a two-dimensional array of uniform nanocells arranged on a regular grid $G$. Each nanocell $(m,n)\in G$ comprises a single nanostructure that locally modulates the transmitted wavefront. In this work, the nanocell size is fixed at $300~\mathrm{nm}\times300~\mathrm{nm}$, and each nanocell contains a centered Silicon Nitride cylindrical nanopillar with a fixed height of $775~\mathrm{nm}$. The metasurface is therefore parameterized by the nanopillar radius $r(m,n)$, which determines the complex modulation function exerted on the wavefront.

We model the modulation function of each nanocell as  
\begin{equation}
C(m,n) = T(m,n)\,e^{j\phi(m,n)},
\end{equation}
where $T(m,n)$ and $\phi(m,n)$ denote the transmittance and phase delay at location $(m,n)$, respectively. For the centered nano-cylinder geometry employed here, the modulation function is fully determined by the nanopillar radius,
\begin{equation}
C(m,n) = f(r(m,n)).
\end{equation}

Direct evaluation of $f(\cdot)$ via full-wave simulation is computationally expensive. To enable efficient metasurface synthesis, we emulate $f(\cdot)$ using a precomputed look-up table (LUT) generated with the Lumerical FDTD solver. The LUT consists of a dense set of mappings between nanopillar radius and modulation function,
\begin{equation}
\{ r_i \rightarrow C_i = T_i e^{j\phi_i} , \quad i = 1,2,\dots,N \}.
\end{equation}

Given a target phase profile $\phi(\mathbf{x},\lambda_0)$ at the design wavelength $\lambda_0$, the desired modulation function at each nanocell center position $\mathbf{x}_{m,n}$ is defined as
\begin{equation}
C(m,n) = e^{j\phi(\mathbf{x}_{m,n},\lambda_0)}.
\end{equation}
The nanopillar radius assigned to nanocell $(m,n)$ is then obtained by solving the following discrete optimization problem:
\begin{equation}
r(m,n) = \underset{\{r_i,\, i=1,\dots,N\}}{\arg\min}\; \big| \angle C_i - \angle C(m,n) \big|.
\label{eq:lutfitting}
\end{equation}
This procedure uniquely determines the nanopillar radius at each nanocell location and yields the complete metasurface layout.

The nanopillar radii are constrained to the range of $50$--$130~\mathrm{nm}$ based on fabrication limits, while the chosen pillar height enables full $2\pi$ phase coverage. The unit-cell response is verified to be angle-insensitive for incident angles from $0^\circ$ to $20^\circ$. Optical and scanning electron microscopy images of a fabricated metasurface are shown in Fig.~\ref{fig:systems}c--d. The metasurface fabrication processes closely follow those reported by Brookshire \emph{et al.}~\cite{Brookshire:24}.

\begin{figure}
\includegraphics[width=\linewidth]{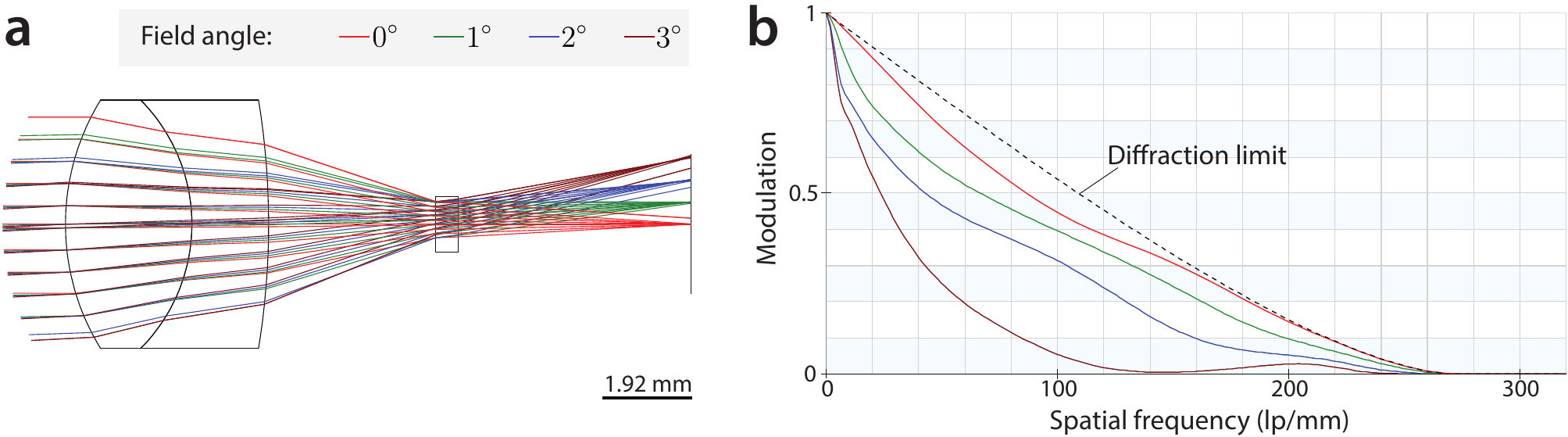}
\caption{Optical performance of the MetaTele prototype in simulation.
(a) Ray-tracing diagram for parallel incident rays at different field angles, illustrating that the optical assembly operates in a Galilean-telescope configuration.
(b) Modulation transfer functions (MTFs) for the corresponding field angles, color-coded to match (a). The system achieves near–diffraction-limited performance on axis.}
\label{fig:system}
\end{figure}

\paragraph{System characterization in simulation.}
Fig.~\ref{fig:system} visualizes the imaging performance of the proposed optical design in Code V. It achieves an Effective Focal Length (EFL) of 30 mm with a Total Track Length (TTL) of 13.2 mm, resulting in a compact telephoto ratio of 0.44.

\subsection{Simulation analysis of raw measurements}

In this section, we analyze the quality of the structure image and the color cue produced by the proposed MetaTele system, and compare them with alternative design choices and prior work through simulation. These studies focus on quantifying the imaging quality of the optics alone, \emph{without} any post-processing.

\paragraph{Comparison with previous works.}

\begin{figure}
    \centering
    \includegraphics[width=0.8\linewidth]{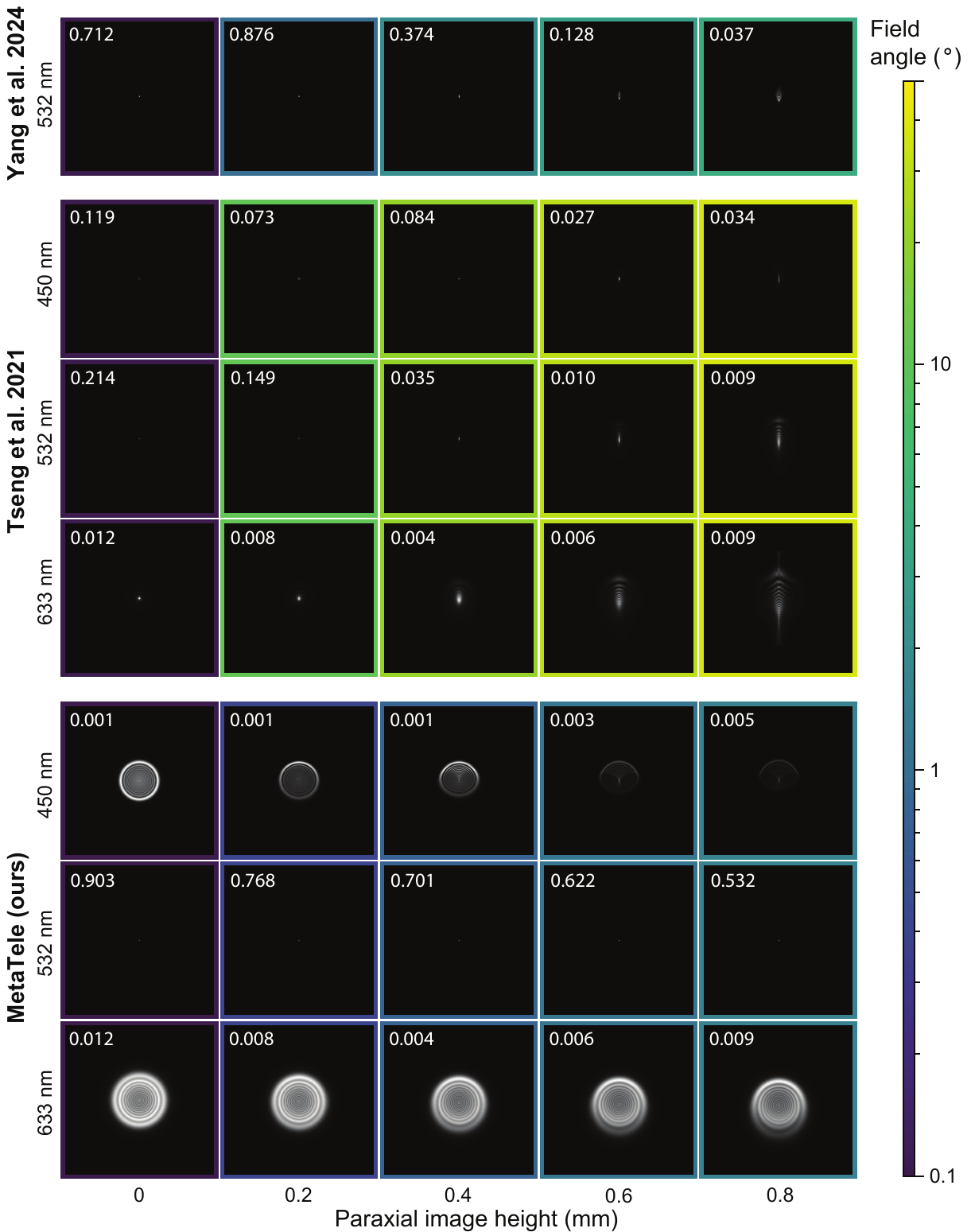}
    \caption{Simulated PSFs of MetaTele and prior metasurface imaging systems~\cite{parfocal}. Since these systems operate with different fields of view (FoVs), we compare their PSFs at identical paraxial image heights, i.e., the sensor-plane locations corresponding to the PSF centroids. The inset numbers report the Strehl ratios. Bounding box colors indicate the corresponding field angles. MetaTele deliberately sacrifices broadband PSF sharpness to optimize image quality at the design wavelength (532 nm), achieving the highest on-design Strehl ratio. For a given sensor area, MetaTele covers a substantially smaller FoV, thereby demonstrating stronger optical magnification and telephoto capability.}
    \label{fig:aoi_sweep}
\end{figure}
To rigorously characterize the spatial and field-dependent behavior of the MetaTele optical system, we synthesize the PSFs using Code V and evaluate the Strehl ratio across field angles and wavelengths, as shown in Fig.~\ref{fig:aoi_sweep}. Unlike conventional achromatic designs that aim for uniform broadband performance, MetaTele intentionally prioritizes diffraction-limited operation at the design wavelength of the structure image (532 nm), thereby maximizing structural detail in the captured structure image.

We compare our system against recent metasurface-based imagers, Yang \emph{et al.}~\cite{parfocal} and Tseng \emph{et al.}~\cite{Tseng2021NeuralNanoOptics}. To ensure a fair comparison across systems with different EFLs, we evaluate PSFs at uniform paraxial image heights (i.e., sensor-plane locations), rather than matching field angles. The baseline systems exhibit rapid off-axis degradation, with focal spots broadening significantly even at modest image heights. In contrast, the 532 nm PSFs of MetaTele remain compact across the full sensor extent, yielding the highest Strehl ratios and enabling structure images with consistently high and spatially uniform visual quality across the entire field of view.

\subsection{System calibration and characterization}

\begin{figure}[ht]
    \centering
    \includegraphics[width=\linewidth]{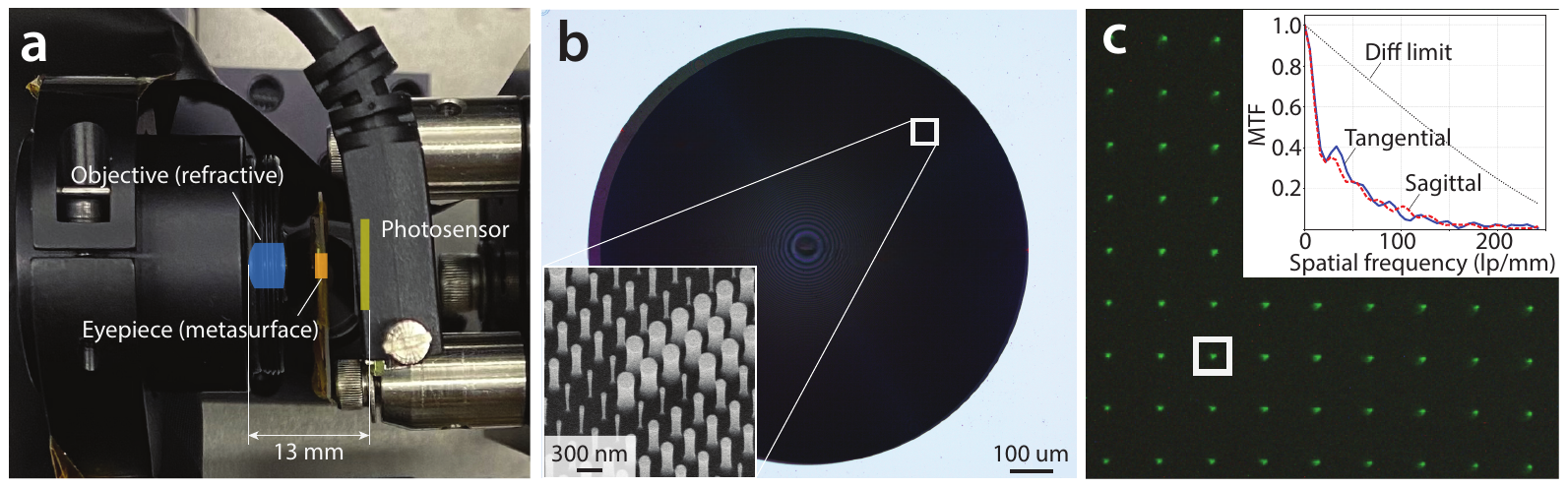}
    \caption{(a) MetaTele optical assembly. The system comprises a Thorlabs AC050-008-A-ML objective lens (f = 7.5 mm, Ø5 mm), a custom-designed metasurface serving as the eyepiece, and a Basler daA3840-45uc RGB (no-mount) sensor as the photosensor. Each component is mounted on multi-axis precision stages to enable accurate alignment and calibration. A 532 nm spectral filter with 10 nm FWHM bandwidth can be inserted into the optical path to capture the structure image. (b) Fabricated metasurface. Optical microscope image of a representative metasurface sample. Inset: Scanning electron microscope (SEM) image of a zoomed-in region at 13,000× magnification. (c) Measured PSFs. Experimentally measured PSFs corresponding to the structure image. Inset: Modulation transfer function (MTF) computed from the PSF at a representative field angle. A related version of this figure appears in the conference paper~\cite{weligampola2025diffusion}.}
    \label{fig:systems}
\end{figure}

\paragraph{Calibration. }
The fully assembled system is shown in Fig.~\ref{fig:systems}a. The objective lens, metasurface, and photosensor were mounted independently in precision stages with 5-axis control (xyz-translation and tip-tilt). First, angular alignment was performed by directing a laser beam through the system and adjusting each stage until the back-reflections coincided with the incident beam, ensuring parallelism between the component planes. The lateral alignment was then achieved by imaging a point grid displayed on a planar target. The objective and eyepiece were translated perpendicular to the optical axis to minimize aberrations at the image center, thereby aligning the optical axes of the elements. Finally, the axial position of the eyepiece were adjusted to focus the target on the sensor. Additionally, we show in Supplement 1 that the hybrid assembly is more robust to lateral/longitudinal decenter than purely refractive systems.

\paragraph{Characterization.} Fig.~\ref{fig:systems}c analyzes the quality of PSF of the real-assembled system by imaging a 2D array of a dotted pattern. According to the measurements, the PSFs remain spatially invariant within the field of view. A typical PSF achieves a cutting-off frequency, defined as MTF $>$ 0.2, of about 50 lp/mm. The real-measured PSF is wider than the simulated ones due to a combination factors. First, the simulation assumed a monochromatic source, whereas the experiment used a 10~nm bandwidth filter; the high chromatic dispersion of the metasurface introduces a focal shift for off-center wavelengths, creating a halo around the central peak. Second, fabrication imperfections in the nanopillars increase background residual lights, reducing the Strehl ratio. Third, slight misalignment between the refractive lens and the metasurface likely introduces coma and astigmatism. More detailed characterization and analysis of the system, including aberration, system robustness to assembly error, are provided in the supplementary.

\subsection{Dataset collection}
\label{sec:dataset}

To fine-tune the computational model and benchmark the imaging performance, we collected a large dataset using the MetaTele prototype we built. The dataset consists of 2,650 scenes, each including a front-parallel displayed image from the Flickr2k dataset \cite{young_etal_2014_flickr30k}. We use the MetaTele prototype to capture a structure image and a color cue for each scene. Sample images of the dataset are shown in Fig.~\ref{fig:qualitative_comparison_algo}. 

We built an automatic data acquisition system to collect the benchmark dataset. The system utilizes a high-resolution display to automatically broadcast pictures randomly sampled from Flickr2k~\cite{young_etal_2014_flickr30k} at a distance of 26.5 inches from the MetaTele prototype. We program the photosensor of the MetaTele prototype to capture a structure image and a color cue for each displayed picture. The structure image is captured with a 10 nm FWHM bandpass filter centered at 532 nm, with a 1-second exposure time and 0 gain. The color cue is captured without the bandpass filter at a 0.1-second exposure time and 0 gain. The dataset consists of 2,650 tuples of structure image, color cue, and ground truth image. 

\subsection{Comparison of computational models}

\begin{figure*}
\centering
\begingroup
\setlength{\tabcolsep}{0pt} 
\renewcommand{\arraystretch}{0.01} 
\resizebox{\linewidth}{!}{
\begin{tabular}{ *{12}{>{\centering\arraybackslash}m{2.2cm}} }
    \includegraphics[width=\linewidth]{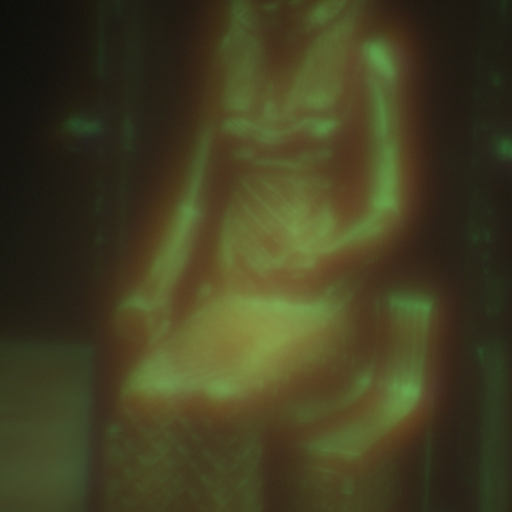} & 
    \includegraphics[width=\linewidth]{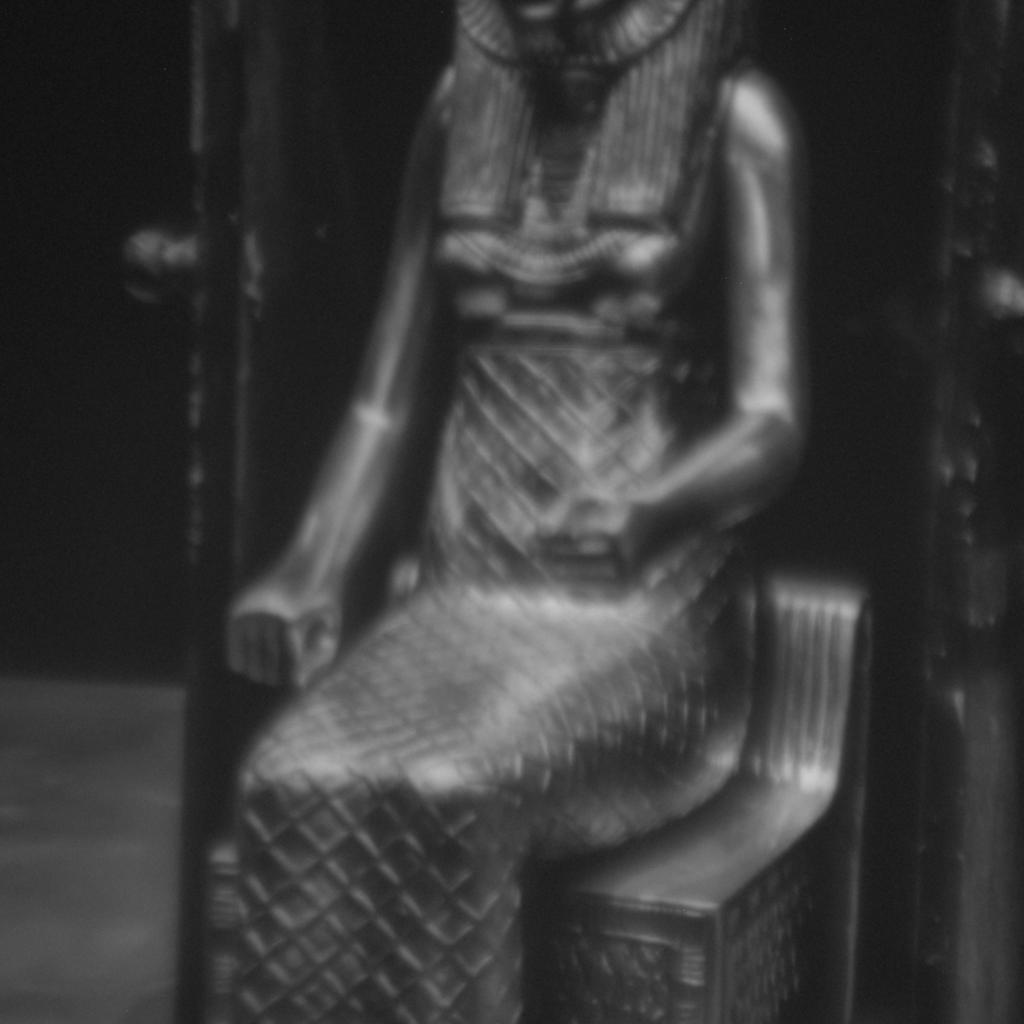} &
    \includegraphics[width=\linewidth]{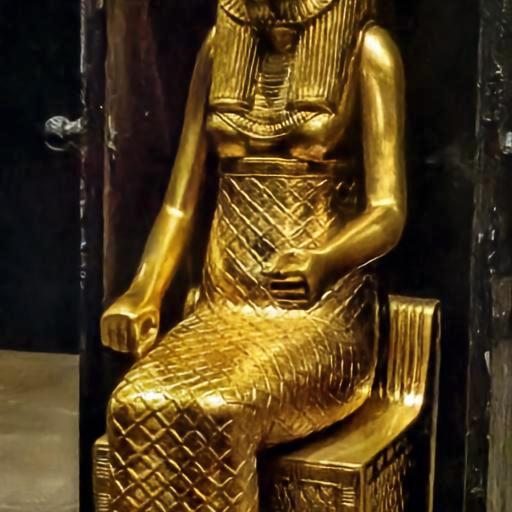} &
    \includegraphics[width=\linewidth]{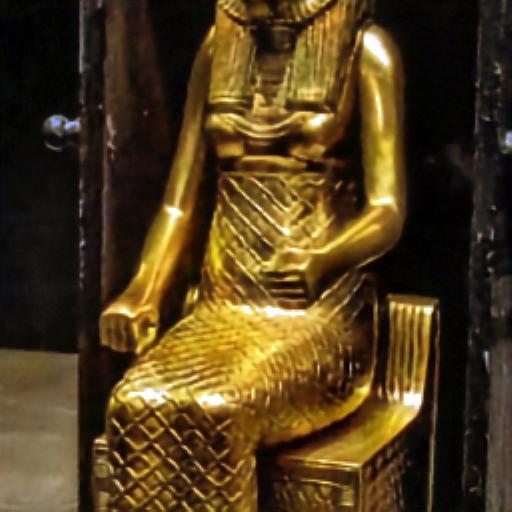} &
    \includegraphics[width=\linewidth]{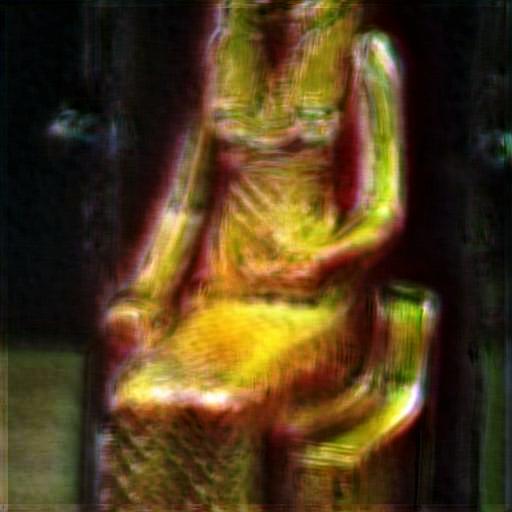} &
    \includegraphics[width=\linewidth]{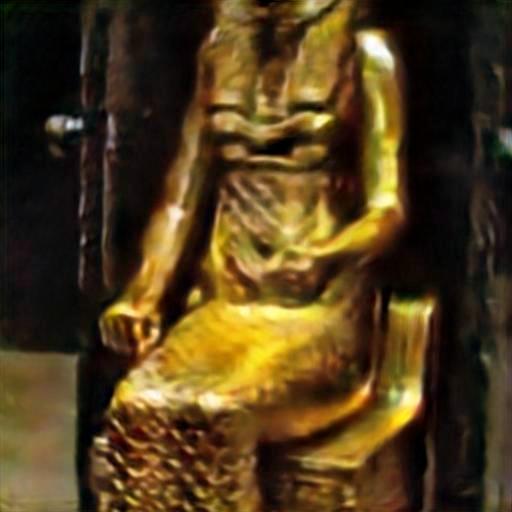} &
    \includegraphics[width=\linewidth]{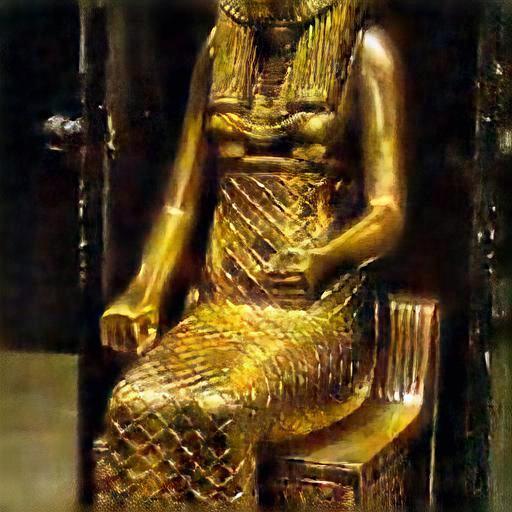} &
    \includegraphics[width=\linewidth]{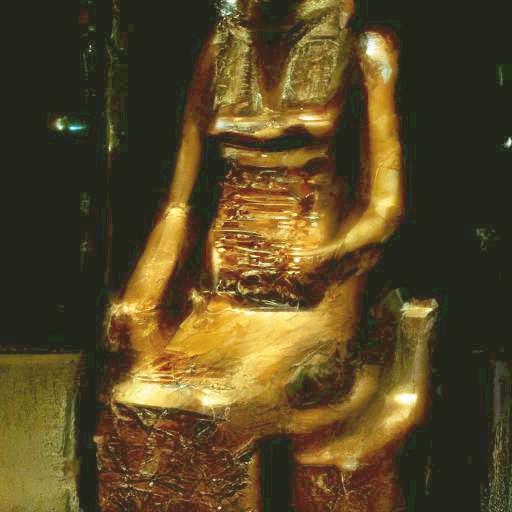} &
    \includegraphics[width=\linewidth]{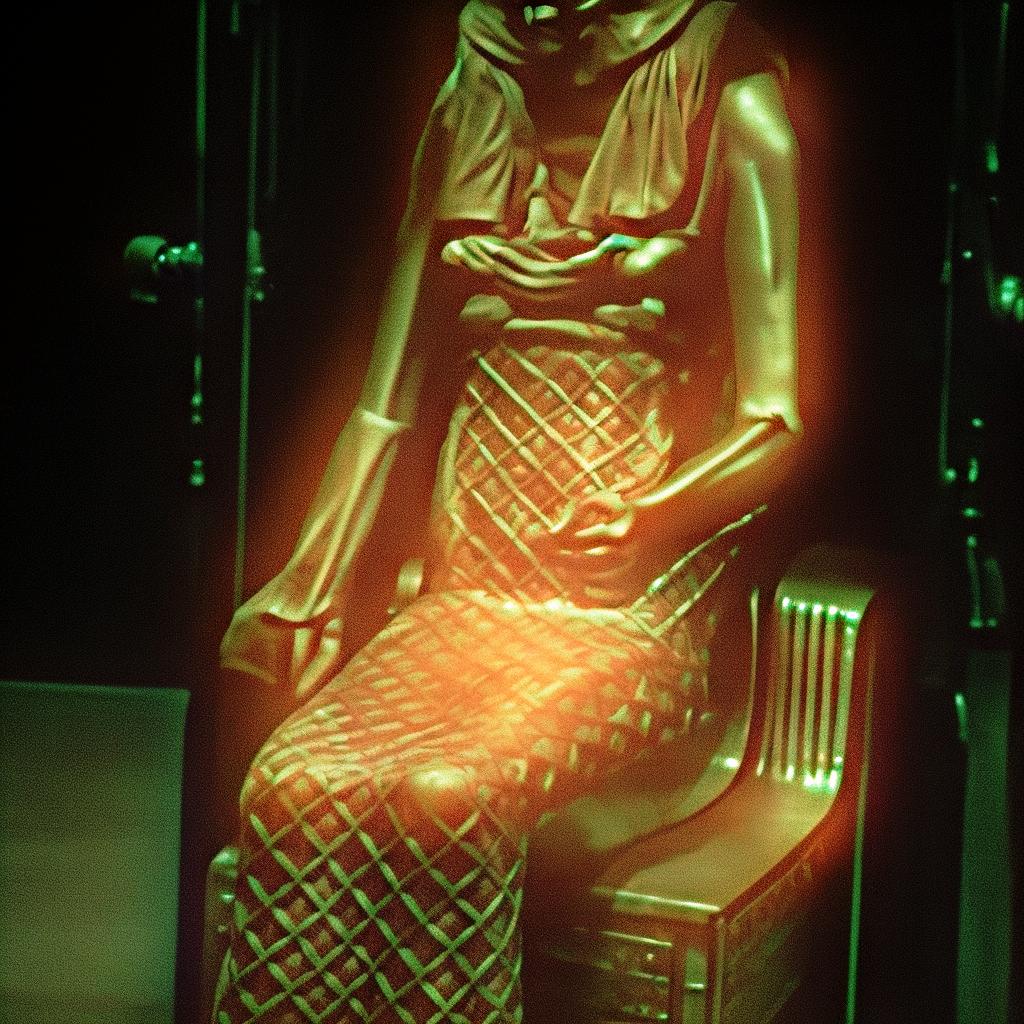} &
    \includegraphics[width=\linewidth]{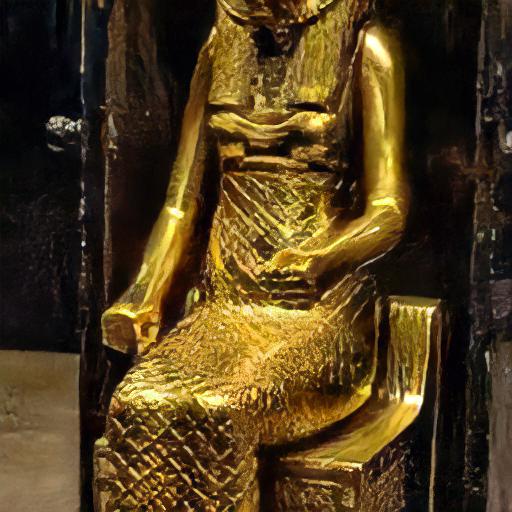} &
    \includegraphics[width=\linewidth]{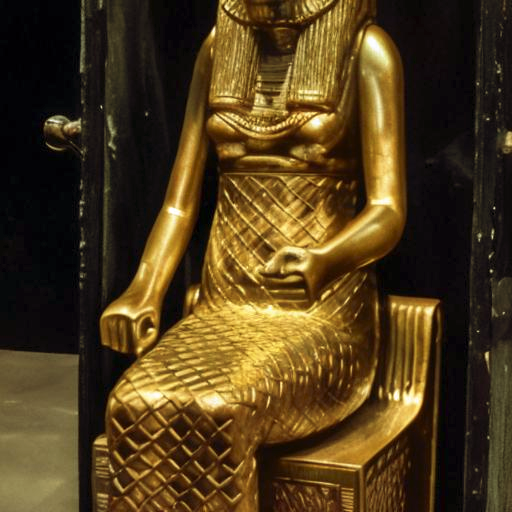} &
    \includegraphics[width=\linewidth]{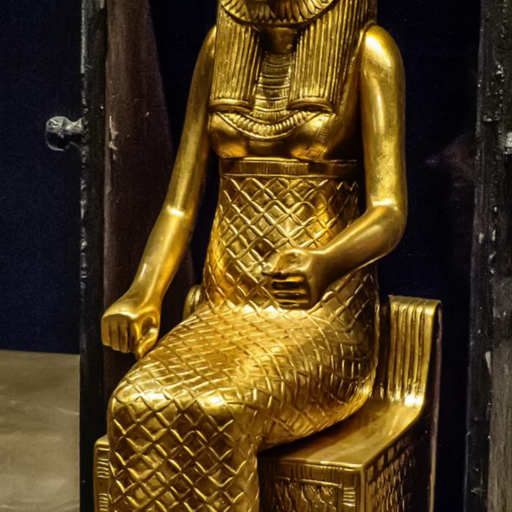} \\
    
    \includegraphics[width=\linewidth]{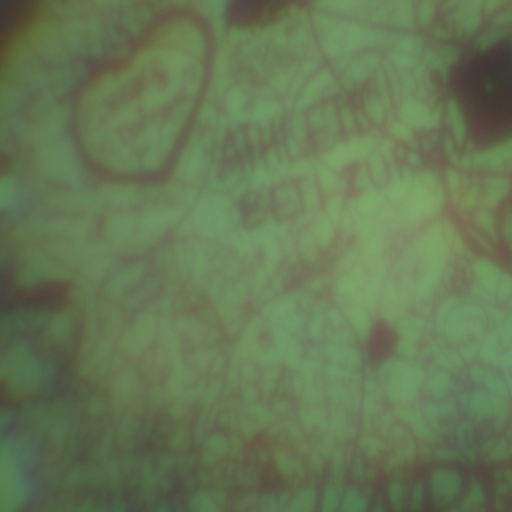} & 
    \includegraphics[width=\linewidth]{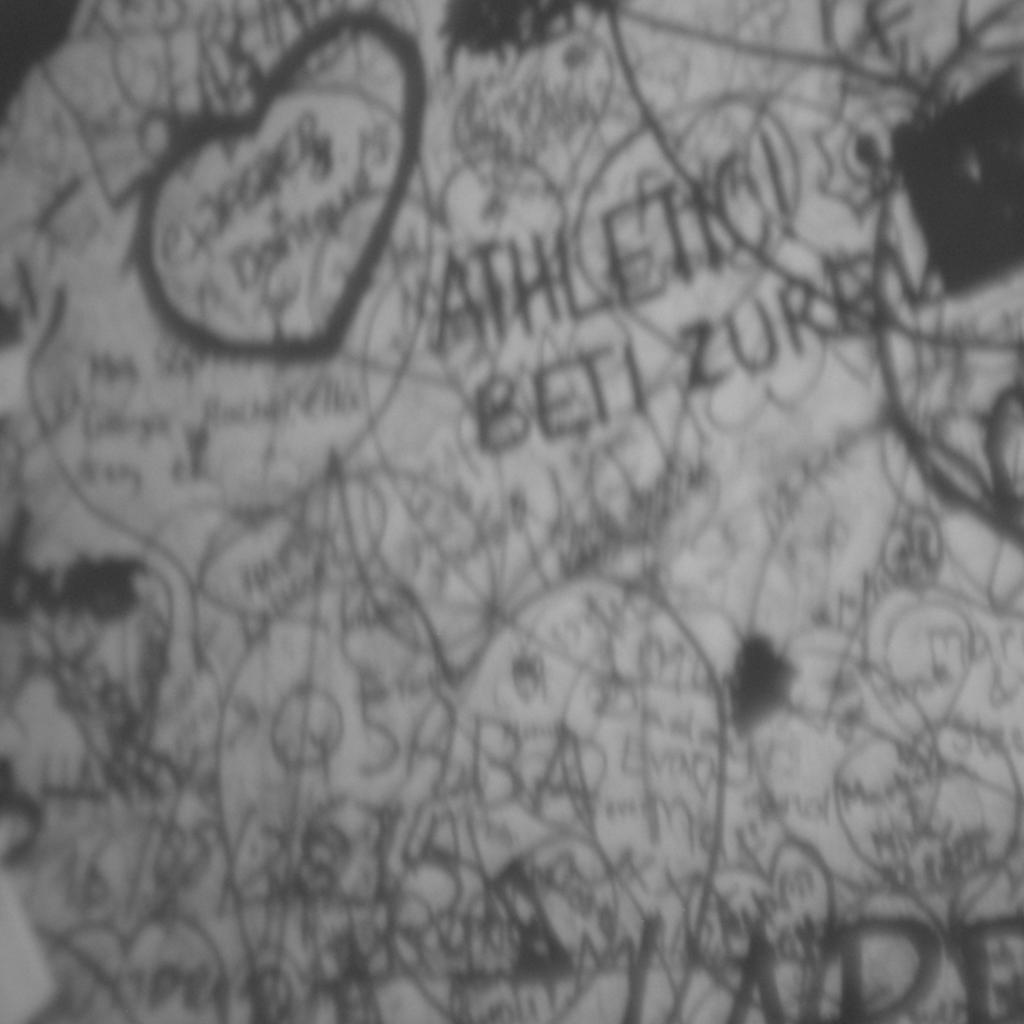} &
    \includegraphics[width=\linewidth]{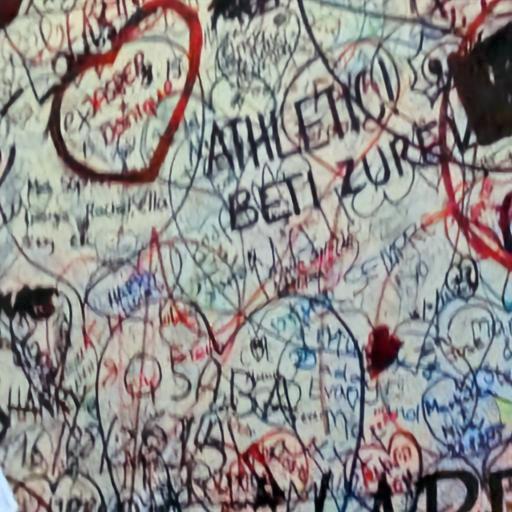} &
    \includegraphics[width=\linewidth]{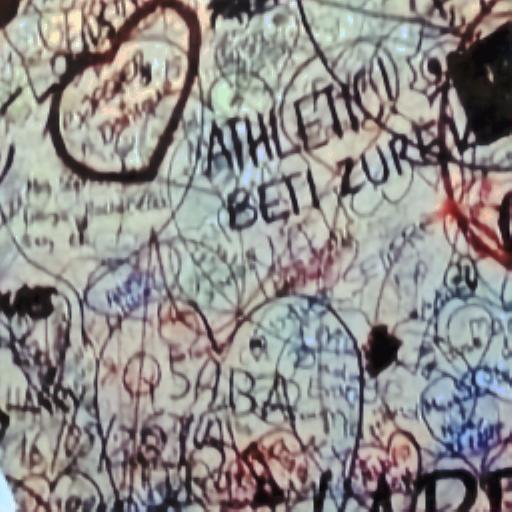} &
    \includegraphics[width=\linewidth]{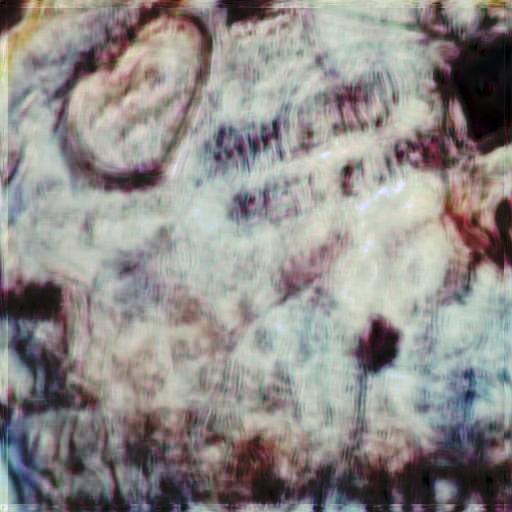} &
    \includegraphics[width=\linewidth]{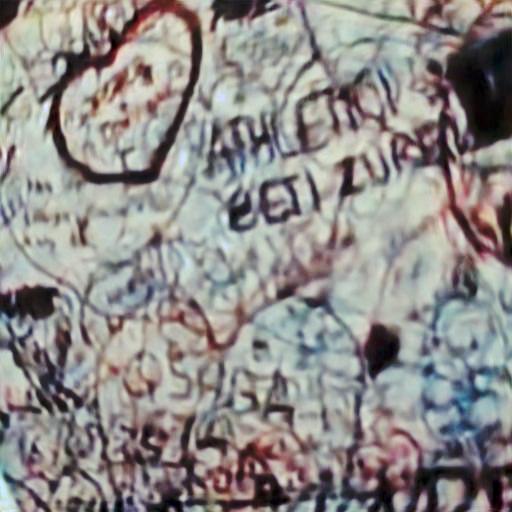} &
    \includegraphics[width=\linewidth]{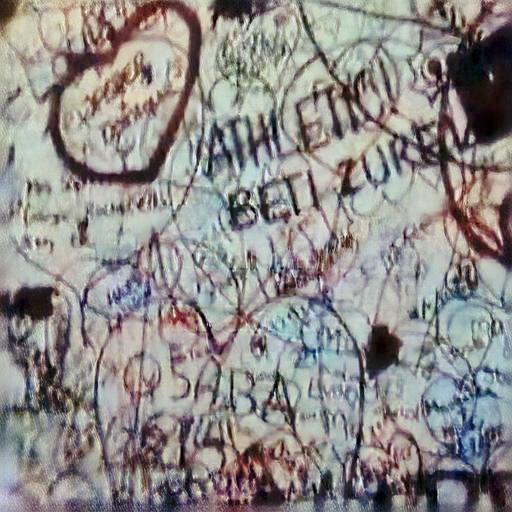} &
    \includegraphics[width=\linewidth]{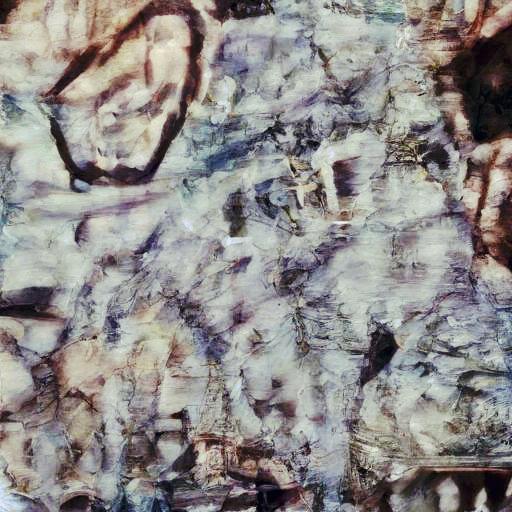} &
    \includegraphics[width=\linewidth]{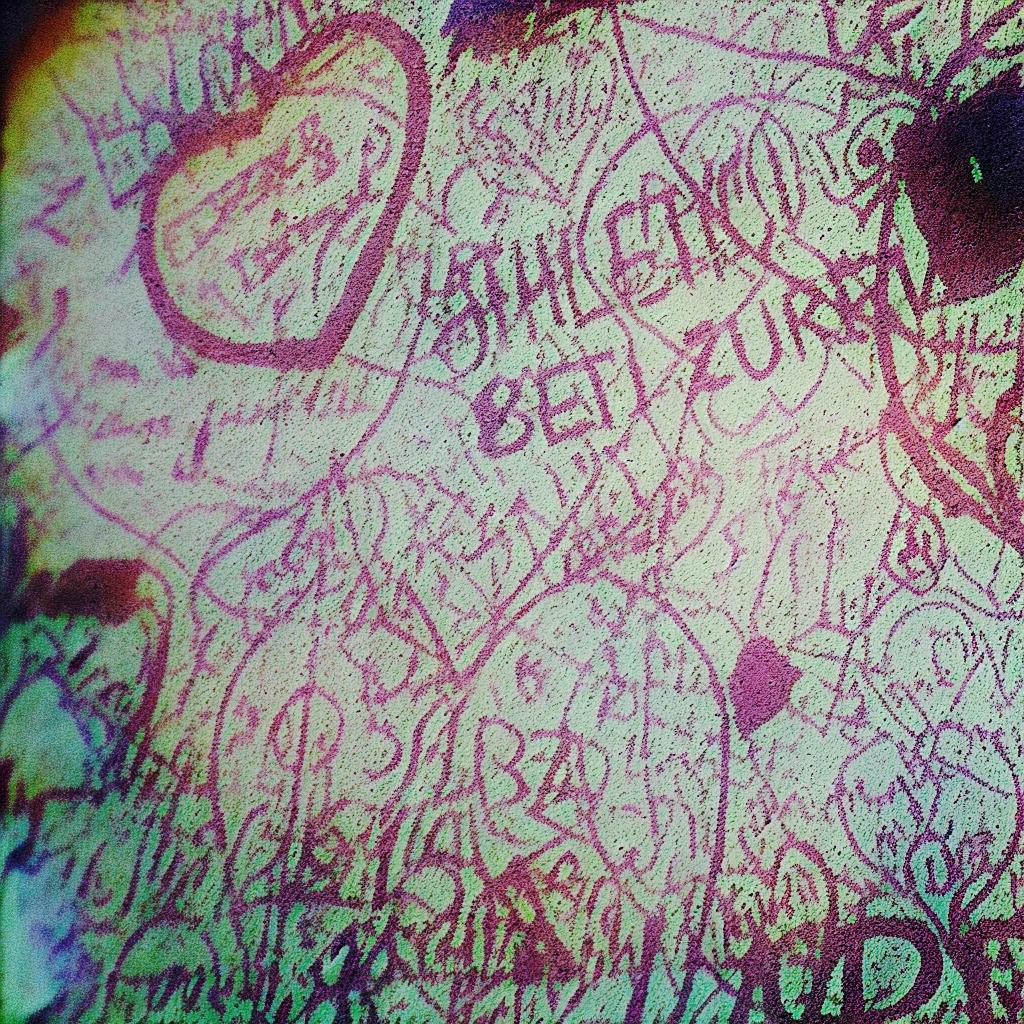} &
    \includegraphics[width=\linewidth]{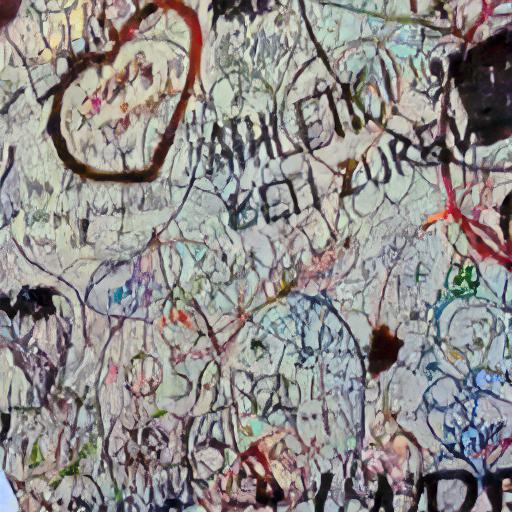} &
    \includegraphics[width=\linewidth]{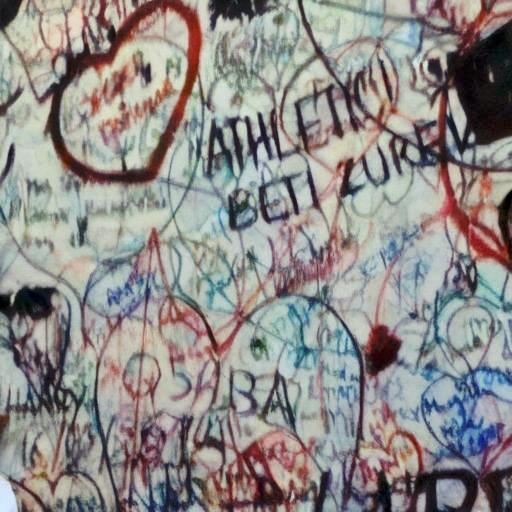} &
    \includegraphics[width=\linewidth]{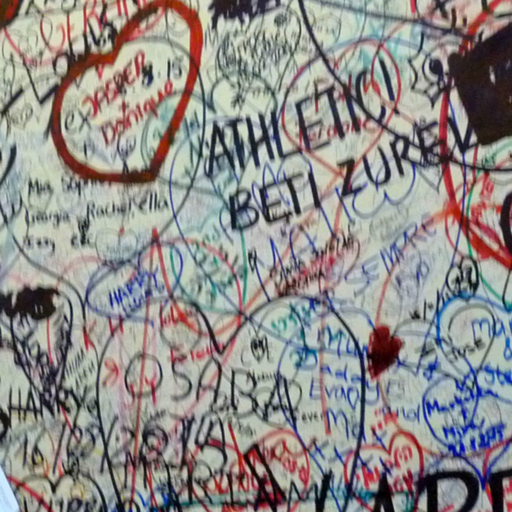} \\
    
    \includegraphics[width=\linewidth]{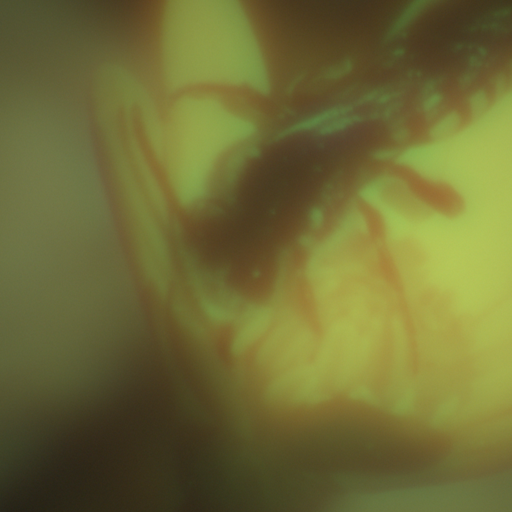} & 
    \includegraphics[width=\linewidth]{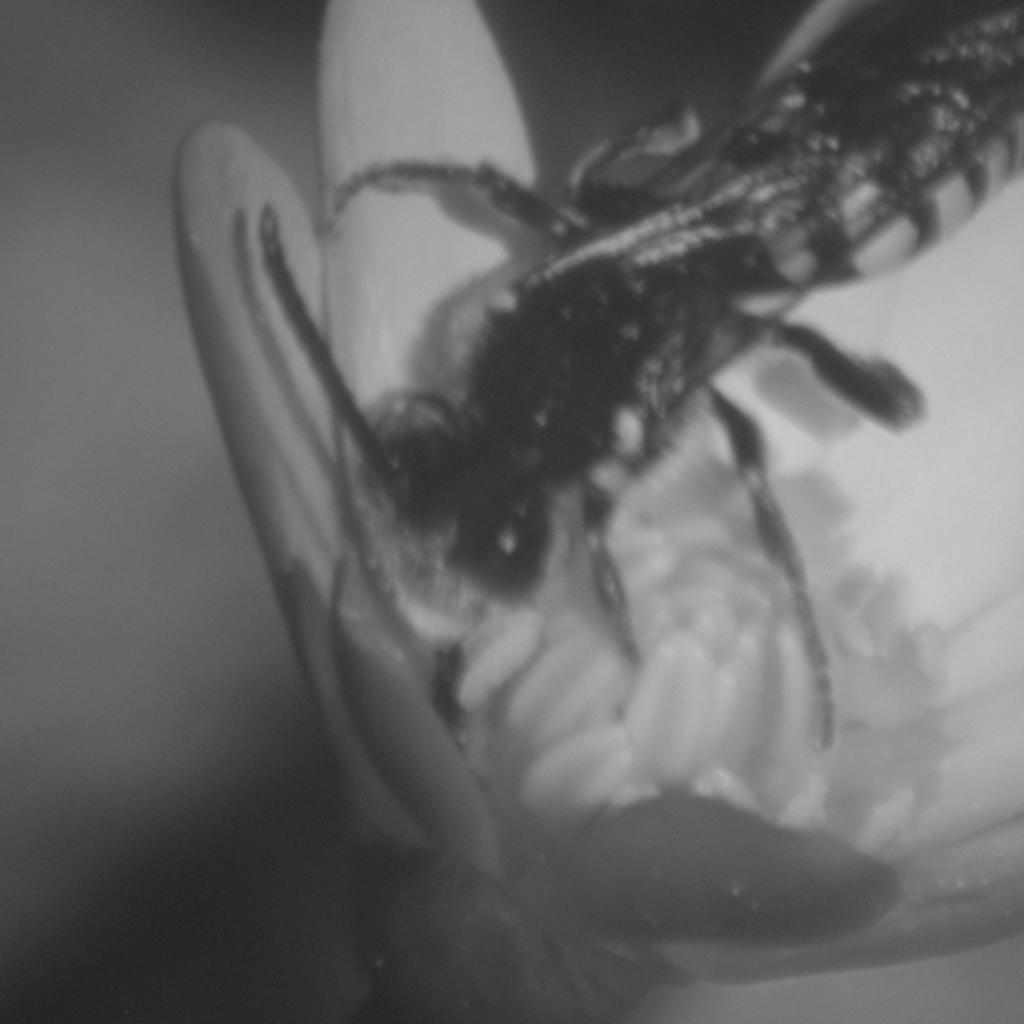} &
    \includegraphics[width=\linewidth]{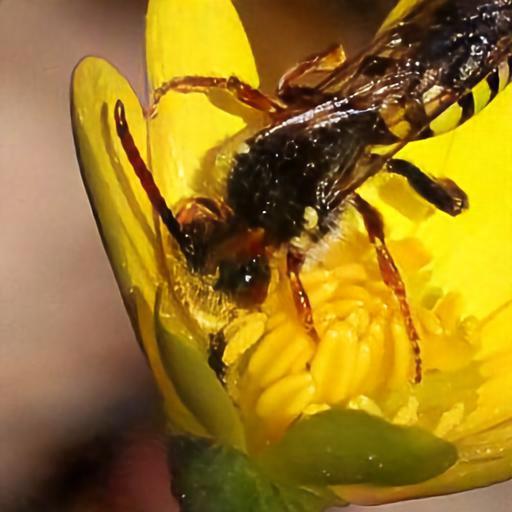} &
    \includegraphics[width=\linewidth]{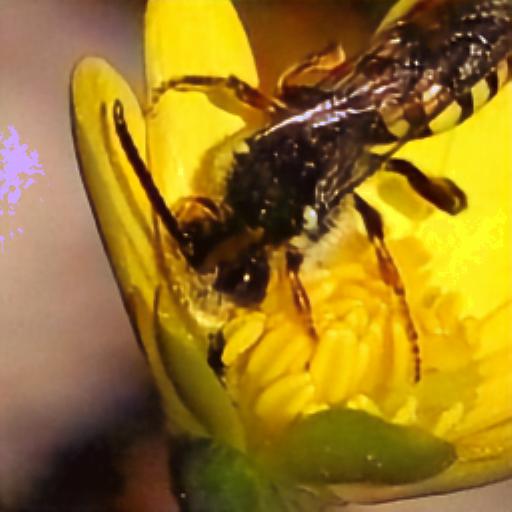} &
    \includegraphics[width=\linewidth]{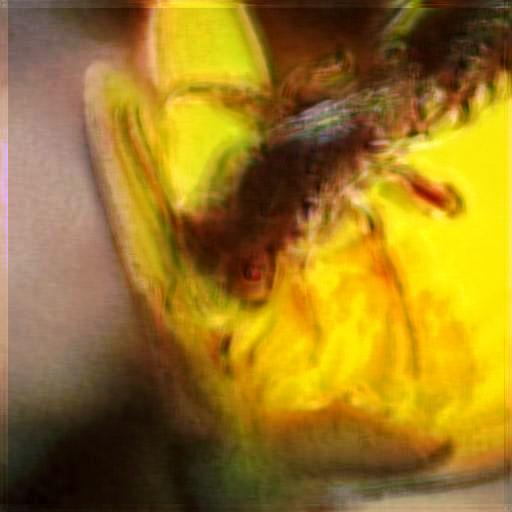} &
    \includegraphics[width=\linewidth]{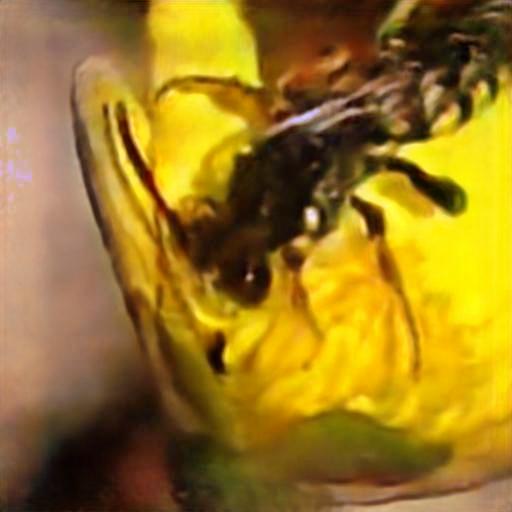} &
    \includegraphics[width=\linewidth]{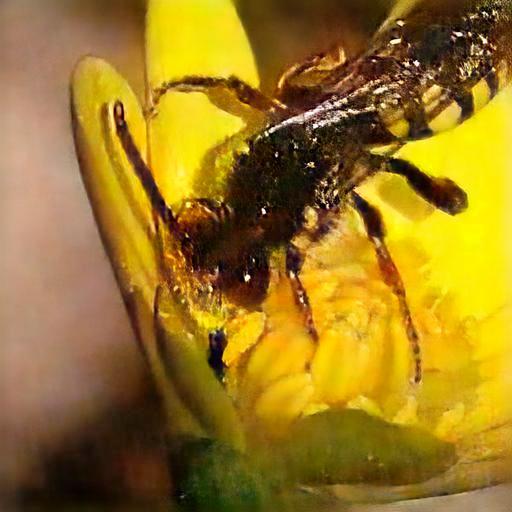} &
    \includegraphics[width=\linewidth]{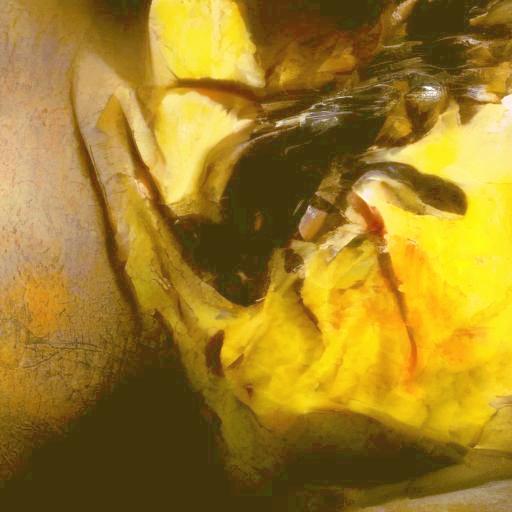} &
    \includegraphics[width=\linewidth]{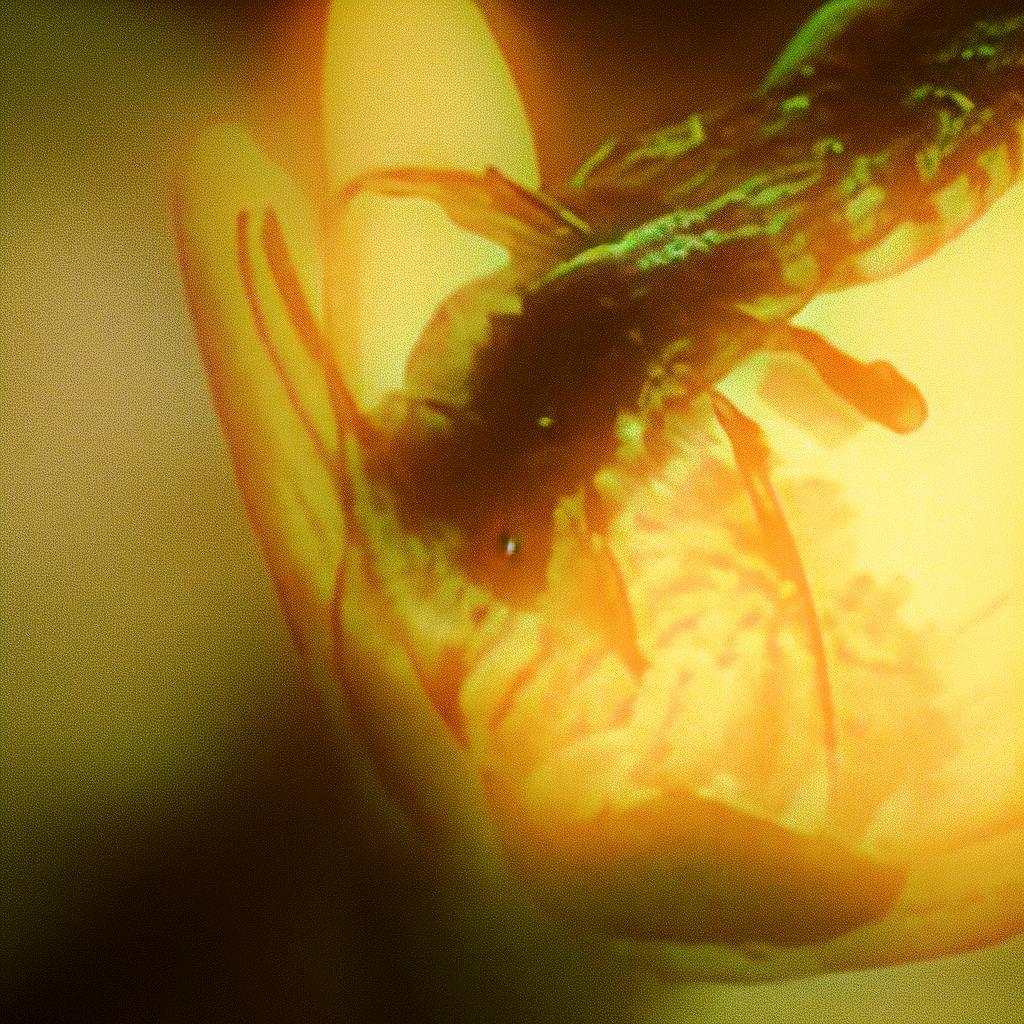} &
    \includegraphics[width=\linewidth]{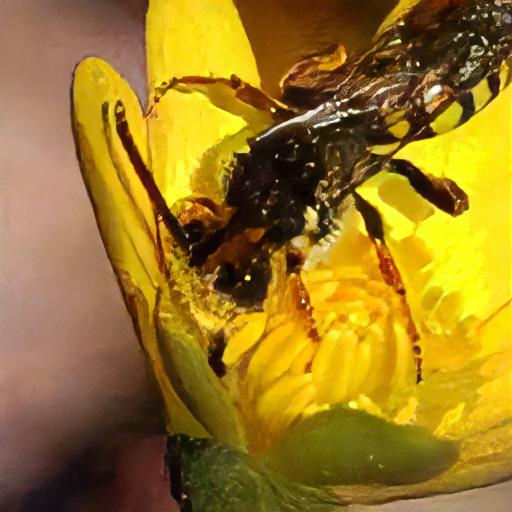} &
    \includegraphics[width=\linewidth]{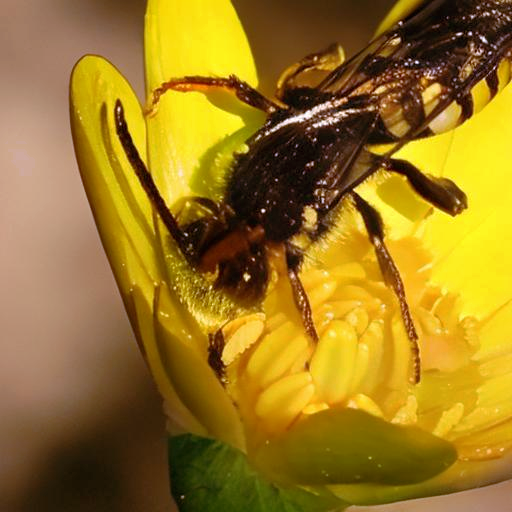} &
    \includegraphics[width=\linewidth]{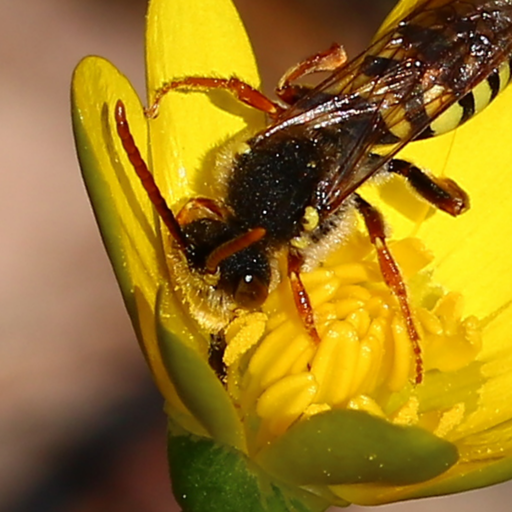} \\
    
    \includegraphics[width=\linewidth]{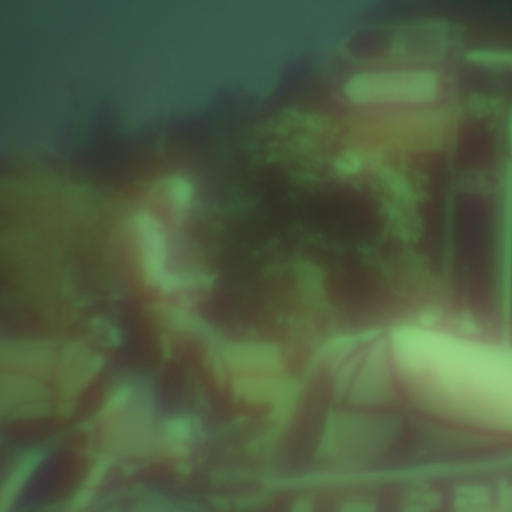} & 
    \includegraphics[width=\linewidth]{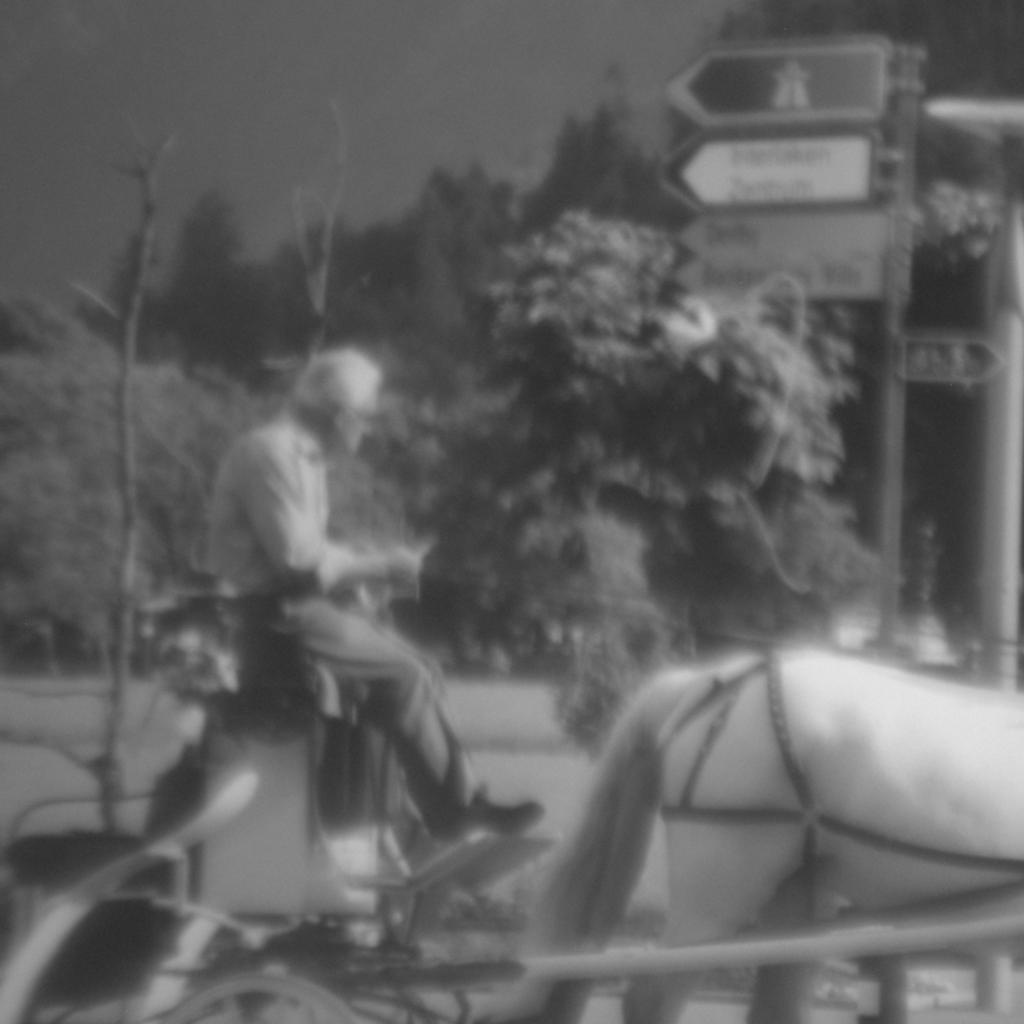} &
    \includegraphics[width=\linewidth]{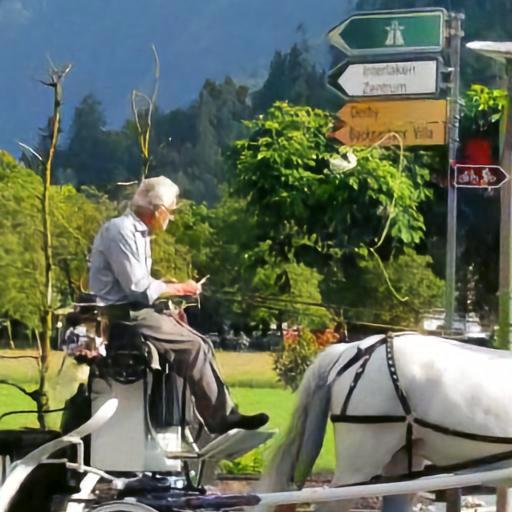} &
    \includegraphics[width=\linewidth]{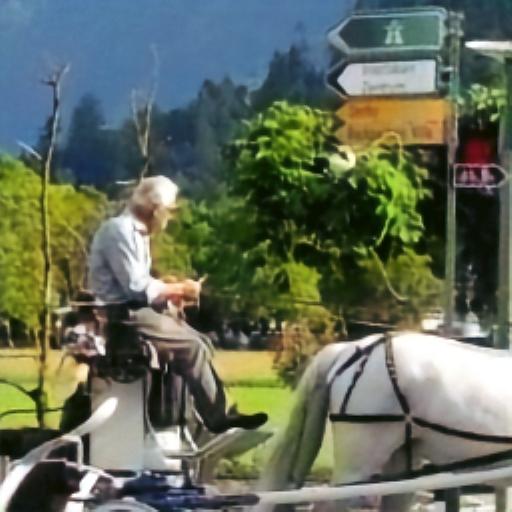} &
    \includegraphics[width=\linewidth]{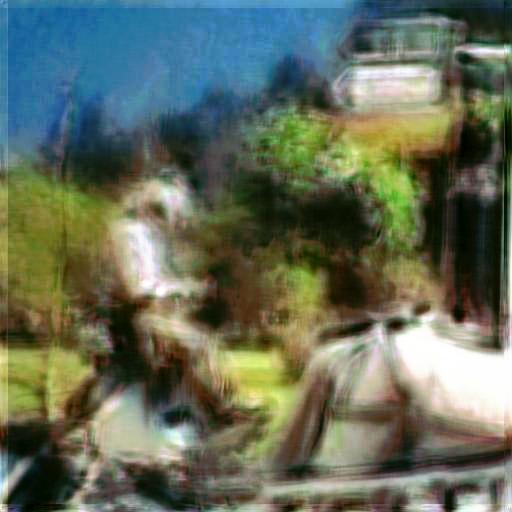} &
    \includegraphics[width=\linewidth]{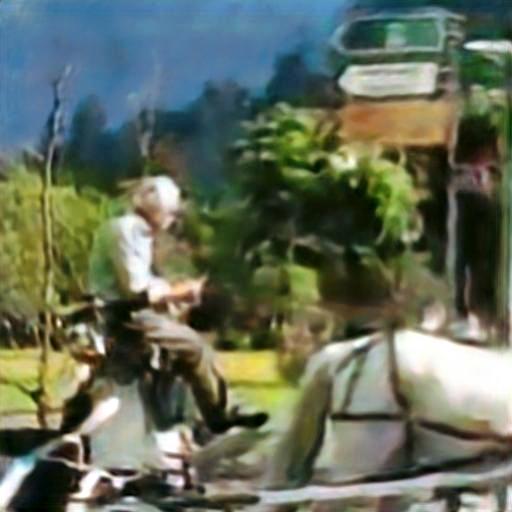} &
    \includegraphics[width=\linewidth]{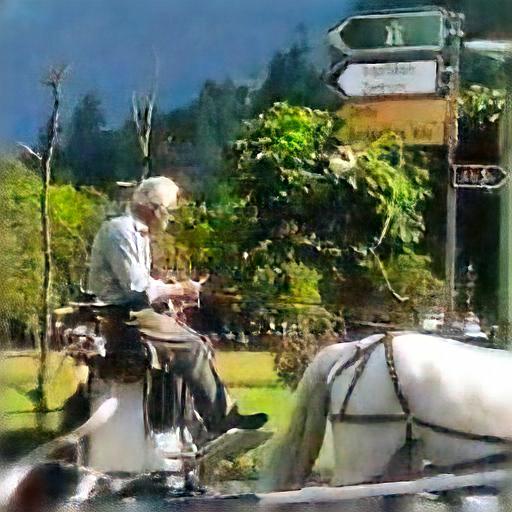} &
    \includegraphics[width=\linewidth]{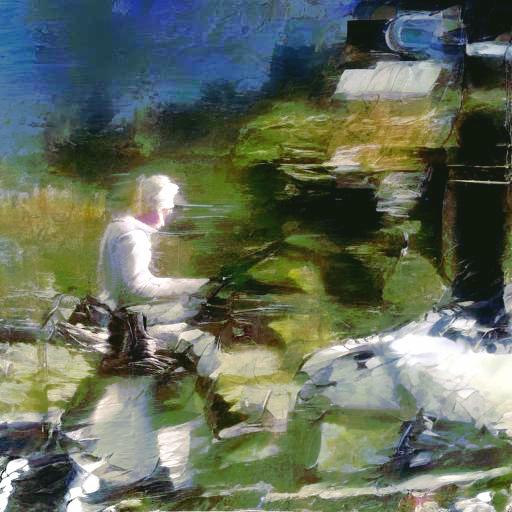} &
    \includegraphics[width=\linewidth]{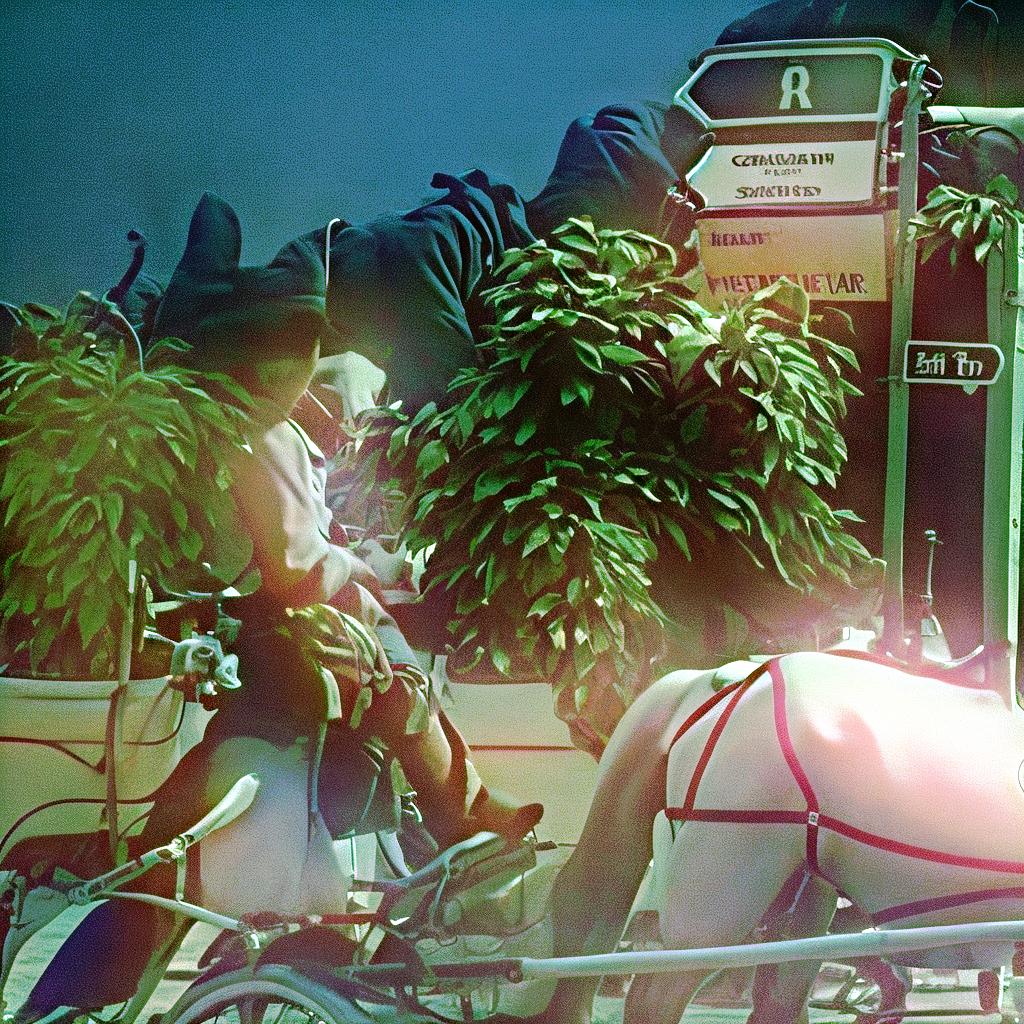} &
    \includegraphics[width=\linewidth]{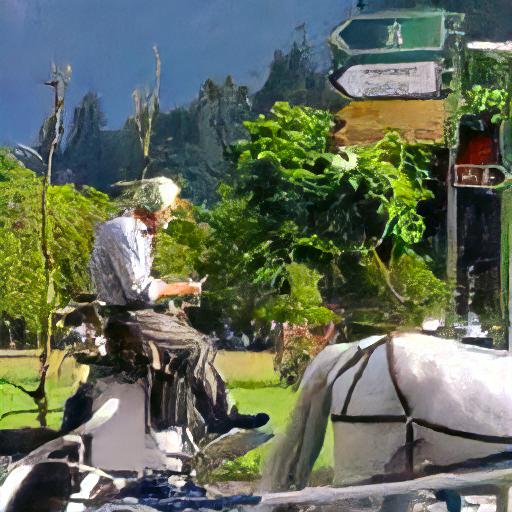} &
    \includegraphics[width=\linewidth]{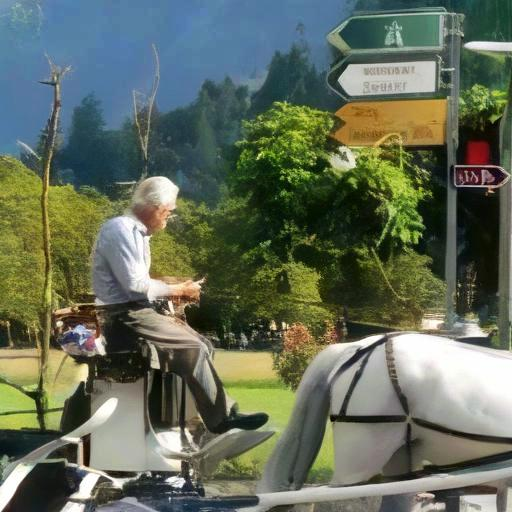} &
    \includegraphics[width=\linewidth]{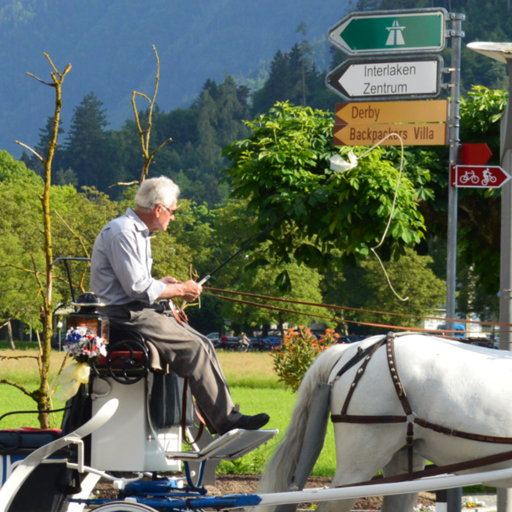} \\
    
    \includegraphics[width=\linewidth]{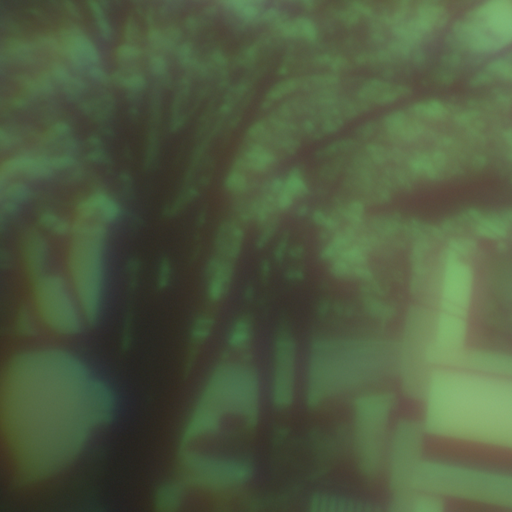} & 
    \includegraphics[width=\linewidth]{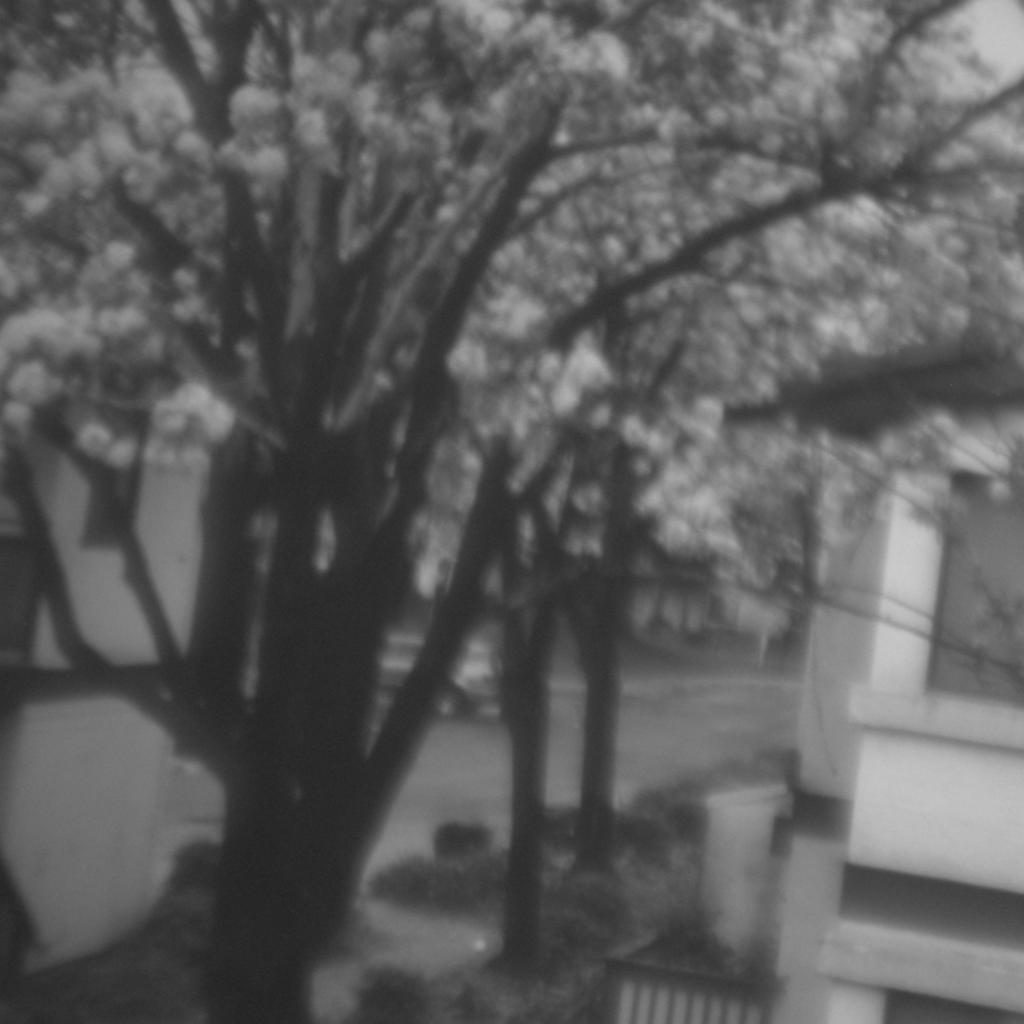} &
    \includegraphics[width=\linewidth]{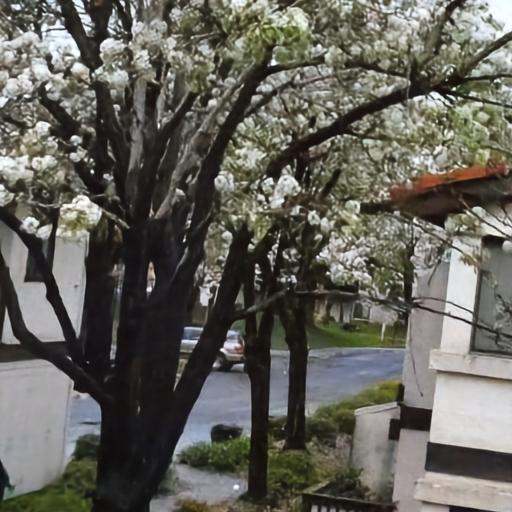} &
    \includegraphics[width=\linewidth]{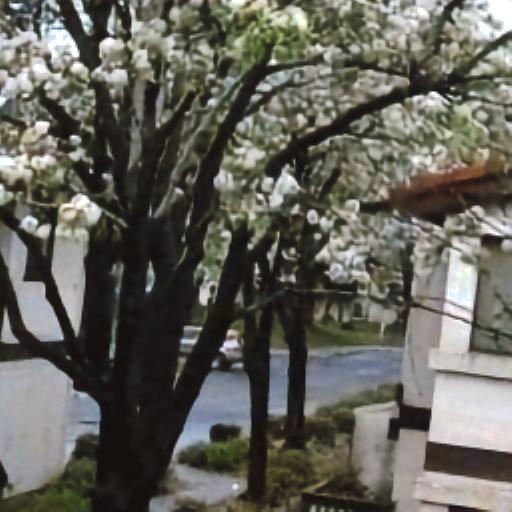} &
    \includegraphics[width=\linewidth]{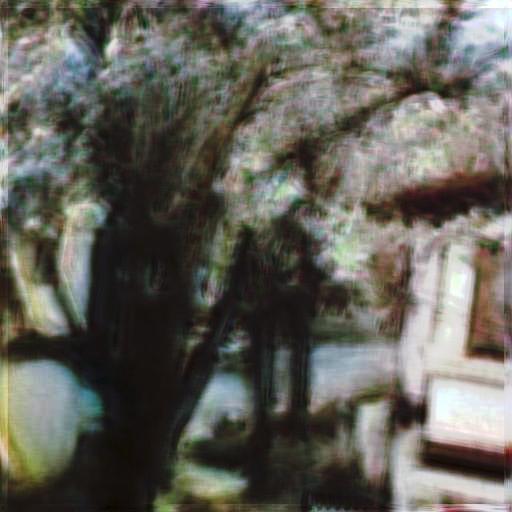} &
    \includegraphics[width=\linewidth]{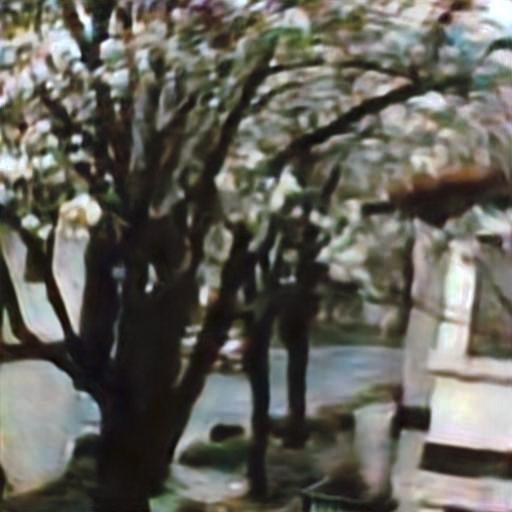} &
    \includegraphics[width=\linewidth]{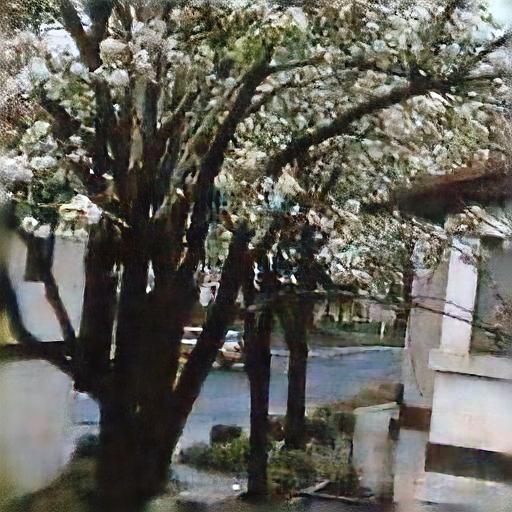} &
    \includegraphics[width=\linewidth]{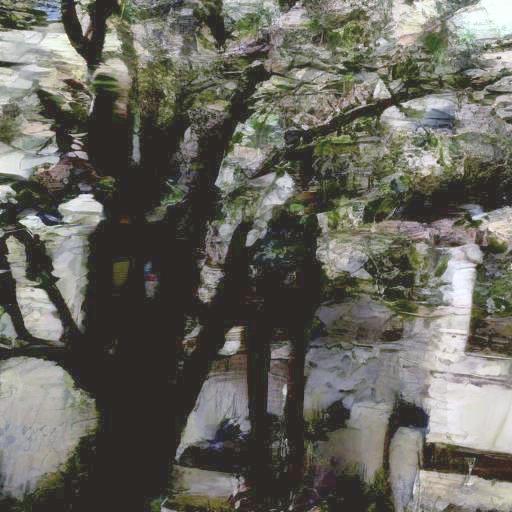} &
    \includegraphics[width=\linewidth]{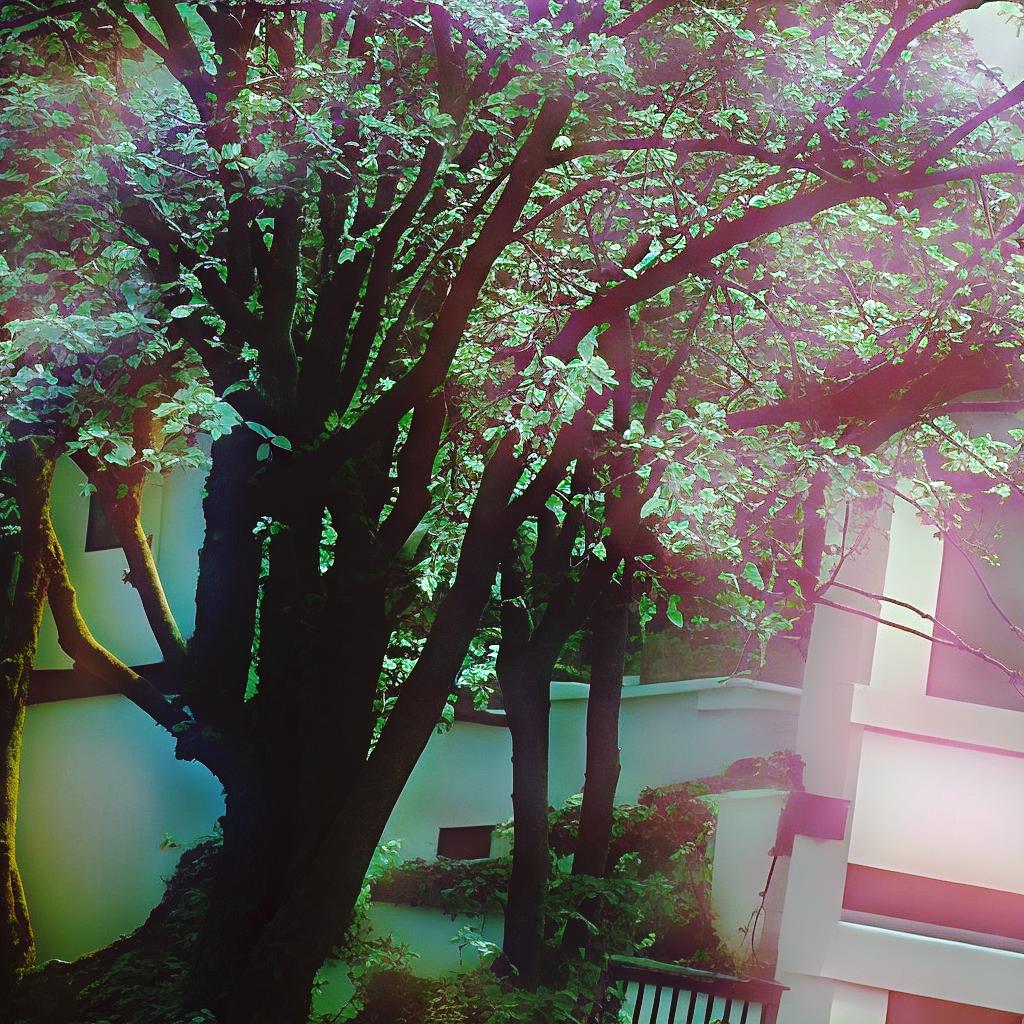} &
    \includegraphics[width=\linewidth]{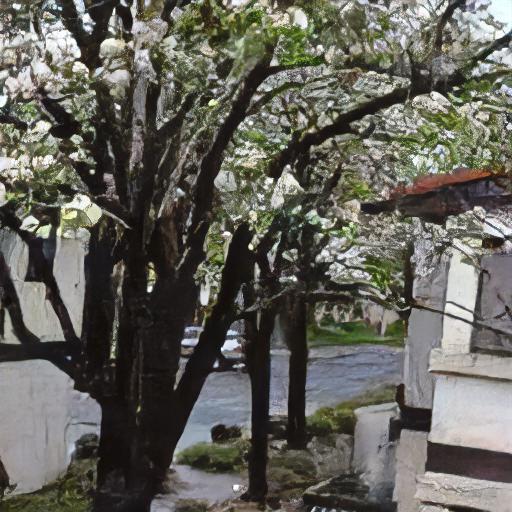} &
    \includegraphics[width=\linewidth]{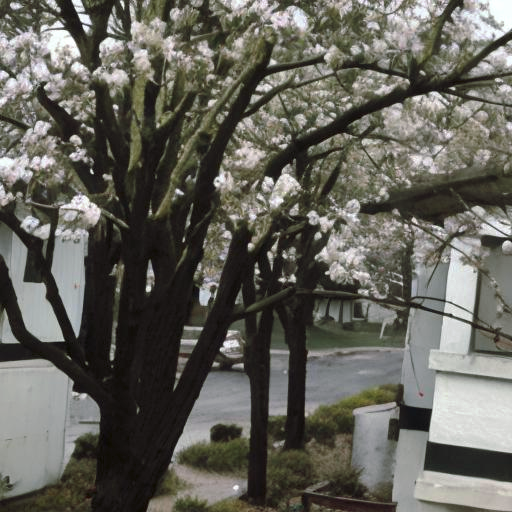} &
    \includegraphics[width=\linewidth]{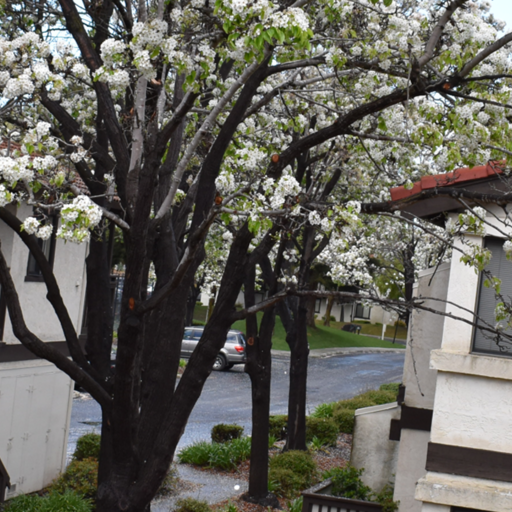} \\
    
    \includegraphics[width=\linewidth]{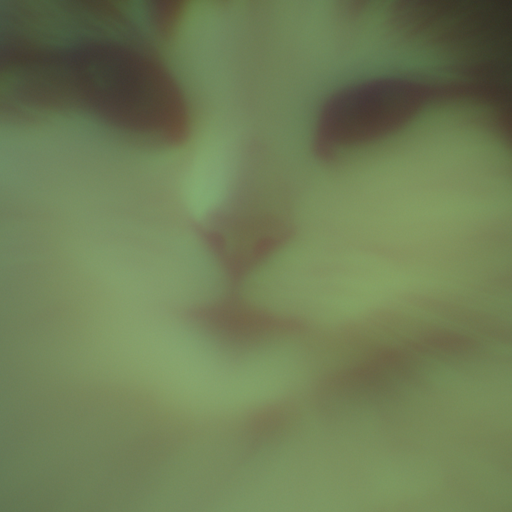} & 
    \includegraphics[width=\linewidth]{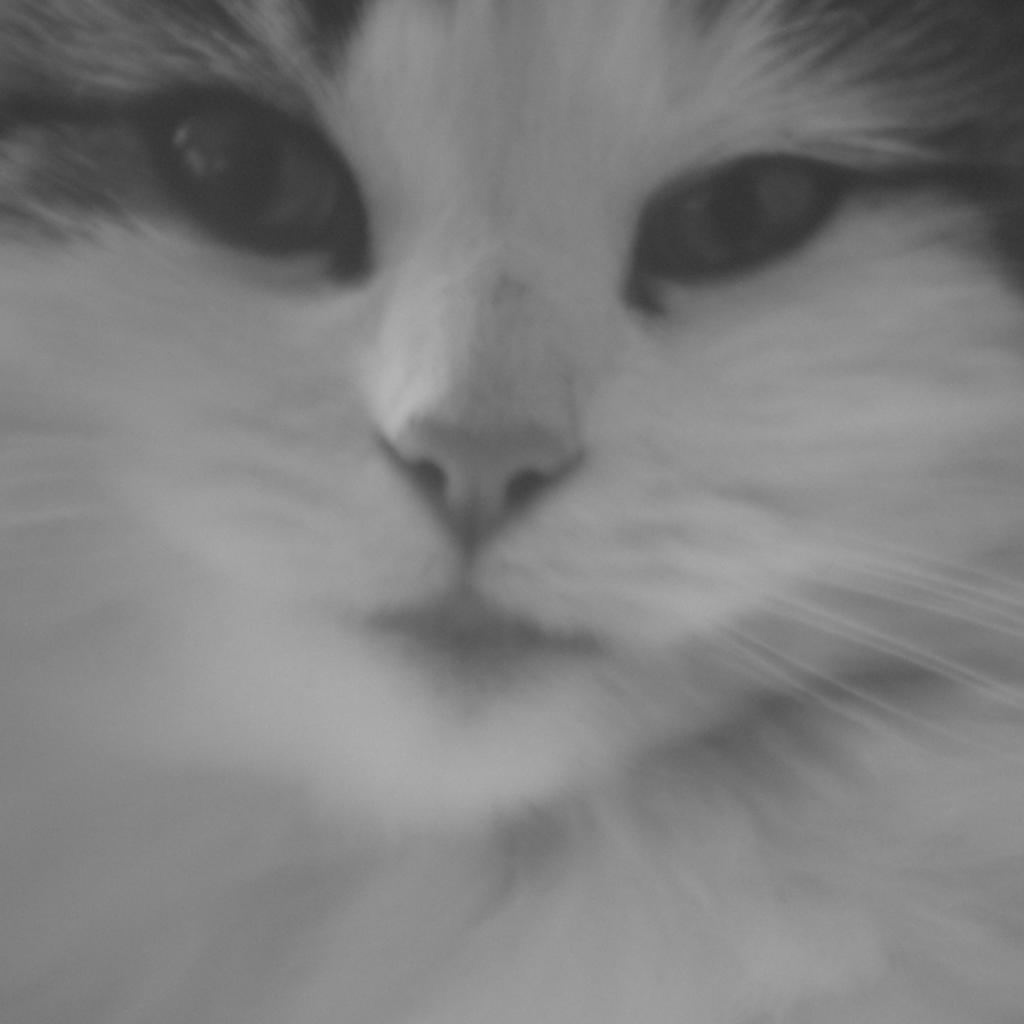} &
    \includegraphics[width=\linewidth]{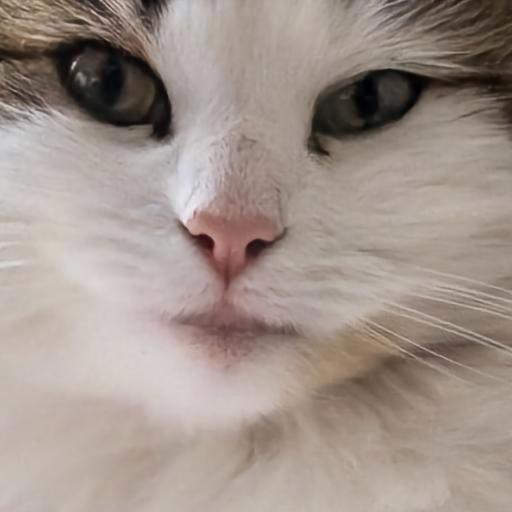} &
    \includegraphics[width=\linewidth]{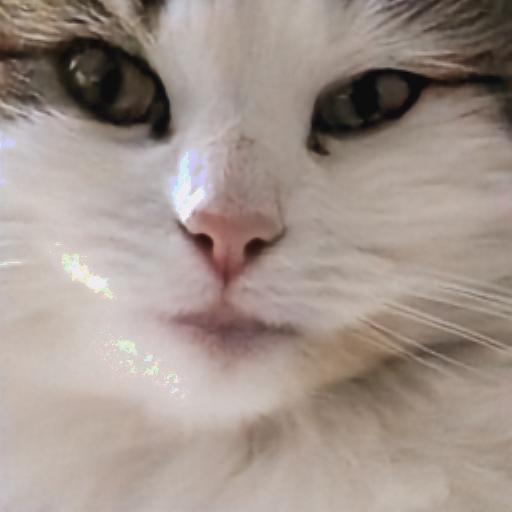} &
    \includegraphics[width=\linewidth]{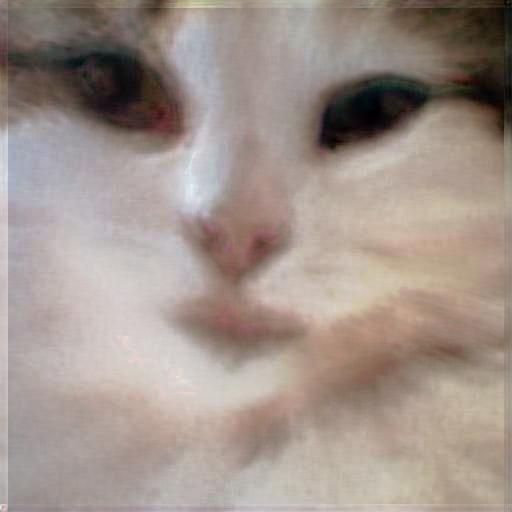} &
    \includegraphics[width=\linewidth]{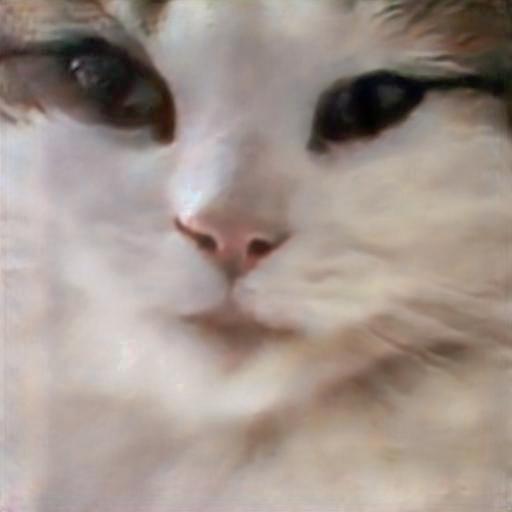} &
    \includegraphics[width=\linewidth]{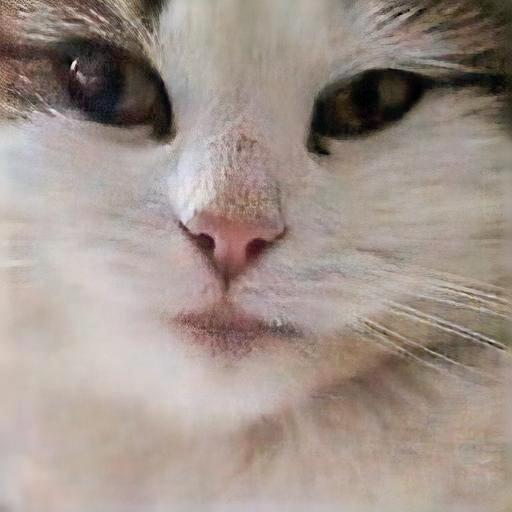} &
    \includegraphics[width=\linewidth]{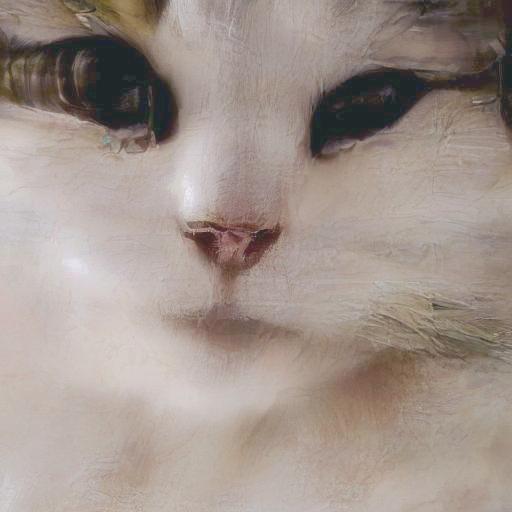} &
    \includegraphics[width=\linewidth]{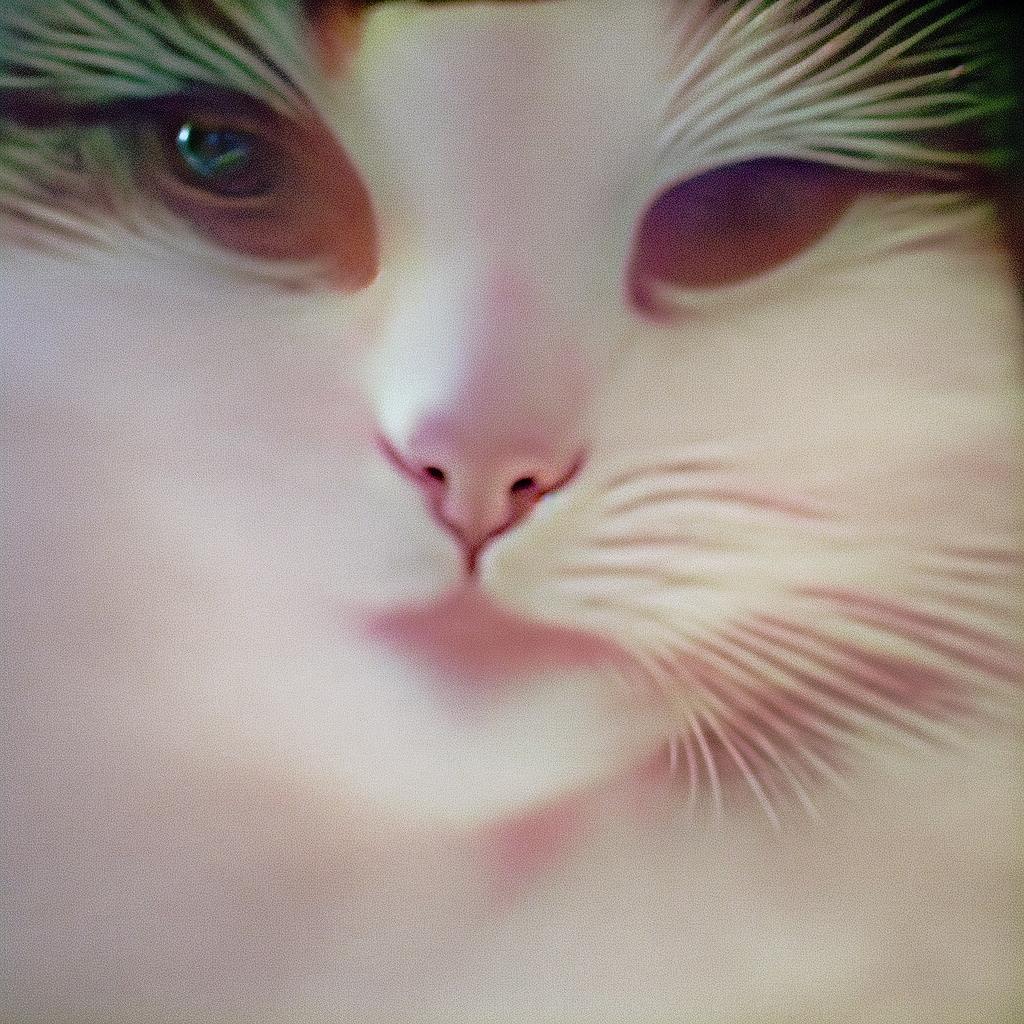} &
    \includegraphics[width=\linewidth]{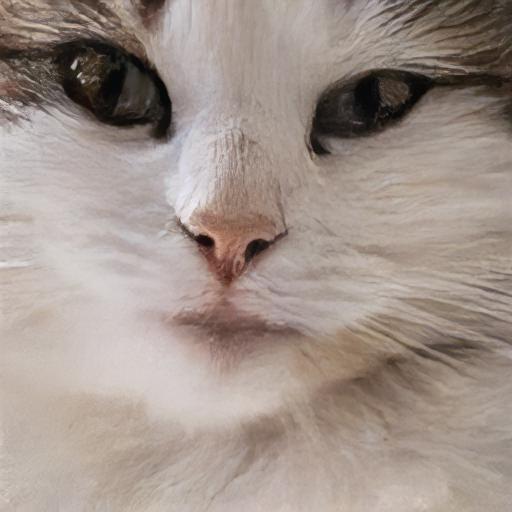} &
    \includegraphics[width=\linewidth]{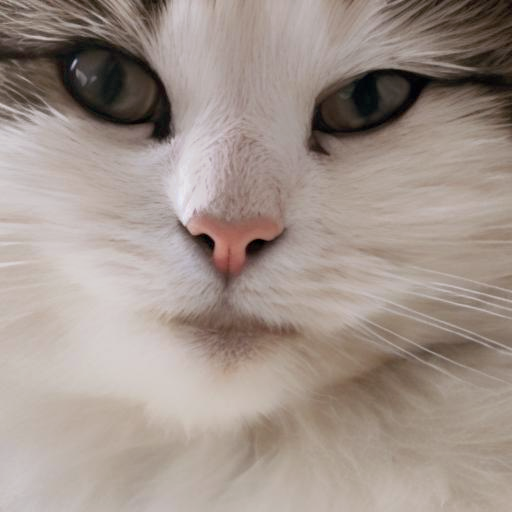} &
    \includegraphics[width=\linewidth]{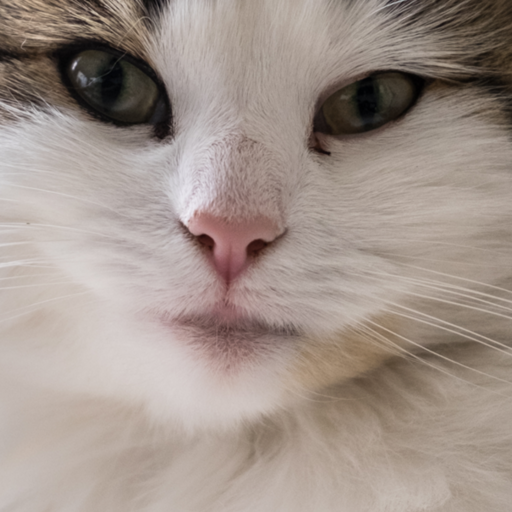} \\
    
    \includegraphics[width=\linewidth]{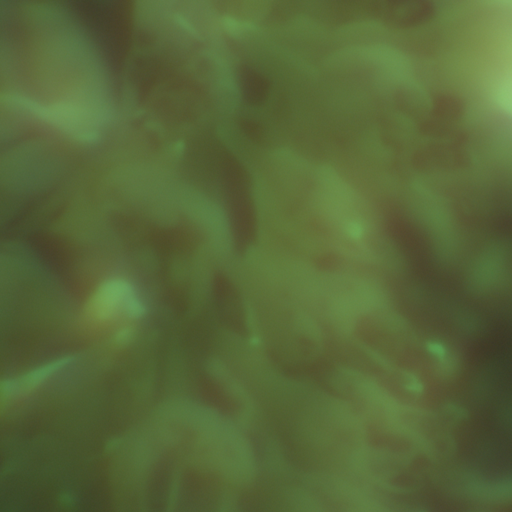} & 
    \includegraphics[width=\linewidth]{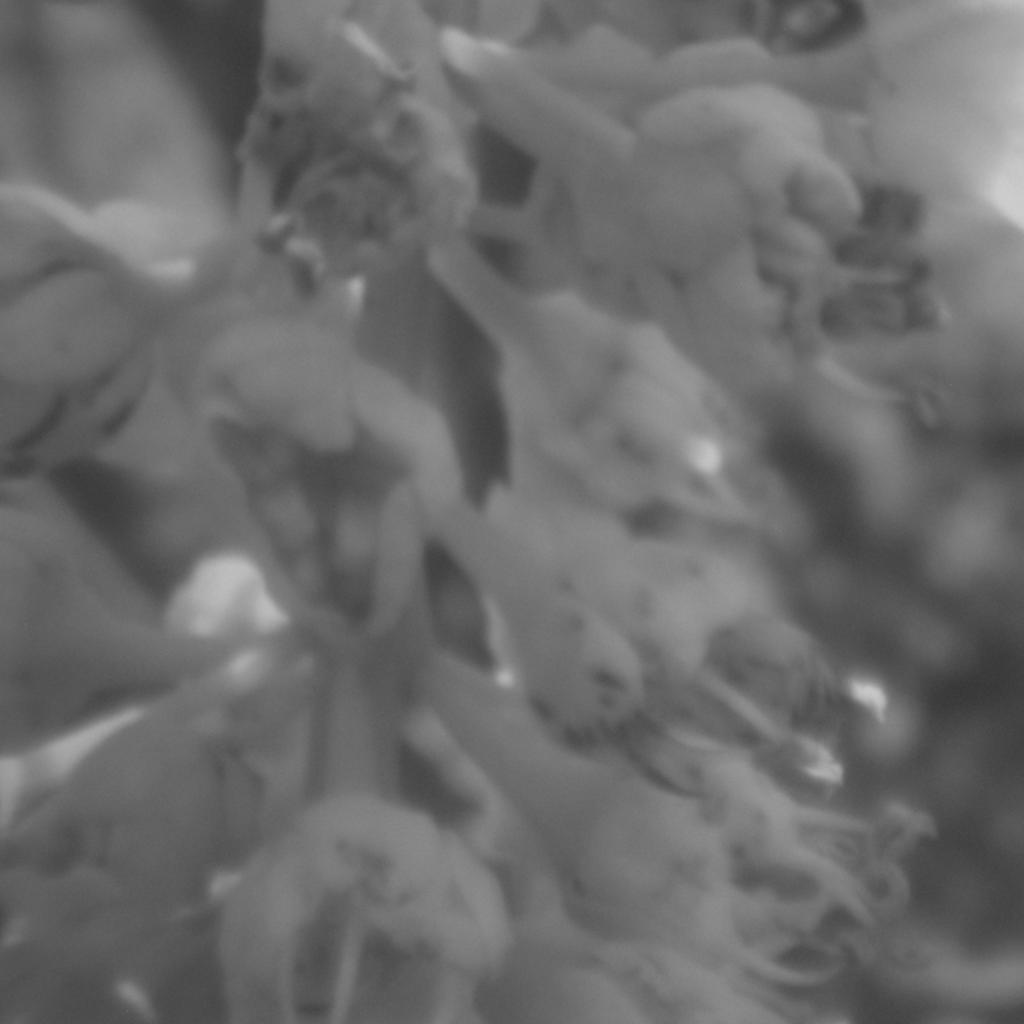} &
    \includegraphics[width=\linewidth]{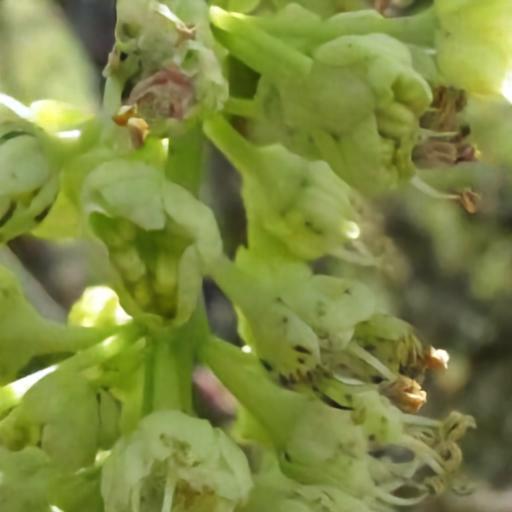} &
    \includegraphics[width=\linewidth]{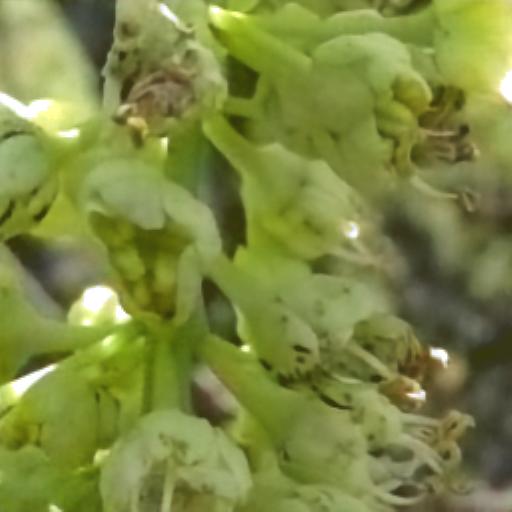} &
    \includegraphics[width=\linewidth]{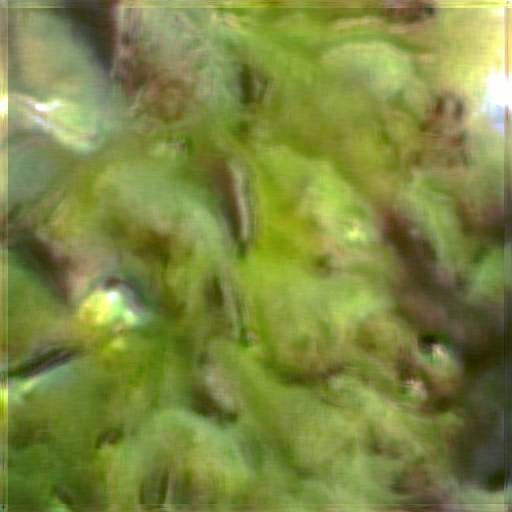} &
    \includegraphics[width=\linewidth]{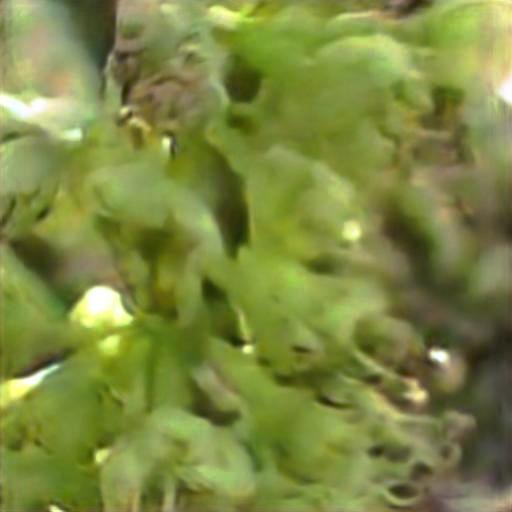} &
    \includegraphics[width=\linewidth]{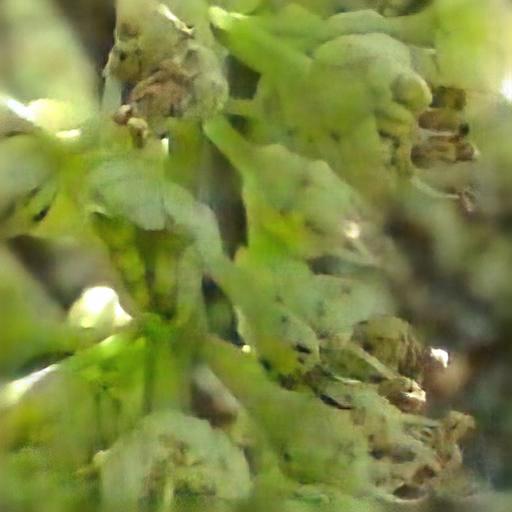} &
    \includegraphics[width=\linewidth]{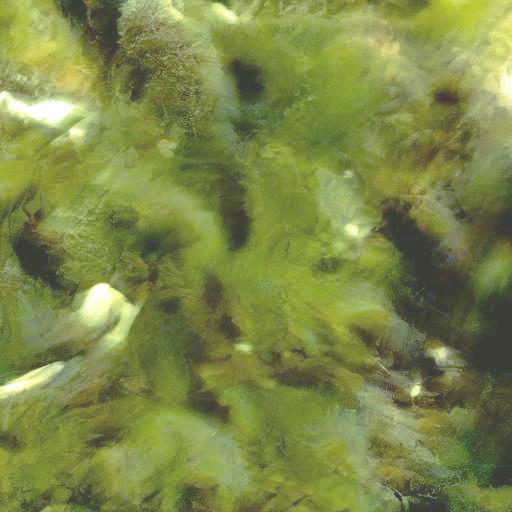} &
    \includegraphics[width=\linewidth]{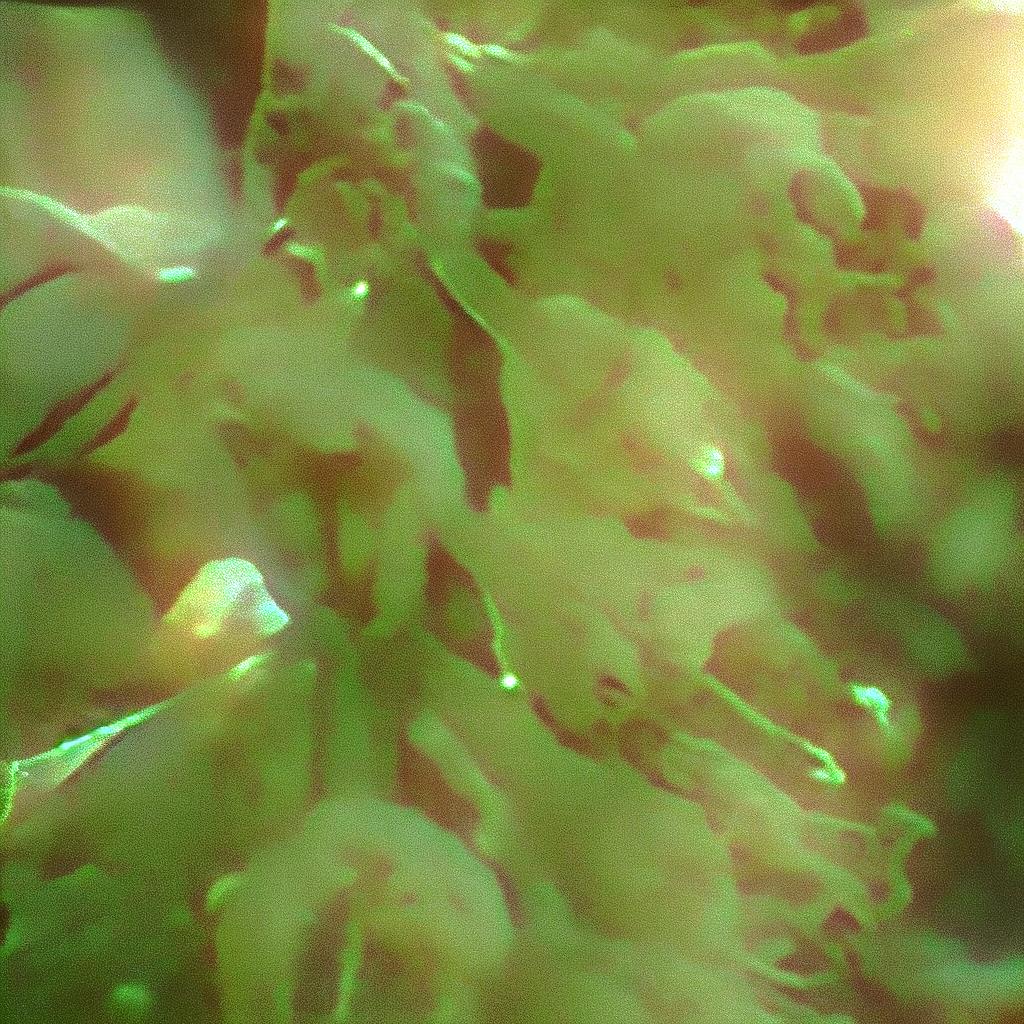} &
    \includegraphics[width=\linewidth]{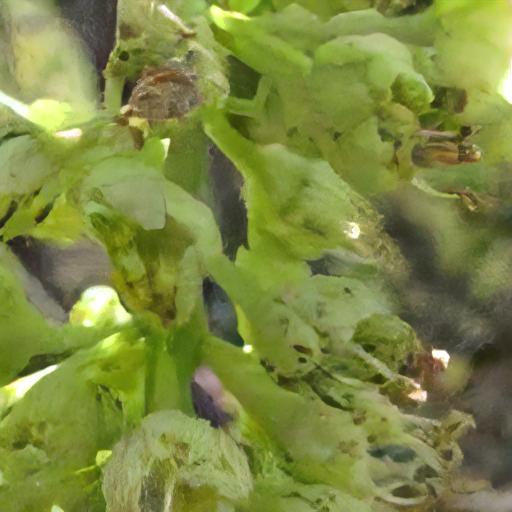} &
    \includegraphics[width=\linewidth]{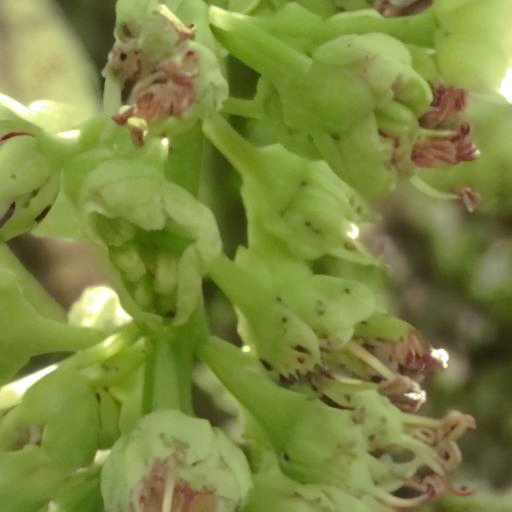} &
    \includegraphics[width=\linewidth]{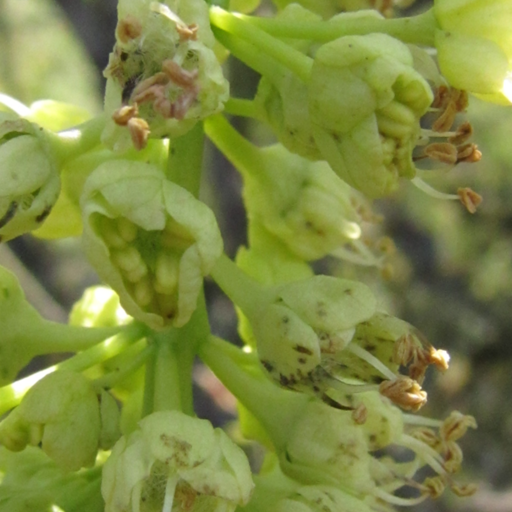} \\
    
    \includegraphics[width=\linewidth]{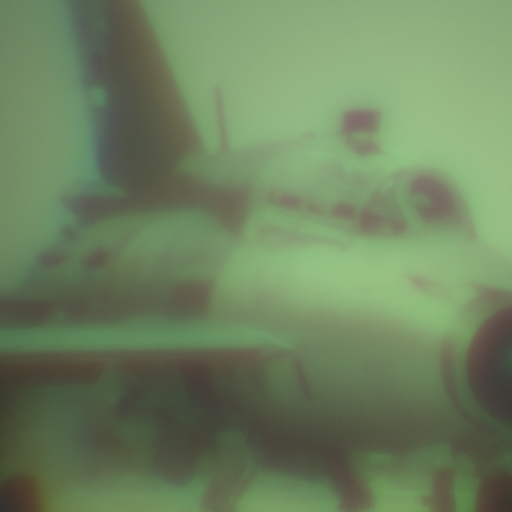} & 
    \includegraphics[width=\linewidth]{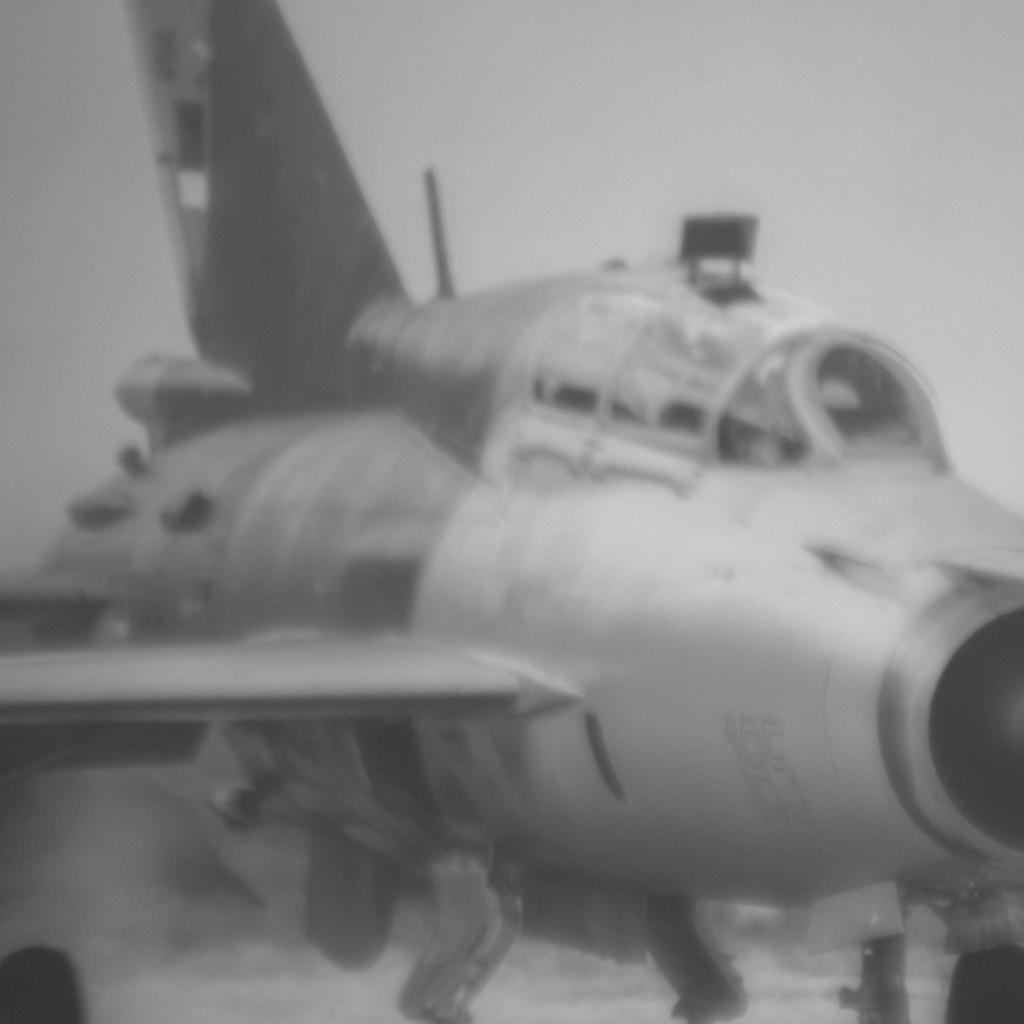} &
    \includegraphics[width=\linewidth]{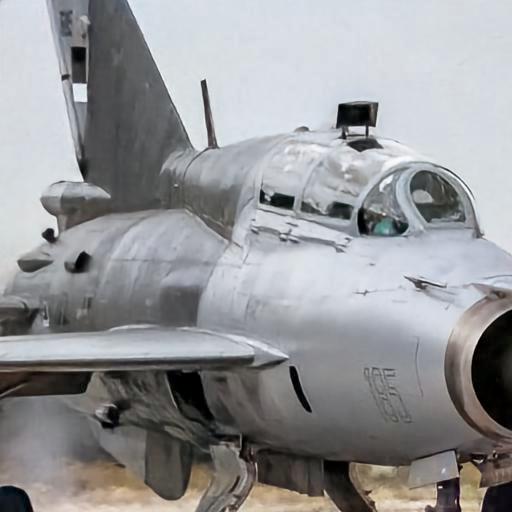} &
    \includegraphics[width=\linewidth]{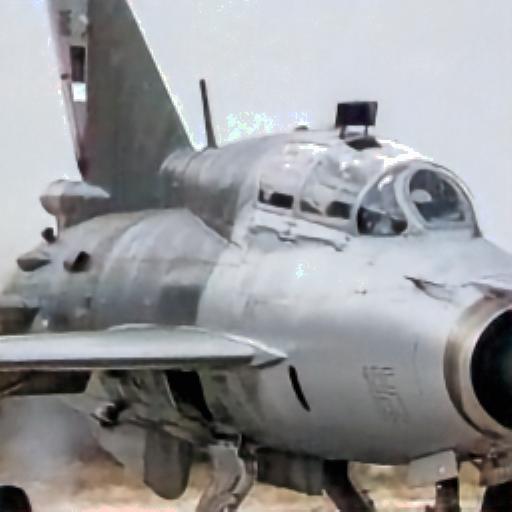} &
    \includegraphics[width=\linewidth]{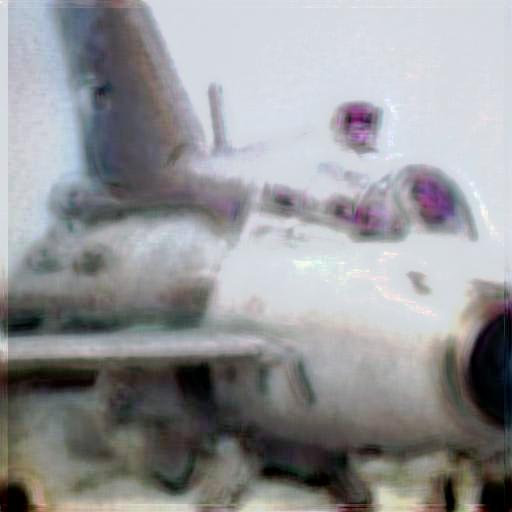} &
    \includegraphics[width=\linewidth]{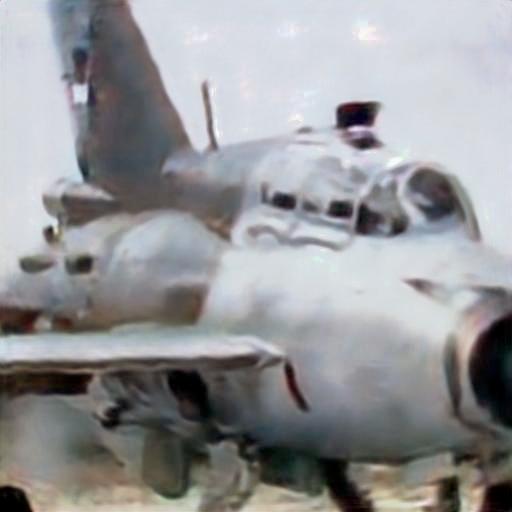} &
    \includegraphics[width=\linewidth]{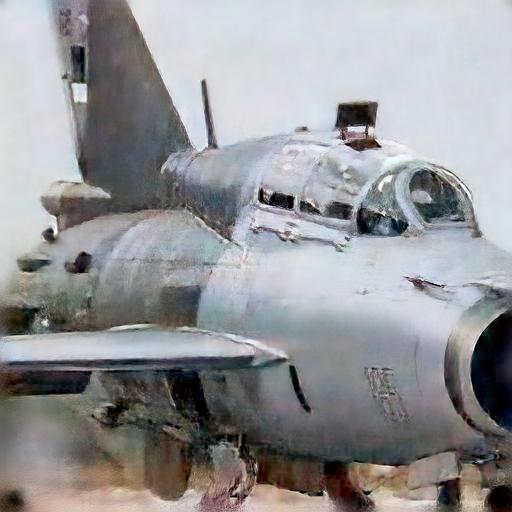} &
    \includegraphics[width=\linewidth]{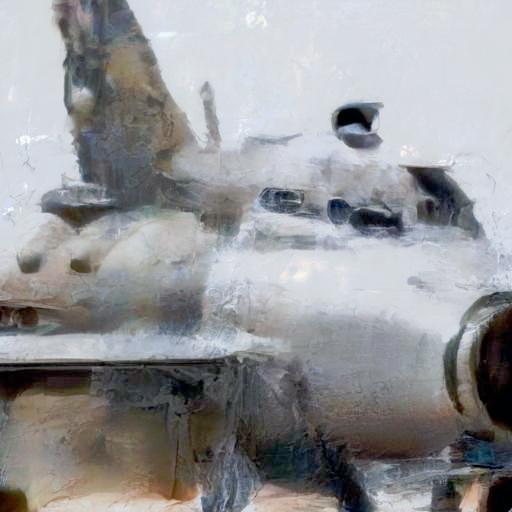} &
    \includegraphics[width=\linewidth]{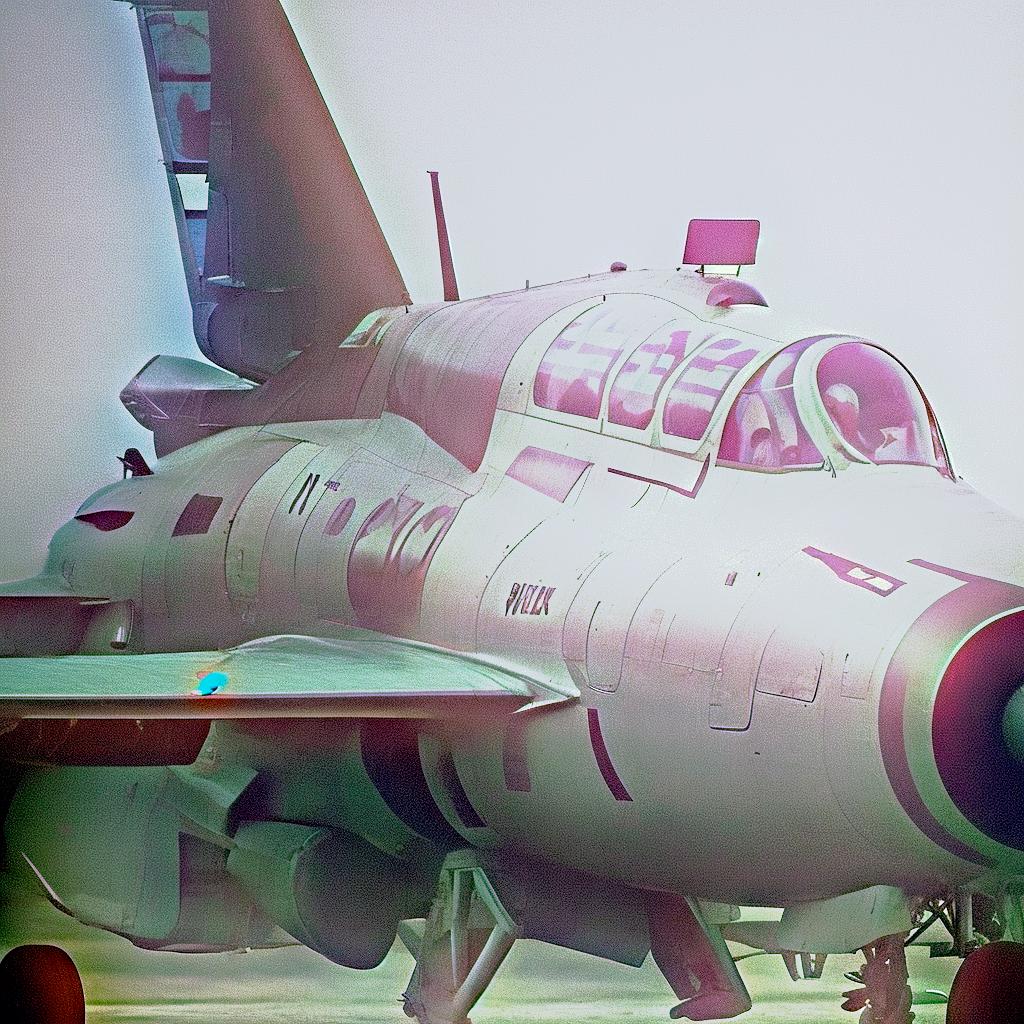} &
    \includegraphics[width=\linewidth]{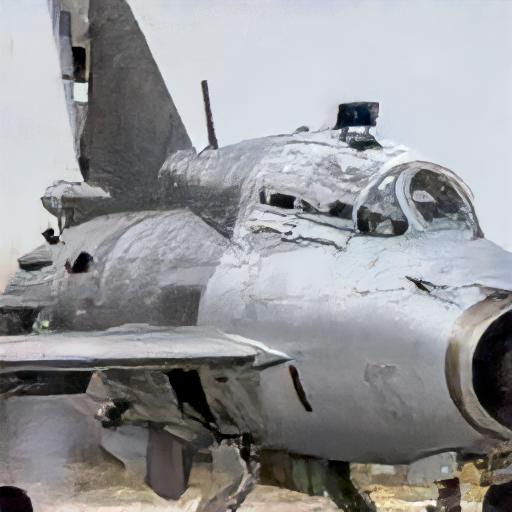} &
    \includegraphics[width=\linewidth]{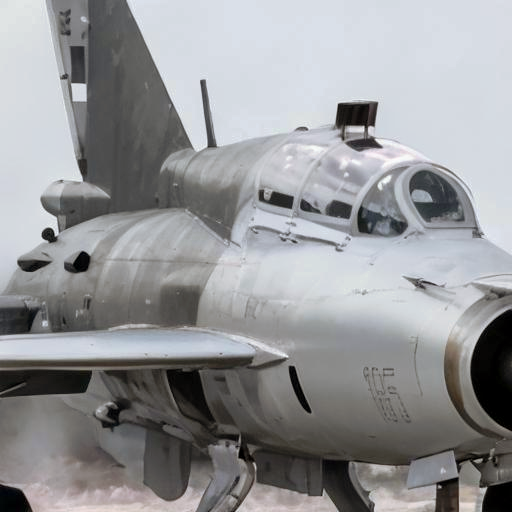} &
    \includegraphics[width=\linewidth]{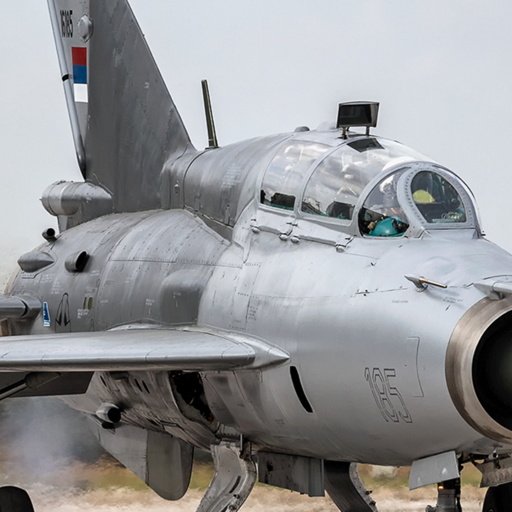} \\[1.5cm]
     $\mathbf{I}_c$ &$\mathbf{I}_s$ & NAFNet \cite{chen2022nafnet} & PanCrafter \cite{do2025pan} & PNN \cite{masi2016pansharpening} & SRPPNN \cite{cai2020srppnn} & Unet \cite{podell_2023_sdxl} & DeblurDiff \cite{kong2025deblurdiff} & DiffBIR \cite{lin2024diffbir} & ResShift \cite{yue2023resshift} &  Ours  & GT\\
\end{tabular}
}
\endgroup

\vspace{2pt}

\resizebox{\linewidth}{!}{
    \begin{tabular}{l| cc| ccc| cccc}
    \toprule
    \textbf{Method} & \textbf{PSNR$\uparrow$} & \textbf{SSIM$\uparrow$} & \textbf{LPIPS$\downarrow$}  & \textbf{DISTS$\downarrow$} &  \textbf{FID$\downarrow$} & \textbf{NIQE$\downarrow$} & \textbf{MUSIQ$\uparrow$} & \textbf{MANIQA$\uparrow$} & \textbf{CLIPIQA$\uparrow$}  \\ 
    \midrule
Structure image & 14.4596 & 0.3691 & 0.5244 & 0.3405 & 194.9119 & 6.8084 & 27.6668 & 0.1518 & 0.2099 \\
Color cue & 13.2701 & 0.3829 & 0.8492 & 0.5270 & 372.5588 & 10.6328 & 17.7413 & 0.1706 & 0.2331 \\
\midrule
NAFNet&\colorbox{red!20}{25.9437}&\colorbox{red!20}{0.8068}&\colorbox{red!20}{0.1862}&\colorbox{magenta!15}{0.1422}&\colorbox{magenta!15}{112.1210}&4.7169&58.1293&0.3095&0.4238\\
PAN-Crafter&18.6192&0.6155&0.3487&0.2179&162.3858&13.2105&40.8314&0.2221&0.2101\\
PNN&17.6193&0.4121&0.6432&0.3671&344.3194&6.0911&17.6830&0.0942&0.1679\\
SRPPNN&20.8644&0.5476&0.5527&0.2954&288.7090&7.2808&24.5899&0.1244&0.1391\\
Unet&21.0032&0.5998&0.3123&0.2439&247.5187&\colorbox{magenta!15}{3.6915}&50.2514&0.2328&0.3104\\
\midrule
DeblurDiff&12.7459&0.2831&0.5718&0.3252&373.0986&\colorbox{red!20}{3.2536}&47.1713&0.2968&0.4787\\
DiffBIR&13.1346&0.3514&0.6467&0.3698&301.6616&6.3770&52.1332&0.3453&0.4780\\
ResShift&19.8529&0.5151&0.2893&0.2365&245.4457&5.6199&59.5060&0.3185&\colorbox{red!20}{0.6467}\\
Ours wo VSD&\colorbox{yellow!10}{21.5626}&\colorbox{yellow!10}{0.6237}&\colorbox{yellow!10}{0.2128}&\colorbox{yellow!10}{0.1444}&\colorbox{yellow!10}{118.4584}&\colorbox{yellow!10}{3.9230}&\colorbox{magenta!15}{62.7171}&\colorbox{magenta!15}{0.3984}&0.5041\\
Ours w VSD&20.4017&0.5933&0.2321&0.1600&131.7168&4.1483&\colorbox{red!20}{64.3806}&\colorbox{red!20}{0.4229}&\colorbox{magenta!15}{0.5398}\\
Ours w HF-VSD&\colorbox{magenta!15}{21.9542}&\colorbox{magenta!15}{0.6294}&\colorbox{magenta!15}{0.2042}&\colorbox{red!20}{0.1397}&\colorbox{red!20}{108.9117}&3.9841&\colorbox{yellow!10}{61.0900}&\colorbox{yellow!10}{0.3747}&\colorbox{yellow!10}{0.5121}\\
    \bottomrule
    \end{tabular}
    }
\caption{Qualitative comparison of the proposed computational model. We train and test different image restoration algorithms using the real captured dataset described in Sec.~\ref{sec:dataset}. Ours achieves the highest visual quality. Table shows a comparison of methods with related reconstruction algorithms using fidelity (PSNR, SSIM), perception quality (LPIPS, DISTS, FID), and no-reference quality (NIQE, MUSIQ, MANIQA, CLIPIQA) metrics. The methods are grouped into non-diffusion-based and diffusion-based approaches to highlight their different training strategies. We highlight the \colorbox{red!20}{best}, \colorbox{magenta!15}{second-best}, and \colorbox{yellow!10}{third-best} values for each metric.}
\label{fig:qualitative_comparison_algo}
\end{figure*}

\paragraph{Comparison with leading image restoration methods.}
We compare the proposed computational model (Sec.~\ref{secsec:comp}) against state-of-the-art spatially varying deblurring methods~\cite{yanny_optica_2022_mwn,lin2024diffbir,kong2025deblurdiff,yue2023resshift} and recent pansharpening approaches~\cite{do2025pan,masi2016pansharpening,cai2020srppnn} on the dataset described in Sec.~\ref{sec:dataset}. Qualitative comparisons and quantitative evaluations, including fidelity-based, perceptual, and no-reference quality metrics, are reported in Fig.~\ref{fig:qualitative_comparison_algo}. Our method consistently achieves the best or second-best performance across all reported metrics and delivers the highest perceptual quality in visual comparisons. These results demonstrate that the proposed computational model achieves state-of-the-art performance for processing the raw measurements captured by MetaTele. Further experiments in Supplement 1 show that the one-step diffusion model is strictly guided by sensor measurements rather than hallucinations. Also, we analyze the reconstruction results using Radially Averaged Power Spectral Density (RAPSD) in Supplement 1 to highlight that our method recovers realistic fine-scale details.

\subsection{Comparison of system in simulation}

This study compares MetaTele with prior metasurface-based imaging systems~\cite{parfocal,Tseng2021NeuralNanoOptics,Pinilla2023} in terms of reconstruction quality. As summarized in Table~\ref{tab:related_works}, MetaTele has a substantially longer EFL than most of the competing systems. To ensure a fair comparison, we evaluate all methods under a unified setting where the imaging target occupies the same sensor area. Consequently, the comparison focuses on reconstruction quality on the sensor plane rather than the effective resolution in object space.

\begin{figure}
    \centering
    \begingroup
    \setlength{\tabcolsep}{1pt} 
    \renewcommand{\arraystretch}{0.5} 
    \resizebox{\linewidth}{!}{
    \begin{tabular}{ *{5}{>{\centering\arraybackslash}m{4.0cm}} }
    \includegraphics[width=\linewidth]{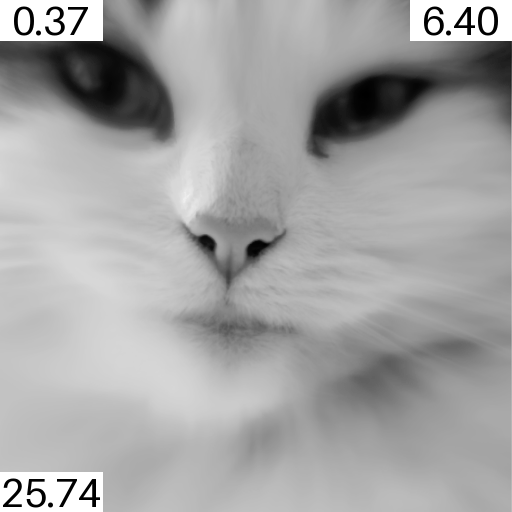} & 
    \includegraphics[width=\linewidth]{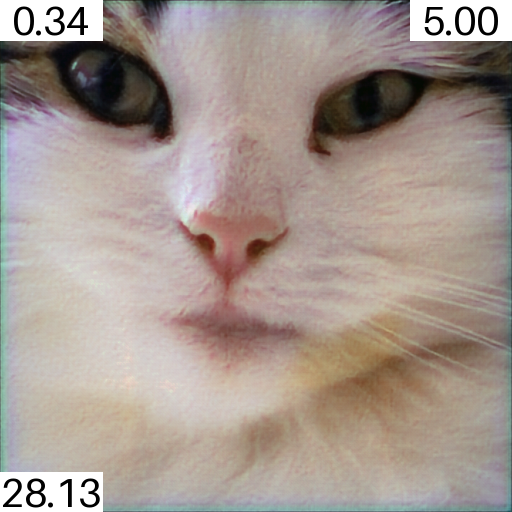} &
    \includegraphics[width=\linewidth]{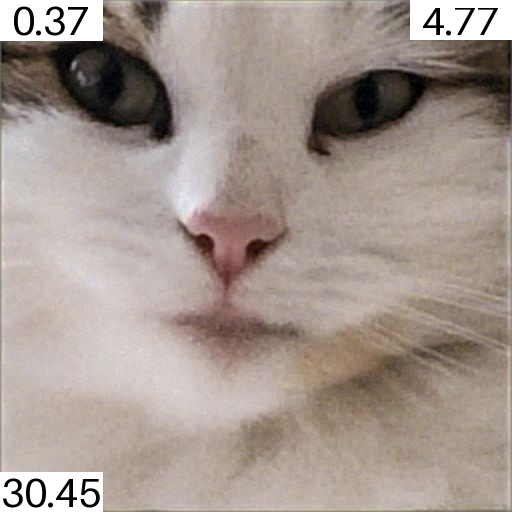} &
    \includegraphics[width=\linewidth]{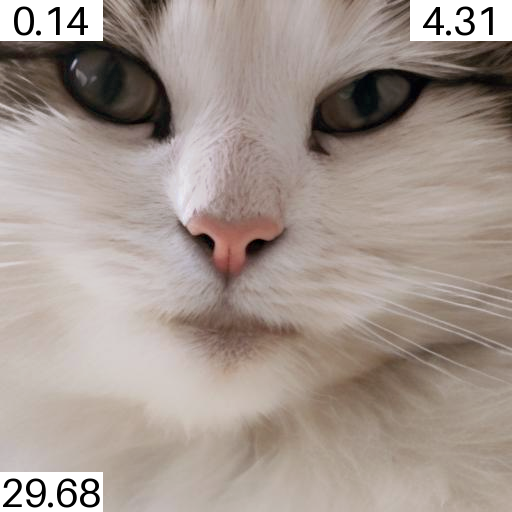} &
    \includegraphics[width=\linewidth]{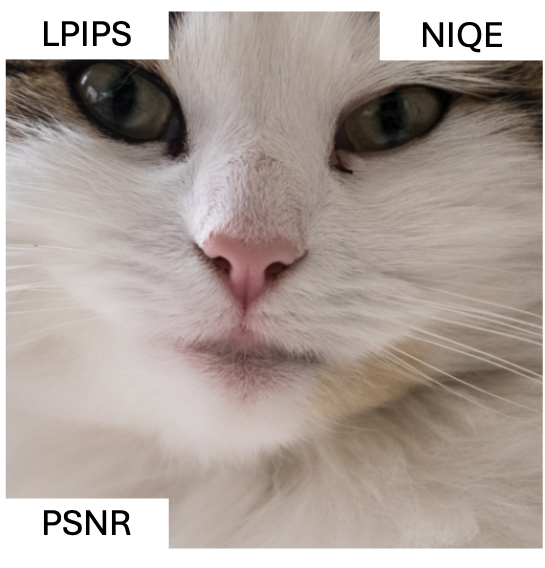} \\
    
    \includegraphics[width=\linewidth]{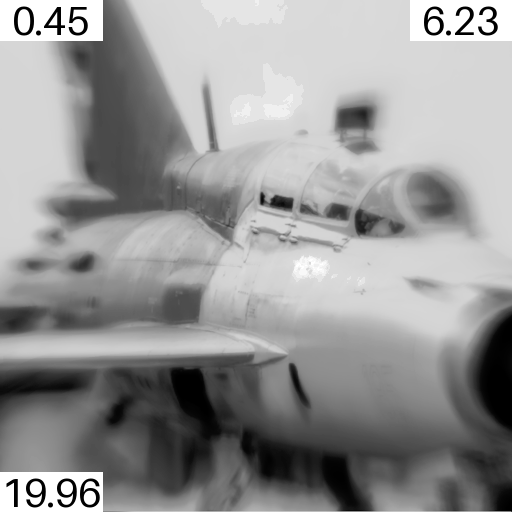} & 
    \includegraphics[width=\linewidth]{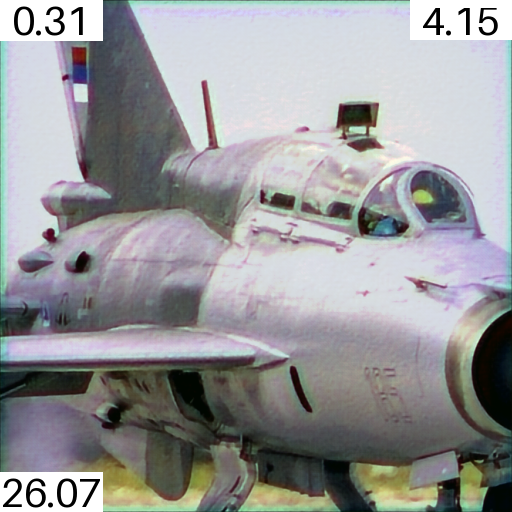} &
    \includegraphics[width=\linewidth]{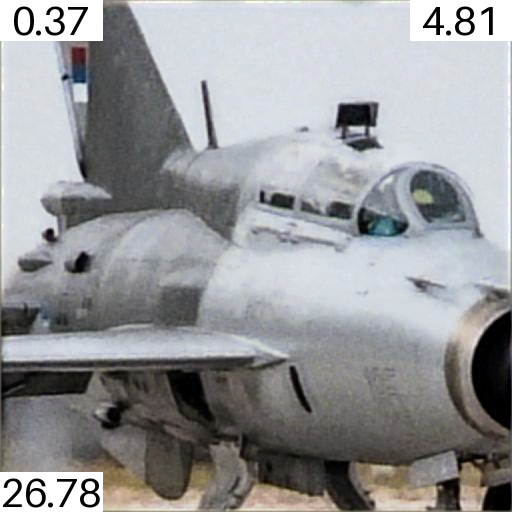} &
    \includegraphics[width=\linewidth]{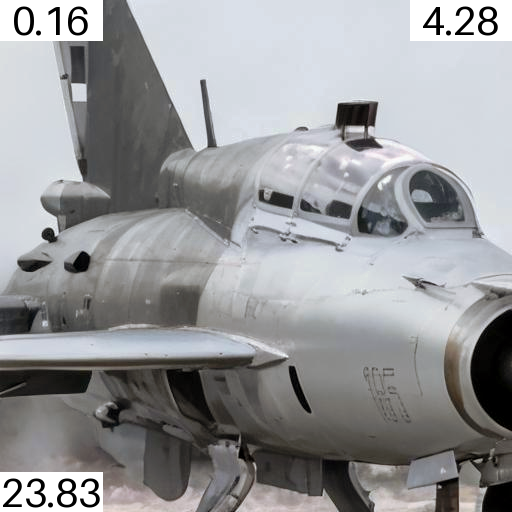} &
    \includegraphics[width=\linewidth]{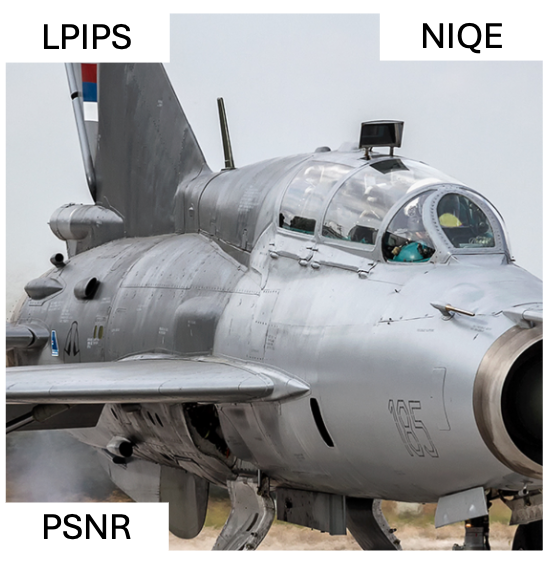} \\
    
    \includegraphics[width=\linewidth]{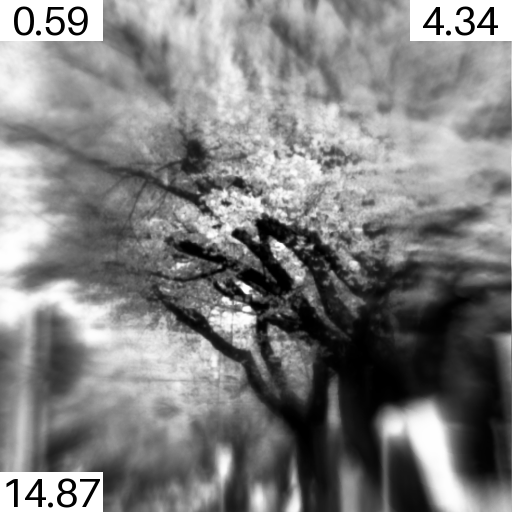} & 
    \includegraphics[width=\linewidth]{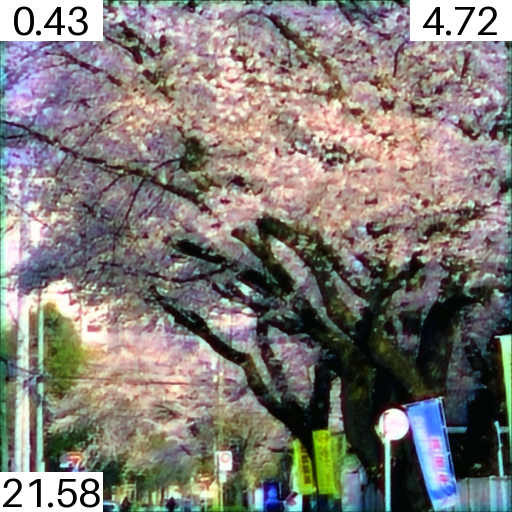} &
    \includegraphics[width=\linewidth]{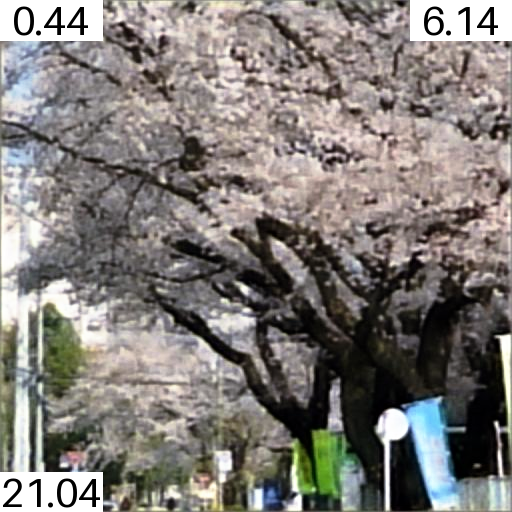} &
    \includegraphics[width=\linewidth]{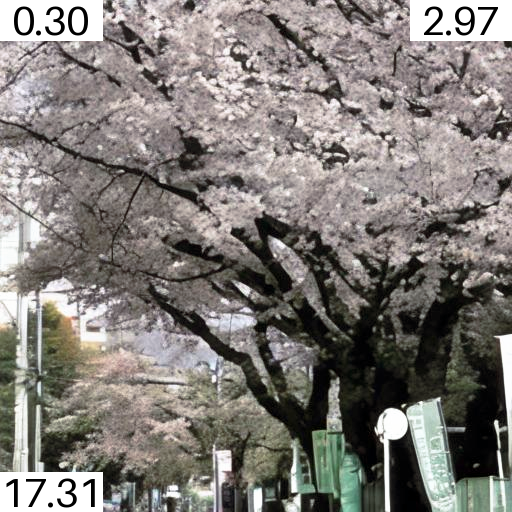} &
    \includegraphics[width=\linewidth]{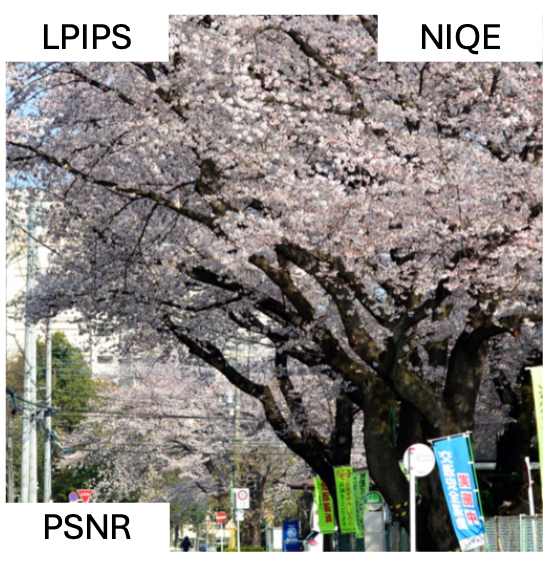} \\
    
    \includegraphics[width=\linewidth]{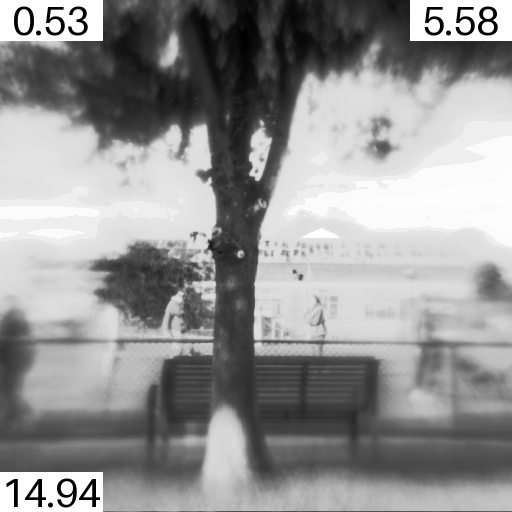} & 
    \includegraphics[width=\linewidth]{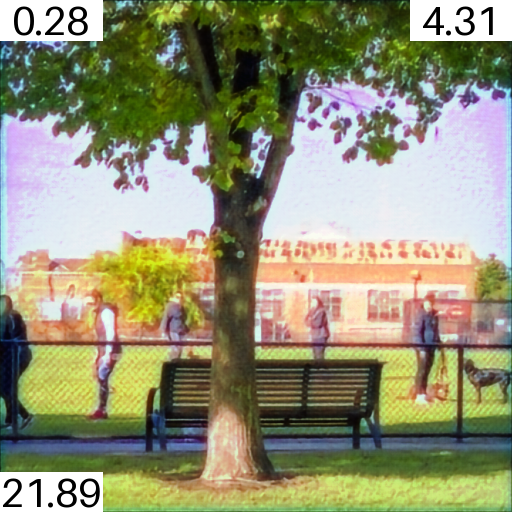} &
    \includegraphics[width=\linewidth]{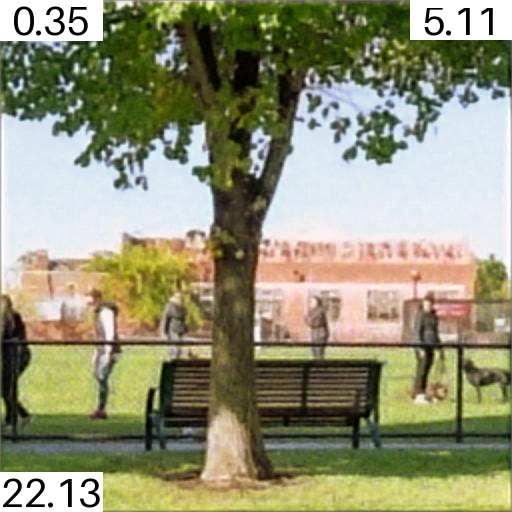} &
    \includegraphics[width=\linewidth]{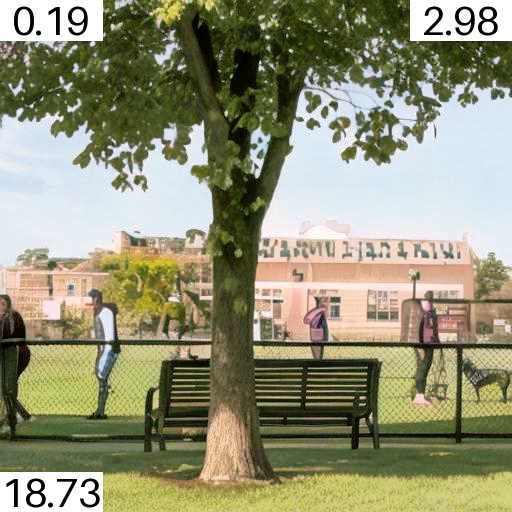} &
    \includegraphics[width=\linewidth]{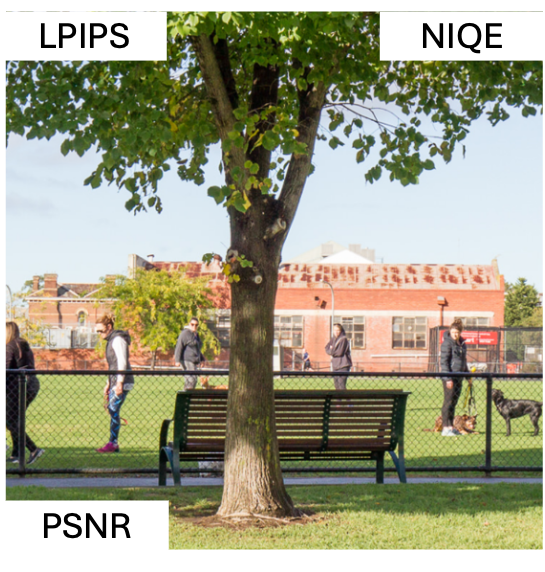} \\

    Yang \textit{et al.} &
    Tseng \textit{et al.} &
    Pinilla \textit{et al.} &
    Ours &
    Ground truth \\

    \end{tabular}
    }
    \endgroup

    \vspace{2pt}
    
    \resizebox{\linewidth}{!}{
    \begin{tabular}{l| cc| ccc| cccc}
    \toprule
     \textbf{Method} & \textbf{PSNR$\uparrow$} & \textbf{SSIM$\uparrow$} & \textbf{LPIPS$\downarrow$}  & \textbf{DISTS$\downarrow$} &  \textbf{FID$\downarrow$} & \textbf{NIQE$\downarrow$} & \textbf{MUSIQ$\uparrow$} & \textbf{MANIQA$\uparrow$} & \textbf{CLIPIQA$\uparrow$}  \\ 
    \hline
    Color cue & 13.2701 & 0.3829 & 0.8492 & 0.5270 & 372.5588 & 10.6328 & 17.7413 & 0.1706 & 0.2331 \\
    Yang \emph{et al.}~\cite{parfocal}&15.0096&0.4429&0.5428&0.3385&192.3649&5.9950&\colorbox{yellow!10}{37.0225}&0.2027&0.2403\\
    Tseng \emph{et al.}~\cite{Tseng2021NeuralNanoOptics}&\colorbox{red!20}{24.5606}&\colorbox{red!20}{0.8032}&\colorbox{magenta!15}{0.3010}&\colorbox{magenta!15}{0.1937}&\colorbox{magenta!15}{124.9688}&\colorbox{magenta!15}{4.7498}&\colorbox{magenta!15}{46.9808}&\colorbox{magenta!15}{0.2378}&\colorbox{magenta!15}{0.3925}\\
    Pinilla \emph{et al.}~\cite{Pinilla2023}&\colorbox{magenta!15}{24.5414}&\colorbox{magenta!15}{0.7069}&\colorbox{yellow!10}{0.3867}&\colorbox{yellow!10}{0.2112}&\colorbox{yellow!10}{163.5929}&\colorbox{yellow!10}{5.2301}&35.2180&\colorbox{yellow!10}{0.2091}&\colorbox{yellow!10}{0.2659}\\
    Ours&\colorbox{yellow!10}{21.9542}&\colorbox{yellow!10}{0.6294}&\colorbox{red!20}{0.2042}&\colorbox{red!20}{0.1397}&\colorbox{red!20}{108.9117}&\colorbox{red!20}{3.9841}&\colorbox{red!20}{61.0900}&\colorbox{red!20}{0.3747}&\colorbox{red!20}{0.5121}\\
    \bottomrule
    \end{tabular}
    } 
    \caption{Comparison with recent metasurface-based imaging systems~\cite{parfocal,Tseng2021NeuralNanoOptics,Pinilla2023} in simulation. Representative reconstruction results from the Flickr2K dataset are shown in the figure, and the accompanying table reports the overall quantitative performance on the same dataset. Our method achieves the highest visual quality among all compared approaches. Although it does not obtain the best scores on fidelity-based metrics, it consistently performs best on perceptual and no-reference quality metrics, indicating superior perceptual reconstruction quality.}
    \label{fig:complete_pipeline}
\end{figure}

We synthesize measurements of front-parallel scenes using textures randomly sampled from the Flickr2K dataset. For each system, we reconstruct the optical model in Code V based on the reported optical parameters and generate the corresponding PSFs (as in Fig.~\ref{fig:aoi_sweep}) to simulate the image formation process. When applicable, the post-processing networks of the baseline methods are retrained on these synthetic measurements. The qualitative and quantitative results are summarized in Fig.~\ref{fig:complete_pipeline}. MetaTele consistently produces reconstructions with the highest perceptual quality and achieves the best overall performance across the evaluated metrics.

\section{Conclusion}\label{sec:conclusion}

Compared to previous works that purely use one or multiple metasurfaces for imaging and only capture one raw measurement, MetaTele presents an alternative using a hybrid refractive-metasurface system and capturing complementary measurements with different domains of information. The hybrid refractive-metasurface system achieves less severe aberration compared to purely metasurface counterparts. It enables the imaging process to be decomposed into complementary measurements that are later integrated through computation.

Two practical challenges remain for this paradigm. The first is the realization of single-shot capture of the complementary measurements. As discussed in this work, this can potentially be addressed through custom sensor architectures, such as spatially multiplexed spectral filter arrays. The second challenge is the relatively long exposure required for the narrowband structure image. One possible solution is to replace the long exposure with a burst of short-exposure measurements and recover the structure image through burst denoising and fusion. Addressing these challenges would further extend the applicability of this hybrid optical–computational framework and open new opportunities for compact, high-performance metasurface-enabled imaging systems.

\begin{backmatter}
\bmsection{Funding} Samsung Research America Global Research Outreach. National Science Foundation Grant No. CCF--2431505.

\bmsection{Acknowledgment}
The metasurface in this work was fabricated by SNOChip Inc. through their custom metasurface fabrication service according to the authors’ specifications. 

\bmsection{Disclosures}
The authors declare no conflicts of interest.

\bmsection{Data Availability Statement} Data underlying the results presented in this paper are available in Ref.~\cite{Weligampola_Chen_Gnanasambandam_Godaliyadda_Sheikh_Chan_Guo_2026}.

\bmsection{Supplemental document}
See Supplement 1 for supporting content.

\end{backmatter}

\bibliography{references}
\bigskip
\newpage


\providecommand{\R}{\ensuremath{\mathbb{R}}}
\providecommand{\C}{\ensuremath{\mathbb{C}}}
\providecommand{\E}{\ensuremath{\mathbb{E}}}
\providecommand{\N}{\ensuremath{\mathbb{N}}}
\providecommand{\Z}{\ensuremath{\mathbb{Z}}}
\providecommand{\Pb}{\ensuremath{\mathbb{P}}}
\providecommand{\I}{\ensuremath{\mathbb{I}}}
\providecommand{\bfP}{\mathbf{P}}
\providecommand{\Var}{\mathrm{Var}}
\providecommand{\Cov}{\mathrm{Cov}}

\providecommand{\abs}[1]{\lvert#1\rvert}
\providecommand{\norm}[1]{\lVert#1\rVert}
\providecommand{\inprod}[1]{\langle#1\rangle}
\providecommand{\bydef}{\overset{\text{def}}{=}}

\renewcommand{\vec}[1]{\ensuremath{\boldsymbol{#1}}}
\providecommand{\mat}[1]{\ensuremath{\boldsymbol{#1}}}

\providecommand{\calA}{\mathcal{A}}
\providecommand{\calB}{\mathcal{B}}
\providecommand{\calC}{\mathcal{C}}
\providecommand{\calD}{\mathcal{D}}
\providecommand{\calE}{\mathcal{E}}
\providecommand{\calF}{\mathcal{F}}
\providecommand{\calG}{\mathcal{G}}
\providecommand{\calH}{\mathcal{H}}
\providecommand{\calI}{\mathcal{I}}
\providecommand{\calJ}{\mathcal{J}}
\providecommand{\calK}{\mathcal{K}}
\providecommand{\calL}{\mathcal{L}}
\providecommand{\calM}{\mathcal{M}}
\providecommand{\calN}{\mathcal{N}}
\providecommand{\calO}{\mathcal{O}}
\providecommand{\calP}{\mathcal{P}}
\providecommand{\calQ}{\mathcal{Q}}
\providecommand{\calR}{\mathcal{R}}
\providecommand{\calS}{\mathcal{S}}
\providecommand{\calT}{\mathcal{T}}
\providecommand{\calU}{\mathcal{U}}
\providecommand{\calV}{\mathcal{V}}
\providecommand{\calW}{\mathcal{W}}
\providecommand{\calX}{\mathcal{X}}
\providecommand{\calY}{\mathcal{Y}}
\providecommand{\calZ}{\mathcal{Z}}

\providecommand{\mA}{\mathbf{A}}
\providecommand{\mB}{\mathbf{B}}
\providecommand{\mC}{\mathbf{C}}
\providecommand{\mD}{\mathbf{D}}
\providecommand{\mE}{\mathbf{E}}
\providecommand{\mF}{\mathbf{F}}
\providecommand{\mG}{\mathbf{G}}
\providecommand{\mH}{\mathbf{H}}
\providecommand{\mI}{\mathbf{I}}
\providecommand{\mJ}{\mathbf{J}}
\providecommand{\mK}{\mathbf{K}}
\providecommand{\mL}{\mathbf{L}}
\providecommand{\mM}{\mathbf{M}}
\providecommand{\mN}{\mathbf{N}}
\providecommand{\mO}{\mathbf{O}}
\providecommand{\mP}{\mathbf{P}}
\providecommand{\mQ}{\mathbf{Q}}
\providecommand{\mR}{\mathbf{R}}
\providecommand{\mS}{\mathbf{S}}
\providecommand{\mT}{\mathbf{T}}
\providecommand{\mU}{\mathbf{U}}
\providecommand{\mV}{\mathbf{V}}
\providecommand{\mW}{\mathbf{W}}
\providecommand{\mX}{\mathbf{X}}
\providecommand{\mY}{\mathbf{Y}}
\providecommand{\mZ}{\mathbf{Z}}

\providecommand{\va}{\mathbf{a}}
\providecommand{\vb}{\mathbf{b}}
\providecommand{\vc}{\mathbf{c}}
\providecommand{\vd}{\mathbf{d}}
\providecommand{\ve}{\mathbf{e}}
\providecommand{\vf}{\mathbf{f}}
\providecommand{\vg}{\mathbf{g}}
\providecommand{\vh}{\mathbf{h}}
\providecommand{\vi}{\mathbf{i}}
\providecommand{\vj}{\mathbf{j}}
\providecommand{\vk}{\mathbf{k}}
\providecommand{\vl}{\mathbf{l}}
\providecommand{\vm}{\mathbf{m}}
\providecommand{\vn}{\mathbf{n}}
\providecommand{\vo}{\mathbf{o}}
\providecommand{\vp}{\mathbf{p}}
\providecommand{\vq}{\mathbf{q}}
\providecommand{\vr}{\mathbf{r}}
\providecommand{\vs}{\mathbf{s}}
\providecommand{\vt}{\mathbf{t}}
\providecommand{\vu}{\mathbf{u}}
\providecommand{\vv}{\mathbf{v}}
\providecommand{\vw}{\mathbf{w}}
\providecommand{\vx}{\mathbf{x}}
\providecommand{\vy}{\mathbf{y}}
\providecommand{\vz}{\mathbf{z}}

\providecommand{\mDelta}{\mat{\Delta}}
\providecommand{\mGamma}{\mat{\Gamma}}
\providecommand{\mLambda}{\mat{\Lambda}}
\providecommand{\mOmega}{\mat{\Omega}}
\providecommand{\mPhi}{\mat{\Phi}}
\providecommand{\mPi}{\mat{\Pi}}
\providecommand{\mPsi}{\mat{\Psi}}
\providecommand{\mSigma}{\mat{\Sigma}}
\providecommand{\mTheta}{\mat{\Theta}}
\providecommand{\mXi}{\mat{\Xi}}

\providecommand{\w}{\omega}
\providecommand{\valpha}{\vec{\alpha}}
\providecommand{\vbeta}{\vec{\beta}}
\providecommand{\vgamma}{\vec{\gamma}}
\providecommand{\vdelta}{\vec{\delta}}
\providecommand{\vepsilon}{\vec{\epsilon}}
\providecommand{\vvarepsilon}{\vec{\varepsilon}}
\providecommand{\vzeta}{\vec{\zeta}}
\providecommand{\veta}{\vec{\eta}}
\providecommand{\vtheta}{\vec{\theta}}
\providecommand{\vkappa}{\vec{\kappa}}
\providecommand{\vlambda}{\vec{\lambda}}
\providecommand{\vmu}{\vec{\mu}}
\providecommand{\vnu}{\vec{\nu}}
\providecommand{\vxi}{\vec{\xi}}
\providecommand{\vpi}{\vec{\pi}}
\providecommand{\vrho}{\vec{\rho}}
\providecommand{\vvarrho}{\vec{\varrho}}
\providecommand{\vsigma}{\vec{\sigma}}
\providecommand{\vtau}{\vec{\tau}}
\providecommand{\vphi}{\vec{\phi}}
\providecommand{\vvarphi}{\vec{\varphi}}
\providecommand{\vchi}{\vec{\chi}}
\providecommand{\vpsi}{\vec{\psi}}
\providecommand{\vomega}{\vec{\omega}}

\providecommand{\wbar}{\overline{w}}
\providecommand{\wtilde}{\widetilde{w}}
\providecommand{\Ytilde}{\widetilde{Y}}
\providecommand{\ptilde}{\widetilde{p}}
\providecommand{\xbar}{\overline{x}}
\providecommand{\ybar}{\overline{y}}
\providecommand{\bbar}{\overline{b}}
\providecommand{\utilde}{\widetilde{u}}
\providecommand{\vtilde}{\widetilde{v}}
\providecommand{\xitilde}{\widetilde{\xi}}
\providecommand{\etatilde}{\widetilde{\eta}}

\providecommand{\mAtilde}{\mat{\widetilde{A}}}
\providecommand{\mDtilde}{\mat{\widetilde{D}}}
\providecommand{\mWtilde}{\mat{\widetilde{W}}}
\providecommand{\vftilde}{\boldsymbol{\widetilde{f}}}
\providecommand{\vptilde}{\boldsymbol{\widetilde{p}}}
\providecommand{\vxbar}{\boldsymbol{\overline{x}}}
\providecommand{\vybar}{\boldsymbol{\overline{y}}}
\providecommand{\vubar}{\boldsymbol{\bar{u}}}
\providecommand{\vxtilde}{\boldsymbol{\widetilde{x}}}
\providecommand{\vvtilde}{\boldsymbol{\widetilde{v}}}
\providecommand{\vthetahat}{\boldsymbol{\widehat{\theta}}}
\providecommand{\vxitilde}{\boldsymbol{\widetilde{\xi}}}

\providecommand{\Ghat}{\widehat{G}}
\providecommand{\fhat}{\widehat{f}}
\providecommand{\zhat}{\widehat{z}}
\providecommand{\yhat}{\widehat{y}}
\providecommand{\Zhat}{\widehat{Z}}
\providecommand{\xhat}{\widehat{x}}
\providecommand{\phat}{\widehat{p}}
\providecommand{\ahat}{\widehat{a}}
\providecommand{\bhat}{\widehat{b}}
\providecommand{\nuhat}{\widehat{\nu}}
\providecommand{\tauhat}{\widehat{\tau}}
\providecommand{\phihat}{\widehat{\phi}}
\providecommand{\Xhat}{\widehat{X}}
\providecommand{\Yhat}{\widehat{Y}}
\providecommand{\thetahat}{\widehat{\theta}}
\providecommand{\muhat}{\widehat{\mu}}
\providecommand{\Thetahat}{\widehat{\Theta}}
\providecommand{\sigmahat}{\widehat{\sigma}}

\providecommand{\vfhat}{\boldsymbol{\widehat{f}}}
\providecommand{\vzhat}{\boldsymbol{\widehat{z}}}
\providecommand{\vyhat}{\boldsymbol{\widehat{y}}}
\providecommand{\vhat}{\widehat{v}}
\providecommand{\mAhat}{\mat{\widehat{A}}}
\providecommand{\mJhat}{\mat{\widehat{J}}}
\providecommand{\mIhat}{\mat{\widehat{I}}}
\providecommand{\vxhat}{\boldsymbol{\widehat{x}}}
\providecommand{\vuhat}{\boldsymbol{\widehat{u}}}
\providecommand{\vvhat}{\boldsymbol{\widehat{v}}}
\providecommand{\vthetahat}{\boldsymbol{\widehat{\theta}}}
\providecommand{\mSigmahat}{\mat{\widehat{\mSigma}}}
\providecommand{\vmuhat}{\mat{\widehat{\vmu}}}
\providecommand{\vbetahat}{\boldsymbol{\widehat{\beta}}}
\providecommand{\mThetahat}{\mat{\widehat{\mTheta}}}

\providecommand{\fstar}{f^*}

\providecommand{\vzero}{\vec{0}}
\providecommand{\vone}{\vec{1}}

\providecommand{\conv}{\ast}
\providecommand{\circonv}[1]{\circledast_{#1}}
\providecommand{\Var}{\mathrm{Var}}
\providecommand{\prox}{\mathrm{prox}}
\providecommand{\to}{\rightarrow}
\providecommand{\from}{\leftarrow}

\newcommand{\defequal}{\mathop{\overset{\mbox{\tiny{def}}}{=}}}
\newcommand{\subjectto}{\mathop{\mathrm{subject\, to}}}
\newcommand{\argmin}[1]{\mathop{\underset{#1}{\mbox{argmin}}}}
\newcommand{\argmax}[1]{\mathop{\underset{#1}{\mbox{argmax}}}}
\newcommand{\minimize}[1]{\mathop{\underset{#1}{\mathrm{minimize}}}}
\newcommand{\maximize}[1]{\mathop{\underset{#1}{\mathrm{maximize}}}}
\newcommand{\diag}[1]{\mathop{\mathrm{diag}\left\{#1\right\}}}
\newcommand{\MSE}{\mathrm{MSE}}
\newcommand{\const}{\mathrm{const}}
\newcommand{\sign}{\mathrm{sign}}
\newcommand{\n}{\boldsymbol{n}}
\newcommand{\x}{\boldsymbol{x}}
\newcommand{\s}{\boldsymbol{u}}

\appendix

\section{Extended System Design Details}

\subsection{Minimal telephoto ratio} 

We study the minimal telephoto ratio optimized under three scenarios: 1) purely using refractive optics for the full visible bandwidth $\lambda \in [380~\text{nm}, 700~\text{nm}]$; 2) purely using refractive optics for a single wavelength $\lambda_0$; and 3) using a metasurface as the eyepiece in the optical assembly for a single wavelength $\lambda_0$. We set $\lambda_0 = 532$~nm, the same as our real experiment. Fig.~\ref{fig:sim}a visualizes the minimal telephoto ratio for each scenario under different numbers of optical elements $N$. While the minimal telephoto ratio reduces as $N$ increases, the improvement becomes less significant after $N=5$. Fig.~\ref{fig:sim}a clearly shows the minimal feasible telephoto ratio improves when dropping the constraint on achromaticity, and it further improves with the inclusion of the metasurface. This result validates the two design ideas of MetaTele: decoupling the achromaticity constraint from the optical design and leveraging metasurfaces for increasing compactness. Fig.~\ref{fig:sim}b demonstrates a sample optimized lens assembly with four refractive lenses and a metasurface, achieving a telephoto ratio of only $0.17$.  

\subsection{Tolerance Analysis}
\label{supp:tolerance}
\begin{figure}
    \centering
    \includegraphics[width=0.9\linewidth]{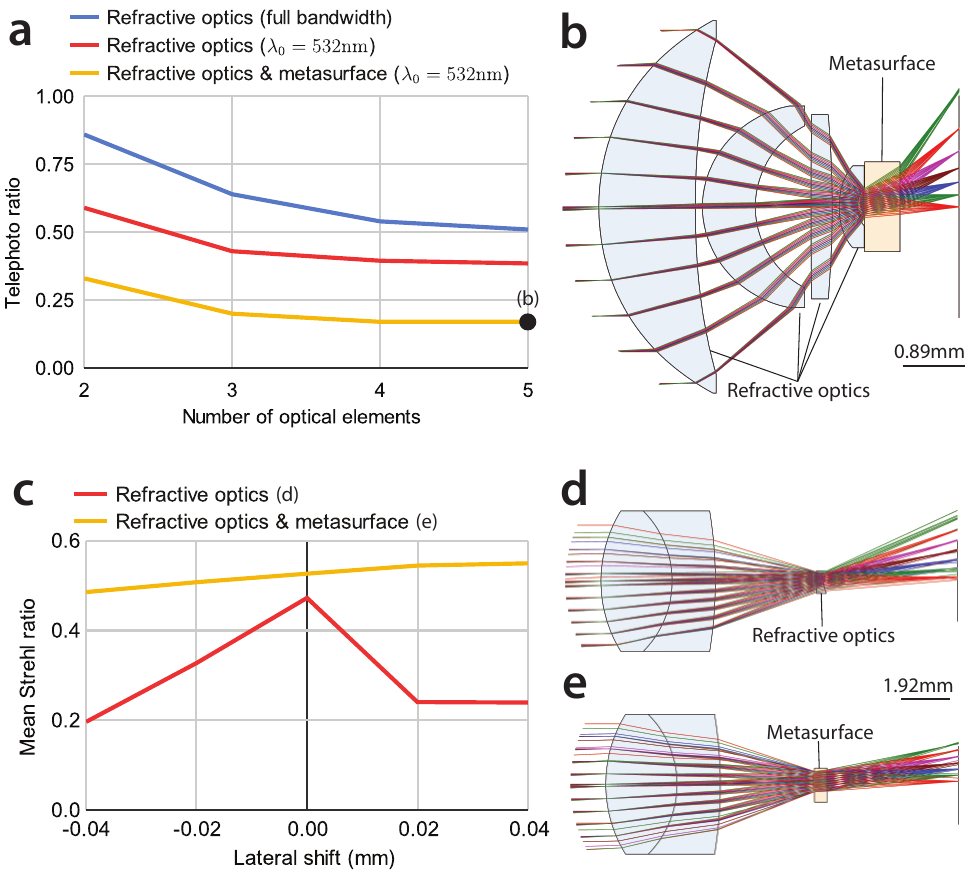}
    \caption{Simulation study of the optical design. (a) Minimal telephoto ratios under different numbers of optical elements $N$. The colors represent three scenarios, as indicated by the legends. By constraining the operating wavelength and incorporating the metasurface in the optical assembly, the minimal feasible telephoto ratio decreases collectively, demonstrating the effectiveness of the MetaZoom design idea. (b) A sample optimized lens assembly with four refractive optics and a metasurface as the eyepiece, achieving a telephoto of 0.17. (c) Optical performance, quantified as the mean Strehl ratio within the field of view, w.r.t. random perturbation on optical element's position. Consider the two-optic lens assemblies in (d) and (e), where they utilize refractive optics and metasurface as the eyepiece, respectively. When the eyepiece shifts laterally by $0.02$~mm, the optical performance of (d) significantly decreases, while (e) remains roughly constant. (d-e) The raytracing diagram when the eyepiece in both assemblies is perturbed downward by $0.04$ mm. The pure refractive assembly (d) suffers from a more severe decrease in optical performance than the hybrid assembly (e). The unperturbed version of (e) is the actual optical design used in the main paper. }
    \label{fig:sim}
\end{figure}

We also examine the optical performance of different lens assemblies when the optical elements are displaced from their ideal positions. Here, we present a special case, where two optical assemblies (Fig~\ref{fig:sim}d-e) share the same amount of lateral perturbation on their eyepiece, the last optical element to the photosensor. The two assemblies have the same objective, i.e., the first lens, and use refractive optics and a metasurface as the eyepiece, respectively. As shown by Fig.~\ref{fig:sim}c, the purely-refractive optical assembly experiences a much more significant loss in imaging quality, quantified by the mean Strehl ratio within a field of view~(FoV) of $6^{\circ}$, than the hybrid one, when the eyepiece's position is perturbed laterally by within $0.04$ mm. This experiment shows the higher tolerance of metasurface than refractive optics as an eyepiece to the lateral positional error. 

We analyzed how the lateral translation of a metasurface or refractive lens from its ideal position could affect the optical performance of the lens system. Here, we provide a more comprehensive analysis of the optical performance's tolerance to the optical element's displacement for purely refractive or hybrid optical assemblies. 

\begin{figure}[ht]
    \centering
    \begin{subfigure}[b]{0.49\linewidth}
        \centering
        \includegraphics[width=\linewidth]{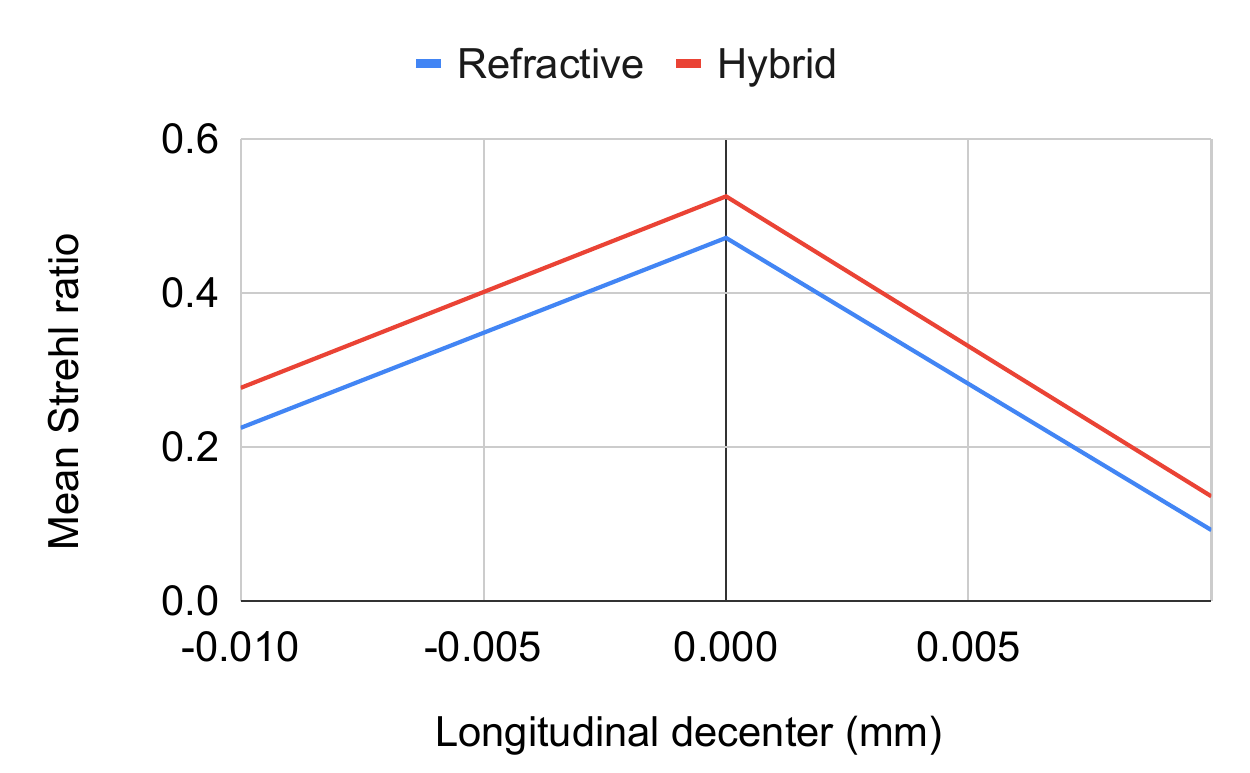}
    \end{subfigure}
    \hfill
    \begin{subfigure}[b]{0.49\linewidth}
        \centering
        \includegraphics[width=\linewidth]{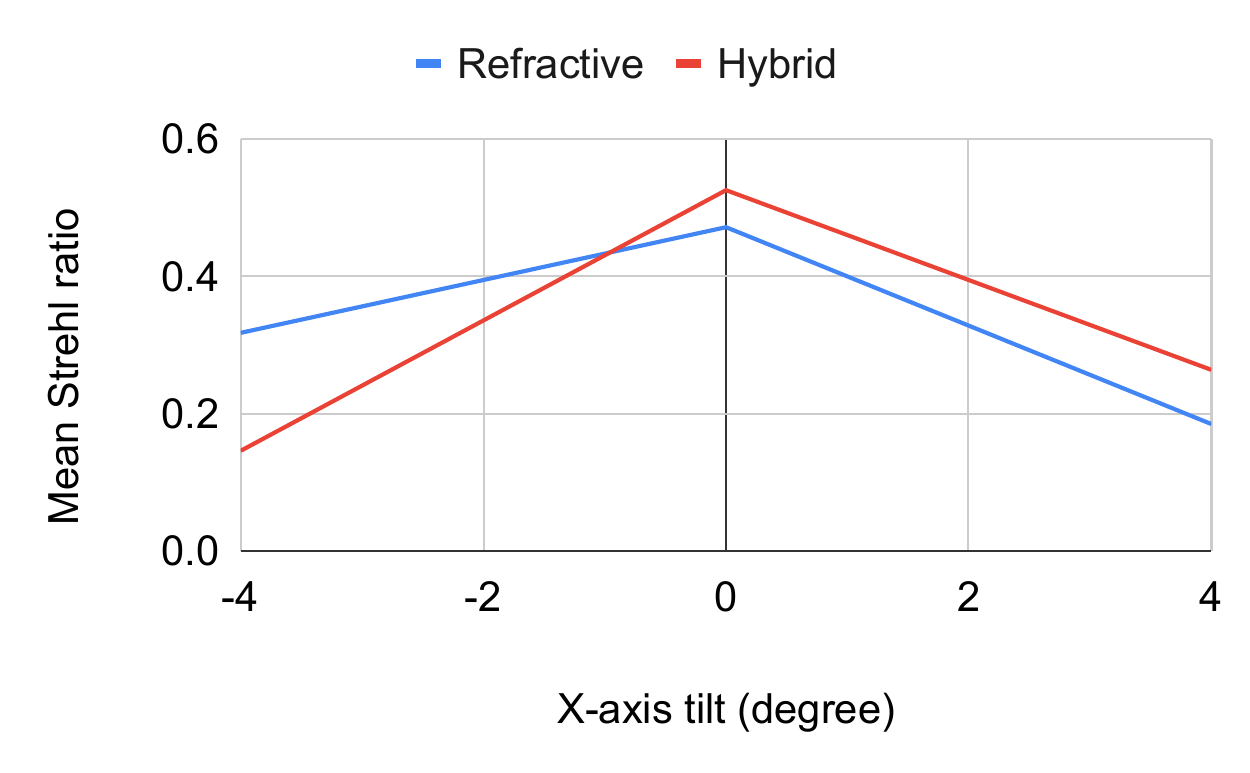}
    \end{subfigure}  
    \hfill
    \vspace{0.3cm}
    \begin{subfigure}[b]{0.51\linewidth}
        \centering
        \includegraphics[width=\linewidth]{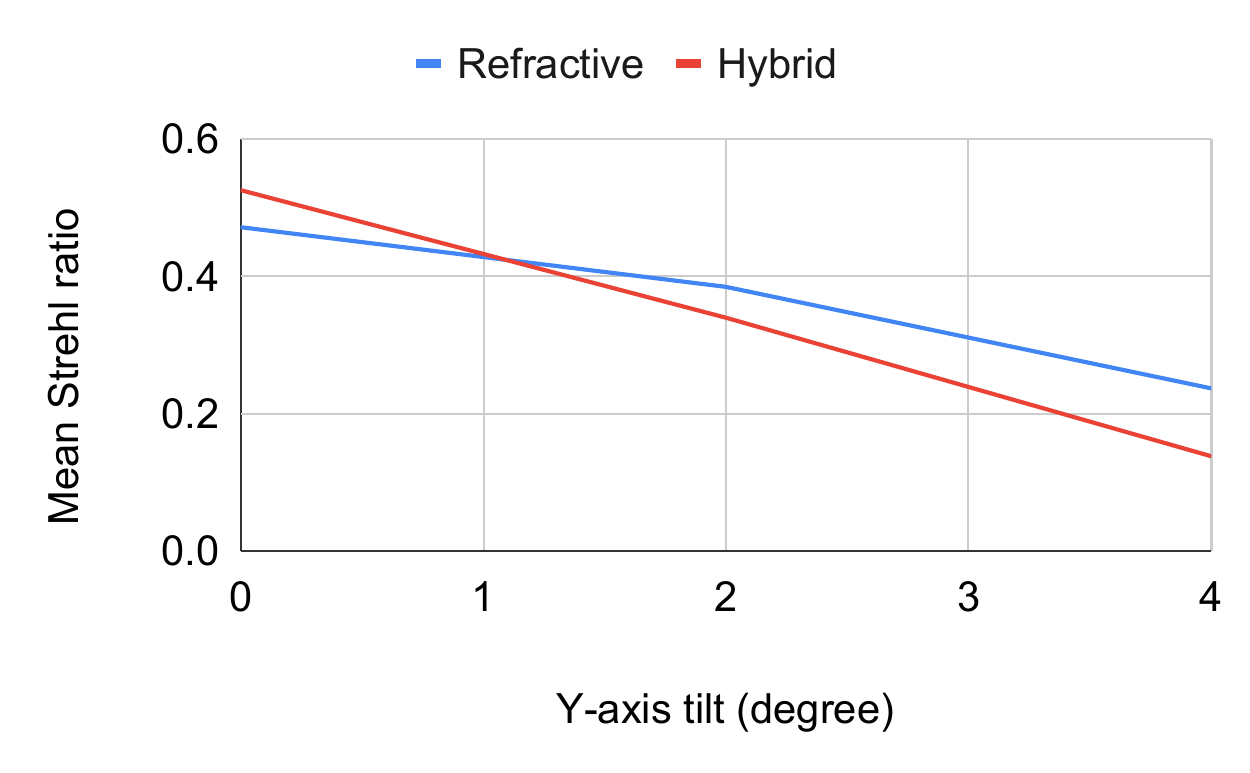}
    \end{subfigure}  
\caption{Tolerance study on different positional perturbations. We analyze the optical performance degradation of the systems in Fig.~4d-e of the main paper when the eyepiece is perturbed with similar displacements. We use the mean Strehl ratio across a $6^\circ$ FoV as the metric for the optical performance. Both systems exhibit comparable tolerance to longitudinal decenter and tilt of different axes. Refer to Fig. 2a of the main paper for the rotation axis.}
\label{fig:perturb}
\end{figure}

We consider the two optical systems in Fig.~\ref{fig:sim}d-e, which share similar imaging qualities and telephoto ratios. By exerting the same amount of displacement, including the longitudinal translation and tilt along lateral axes, to the refractive (Fig.~\ref{fig:sim}d) and metasurface (Fig.~\ref{fig:sim}e) eyepiece, we analyze the degradation of the optical performance as quantized by the mean Strehl ratio across a FoV of $6^{\circ}$. As shown in Fig.~\ref{fig:perturb}, both systems show a comparable decrease in optical performance, demonstrating that both systems share similar tolerance to these displacements. We show that the hybrid system is more, if not similarly, robust to perturbation on the eyepiece than the purely refractive system.

\subsection{Effect of different metasurface designs}
\begin{figure}[h!]
    \centering
    \includegraphics[width=0.9\linewidth]{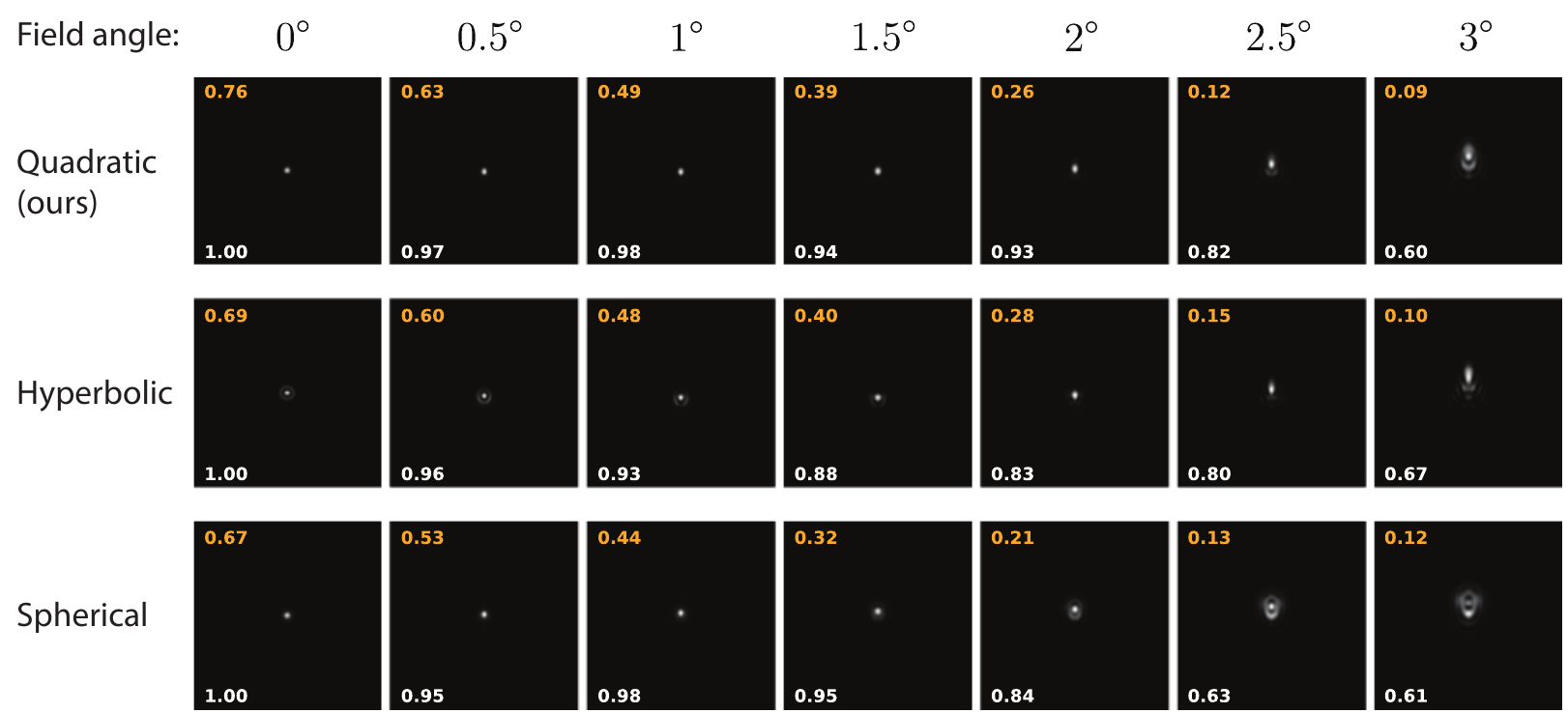}
    \caption{Simulated point spread functions (PSFs) at different field angles for the MetaTele prototype using different metasurface phase profiles (Eqs.~19 in the main paper, \ref{eq:hyper_phase}, and \ref{eq:sphr_phase}). For each PSF, the inset values in the top-left and bottom-left corners indicate the Strehl ratio and the cross-correlation coefficient with respect to the on-axis (\(0^\circ\)) PSF, respectively.}
    \label{fig:metasurface_comparison}
\end{figure}

As shown in Fig.~\ref{fig:metasurface_comparison}, we visualize the PSFs of the structure image at different field angles for metasurfaces with different phase delay profiles $\phi(\mathbf{x}, \lambda_0)$. In addition to the quadratic phase profile used in MetaTele (Eq.~19 in the main paper), we also consider the hyperbolic and spherical phase profiles listed below:
\begin{align}
    \phi_{\text{hyperbolic}}(\mathbf{x},\lambda_0) &= \frac{2\pi}{\lambda_0}\left(f -\sqrt{\lVert \mathbf{x} \rVert^2 + f^2}\right), \label{eq:hyper_phase}\\
    \phi_{\text{spherical}}(\mathbf{x},\lambda_0) &= \frac{2\pi}{\lambda_0}\left(\sqrt{f^{2}-\lVert \mathbf{x} \rVert^2} - f\right). \label{eq:sphr_phase}
\end{align}
We set the focal length $f=-2$~mm for all three designs. We chose the hyperbolic and spherical phase profiles to compare as they have been frequently utilized as baselines in prior work~\cite{groever2017meta, martins2020metalenses}. As shown in Fig.~\ref{fig:metasurface_comparison}, all three designs yield comparable imaging quality, as quantified by the Strehl ratio. However, the quadratic phase profile adopted in MetaTele produces a slightly more uniform PSF across field angles, which is desirable for maintaining consistent spatial fidelity over the field of view.

\subsection{Optimized metasurface}
\begin{figure}
    \centering
    \includegraphics[width=0.45\linewidth]{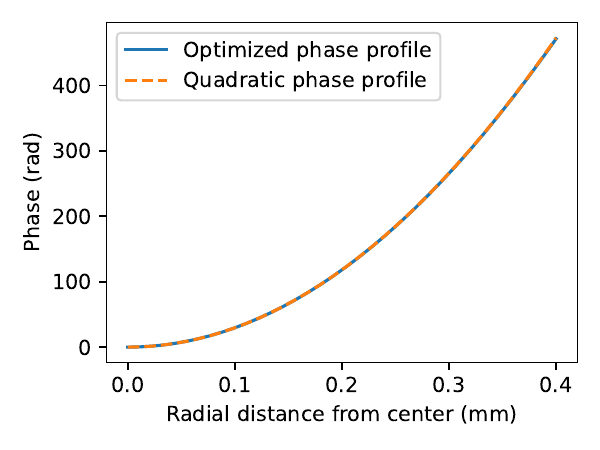}
    \caption{Phase delay profile of the optimized metasurface. Note that it matches a quadratic function.}
    \label{fig:phase_radial}
\end{figure}

The metasurface phase profile is parameterized using radially symmetric even-order polynomials up to the fourteenth order:
\begin{equation}
\label{eq:quad_phase_profile_s}
    \phi(\mathbf{x},\lambda_0) = \frac{2\pi}{\lambda_0}\sum_{i=1}^{7} c_i \|\mathbf{x}\|^{2i}.
\end{equation}
The converged metasurface phase profile $\Tilde{\phi}(\vx,\lambda_0)$ closely resembles a quadratic function, corresponding to a diverging lens:
\begin{align}
    \Tilde{\phi}(\vx,\lambda_0) \approx -\frac{2\pi}{\lambda_0}\frac{\|\mathbf{x}\|^2}{2f}, 
    \label{eq:quad_phase_s}
\end{align}
with a focal length $f=-2~\text{mm}$.
We list the converged coefficients, $c_{1-7}$, of Eq.~\ref{eq:quad_phase_s} in Table~\ref{tab:coeffs}. As shown in Fig.~\ref{fig:phase_radial}, the phase delay profile closely matches a quadratic function.     
\begin{table}
    \centering
    \caption{Polynomial coefficients of the optimized metasurface phase profile according to Eq.(19) in the main text.}
    \label{tab:coeffs}
    \begin{tabular}{ccccccc}
        \toprule
        $c_1$ &$c_2$ &$c_3$ &$c_4$ &$c_5$ &$c_6$ &$c_7$ \\ \hline
        0.25  & -0.0156  & 0.2133  & -0.6931  & -1.5622  & -0.0633 & 10.8101   \\
        \bottomrule
    \end{tabular}
    
\end{table}

\begin{table}
    \centering
    \caption{Spot diagram RMS values of MetaTele's PSFs at different field angles in simulation.}
    \label{tab:RMS}
    \setlength{\tabcolsep}{0pt} 
    \renewcommand{\arraystretch}{1.2} 
    
    \begin{tabular*}{\textwidth}{@{\extracolsep{\fill}}cccccccc}
        \hline
        Field & 0$^\circ$  & 0.5$^\circ$ & 1$^\circ$ & 1.5$^\circ$ & 2$^\circ$ & 2.5$^\circ$ & 3$^\circ$  \\
        \hline
        Spot diagram RMS~($\mu m$) & 10.0 & 21.4 & 48.6 & 85.9 & 94.9 & 107.9 & 112.8 \\ 
        \hline
    \end{tabular*}

\end{table}

During optimization, we set the metasurface aperture diameter to 1 mm and then reduce it to 0.8 mm to attenuate off-axis aberrations, at the cost of slightly reduced modulation transfer functions (MTFs) for the normal incident field. We also report RMS values of the optimized PSFs in Table~\ref{tab:RMS}.

\begin{figure}
  \centering
  \begin{subfigure}{0.49\linewidth}
    \centering
    \includegraphics[height=8.4cm]{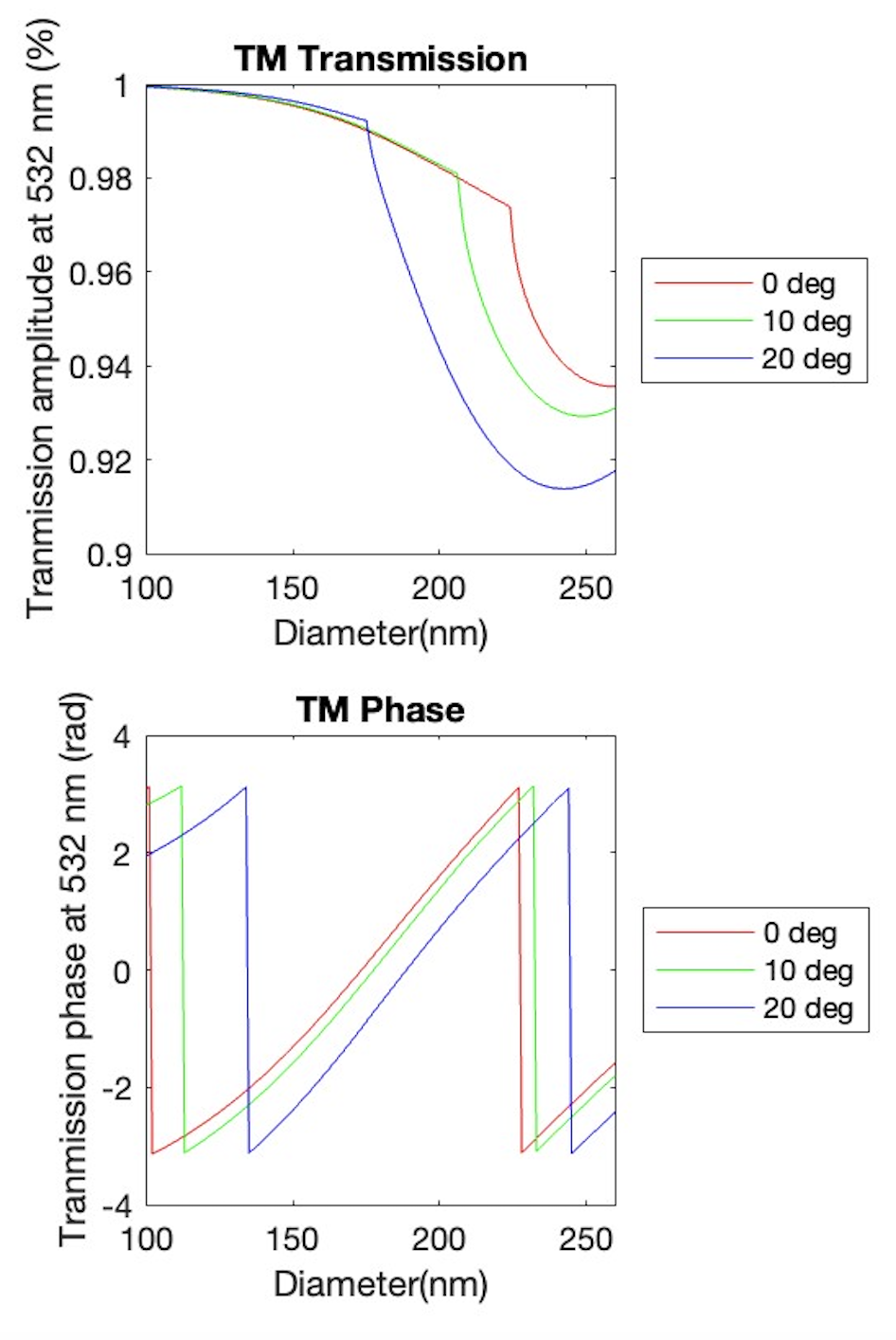}
    \caption{TM polarization.}
    \label{fig:TM}
  \end{subfigure}
  \hfill
  \begin{subfigure}{0.49\linewidth}
    \centering
    \includegraphics[height=8.4cm]{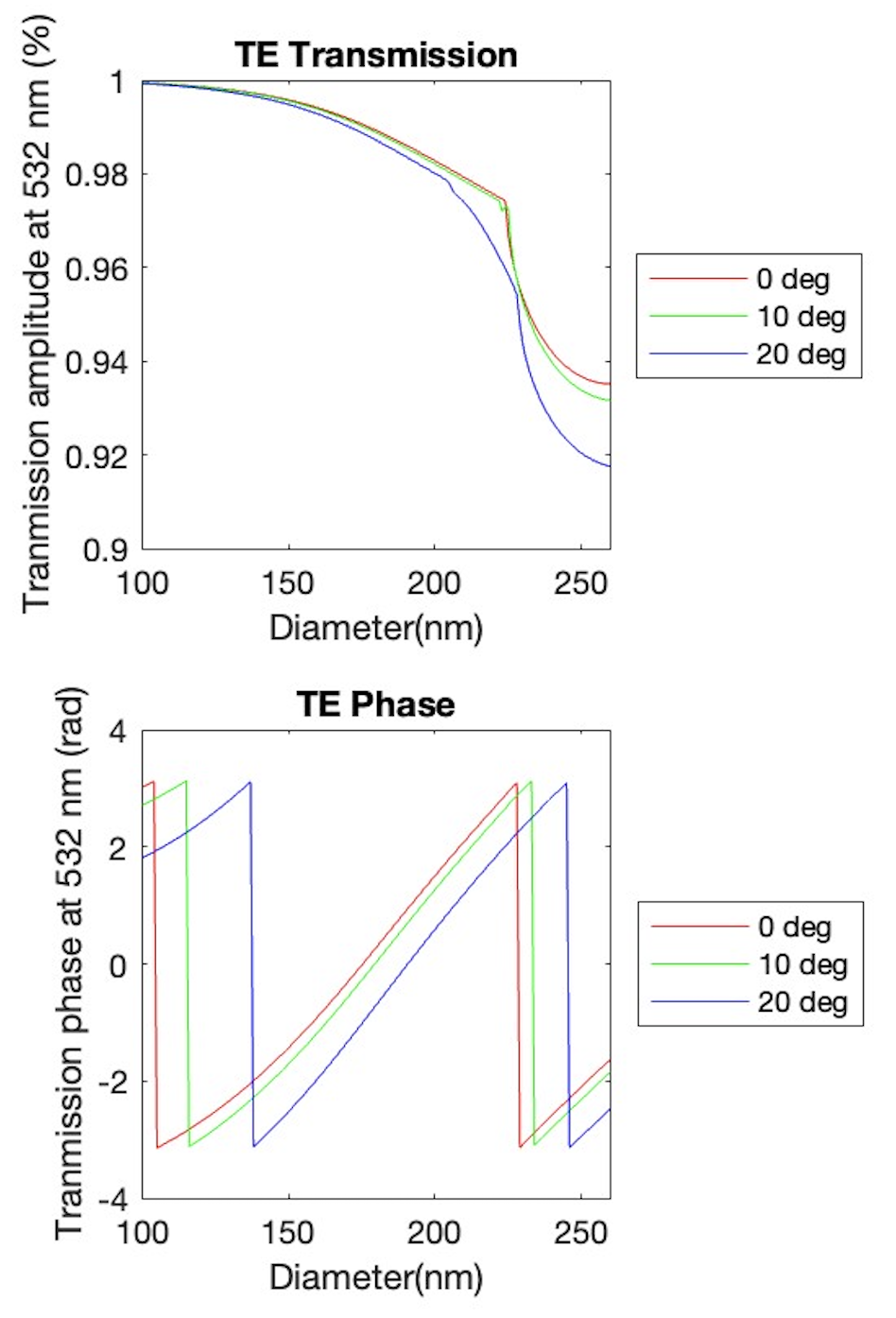}
    \caption{TE polarization.}
    \label{fig:TE}
  \end{subfigure}
  \caption{The nanocells we use demonstrate insensitivity to the incident angle of the light according to our simulation.}
  \label{fig:angle dependence}
\end{figure}

The properties of the nanocells we used for fabricating the metasurface is shown in Fig.~\ref{fig:angle dependence}.

\subsection{Estimating hyperfocal distance}
To characterize the operational range of the proposed telephoto system, we calculate the hyperfocal distance $H$. The system parameters are summarized in Table~\ref{tab:params}. Given the operational F-number of $f/6.0$, the diffraction-limited spot size (Airy disk diameter) determines the system's resolution limit:
\begin{equation}
    d_{Airy} = 2.44 \cdot \lambda \cdot N \approx 2.44 \cdot (0.532\,\mu\text{m}) \cdot 6.0 \approx 7.8\,\mu\text{m}
\end{equation}
Since $d_{Airy} > p$, the circle of confusion is dictated by diffraction rather than the detector pixel pitch. Consequently, the effective hyperfocal distance $H_{eff}$ is calculated as:
\begin{equation}
    H_{eff} \approx \frac{f^2}{N \cdot d_{Airy}} = \frac{(30\,\text{mm})^2}{6.0 \cdot 0.0078\,\text{mm}} \approx 19.2\,\text{m}
\end{equation}
Based on the physical diffraction limit, the effective hyperfocal distance is approximately \textbf{19.2\,m}. When focused at this distance, the system maintains diffraction-limited performance from roughly 9.6\,m to infinity. This metric provides a physically rigorous definition of the system's depth of field, accounting for the constraints of the $f/6.0$ aperture.

\begin{table}[h!]
\centering
\caption{System Parameters for Depth of Field Calculation}
\label{tab:params}
\begin{tabular}{lc}
\toprule
\textbf{Parameter} & \textbf{Value} \\
\midrule
Entrance Pupil Diameter (EPD) & $5.0$\,mm \\
Effective Focal Length ($f$) & $30.0$\,mm \\
F-number ($N$) & $f/6.0$ \\
Pixel Pitch ($p$) & $2.0\,\mu$m \\
Design Wavelength ($\lambda$) & $532$\,nm \\
\bottomrule
\end{tabular}
\end{table}

\subsection{Autofocus and continuous zoom}
\begin{figure}[ht]
    \centering
    \begin{subfigure}[b]{0.49\linewidth}
        \centering
        \includegraphics[width=\linewidth]{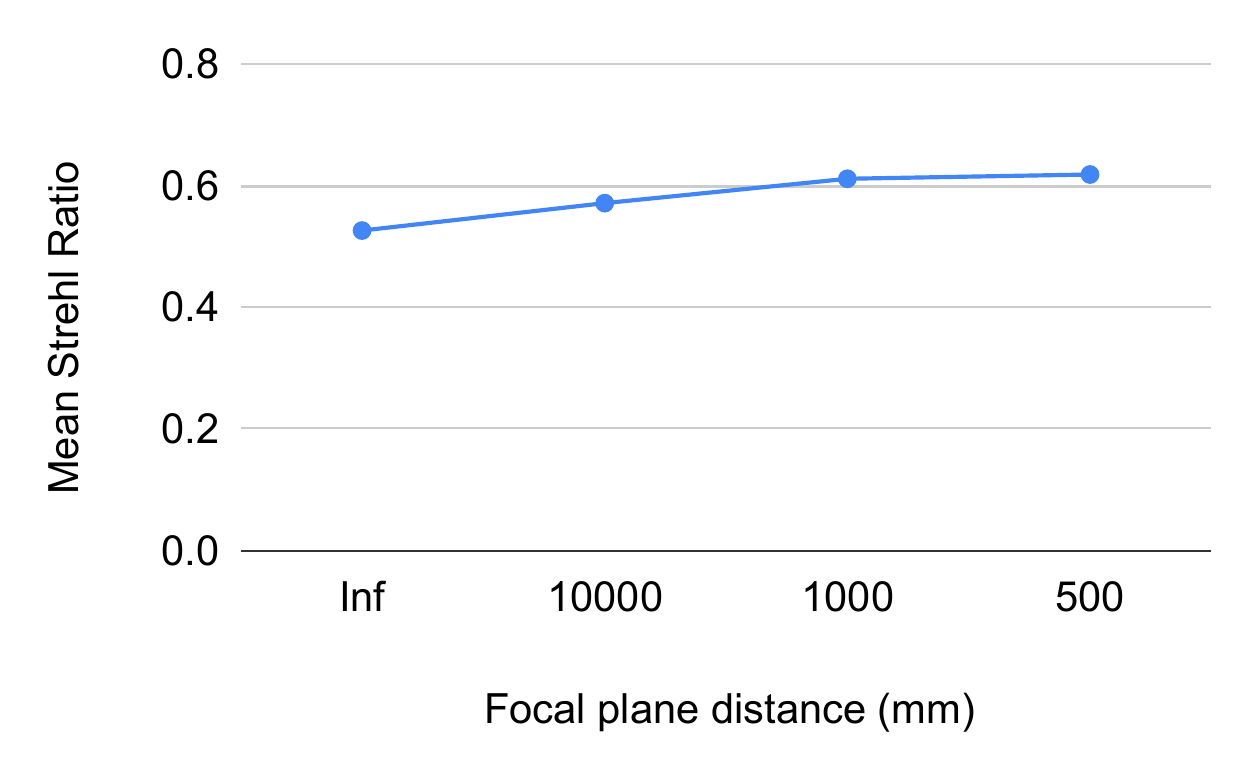}
        \caption{Autofocus.}
        \label{fig:autofocus}
    \end{subfigure}
    \vspace{0.3cm}
    \begin{subfigure}[b]{0.49\linewidth}
        \centering
        \includegraphics[width=\linewidth]{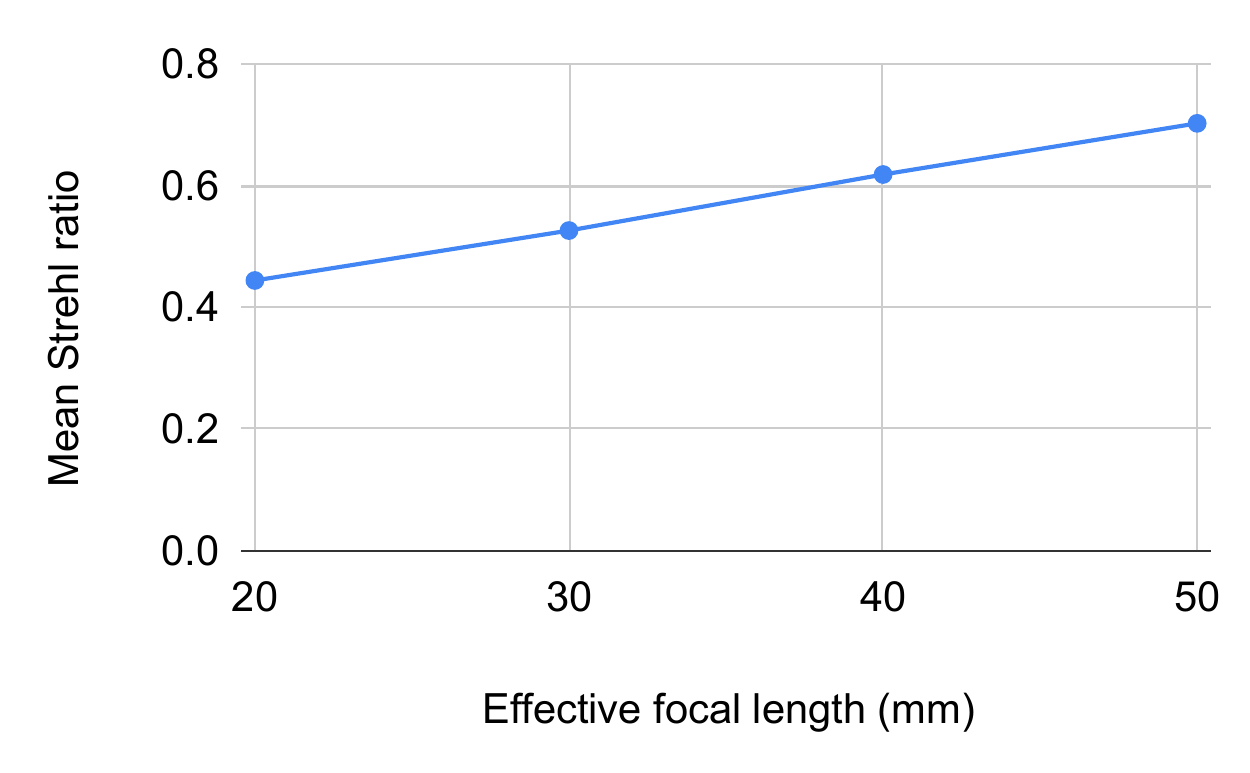}
        \caption{Continuous zoom.}
        \label{fig:conzoom}
    \end{subfigure}  
\caption{Simulation study of the MetaTele prototype. (a) The system can vary its focal plane to achieve autofocus by adjusting the distance between the metasurface eyepiece and the photosensor. Meanwhile, the optical performance, quantified as the mean Strehl ratio, stays approximately constant. (b) By adjusting the distance between the refractive objective and the metasurface, the effective focal length (EFL) can be varied smoothly to achieve continuous zoom with overall satisfactory optical performance. }
\end{figure}

The MetaTele prototype we designed has the advantage of performing autofocus and continuous zoom. By adjusting the distance between the metasurface eyepiece and the photosensor, the focal plane of the optical system can vary. Fig.~\ref{fig:autofocus} shows that the optical performance stays constant at different focal distances in simulation. Furthermore, the system can continuously adapt its optical zoom. Its effective focal length (EFL) can be smoothly adjusted by varying the distance between the refractive objective and the metasurface eyepiece. Fig.~\ref{fig:conzoom} demonstrates that the optical performance remains satisfactory when the EFL of the same optical assembly is adjusted between 20 and 50 mm in simulation. 

\section{Ablation study on the post-processing algorithm}

\subsection{Effectiveness of the HF-VSD loss.} 
\begin{figure}[t]
  \centering
  \begingroup
    \setlength{\tabcolsep}{1pt} 
  \renewcommand{\arraystretch}{1} 
  \resizebox{\linewidth}{!}{%
  \begin{tabular}{cccc}
    \begin{subfigure}{0.25\linewidth}
    \begin{tikzpicture}[
        spy using outlines={%
            rectangle, 
            green, 
            magnification=4, 
            size=0.5\linewidth,        
            connect spies    
            }
        ]
        \node (main_image) at (0,0) {
            \includegraphics[width=\linewidth]{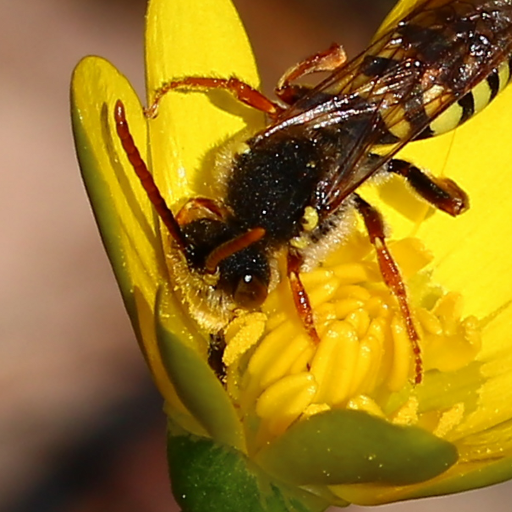} 
        };
        
        \spy on (-0.3, -0.2) in node at (0.8, -0.8); 
    \end{tikzpicture}
    \caption{GT}
    \end{subfigure} &
    
    \begin{subfigure}{0.25\linewidth}
    \begin{tikzpicture}[
        spy using outlines={%
            rectangle, 
            green, 
            magnification=4, 
            size=0.5\linewidth,        
            connect spies    
            }
        ]
        \node (main_image) at (0,0) {
            \includegraphics[width=\linewidth]{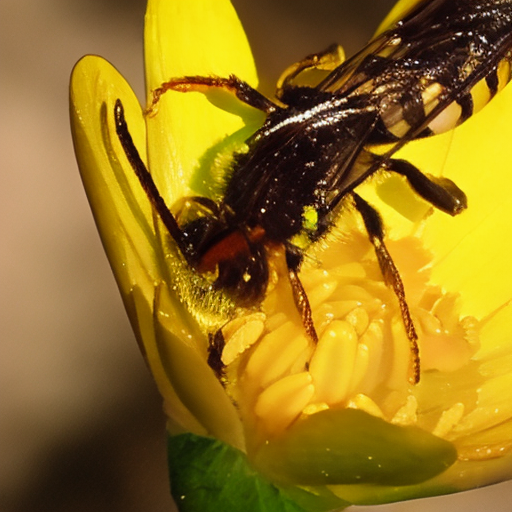} 
        };
        
        \spy on (-0.3, -0.2) in node at (0.8, -0.8); 
    \end{tikzpicture}
    \caption{Ours (w/o VSD)}
    \end{subfigure} &
    
    \begin{subfigure}{0.25\linewidth}
    \begin{tikzpicture}[
        spy using outlines={%
            rectangle, 
            green, 
            magnification=4, 
            size=0.5\linewidth,        
            connect spies    
            }
        ]
        \node (main_image) at (0,0) {
            \includegraphics[width=\linewidth]{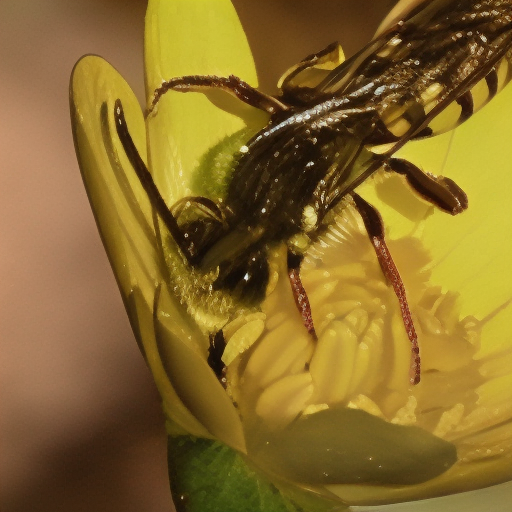} 
        };
        
        \spy on (-0.3, -0.2) in node at (0.8, -0.8); 
    \end{tikzpicture}
    \caption{Ours (w/ VSD)}
    \end{subfigure} &

    \begin{subfigure}{0.25\linewidth}
    \begin{tikzpicture}[
        spy using outlines={%
            rectangle, 
            green, 
            magnification=4, 
            size=0.5\linewidth,        
            connect spies    
            }
        ]
        \node (main_image) at (0,0) {
            \includegraphics[width=\linewidth]{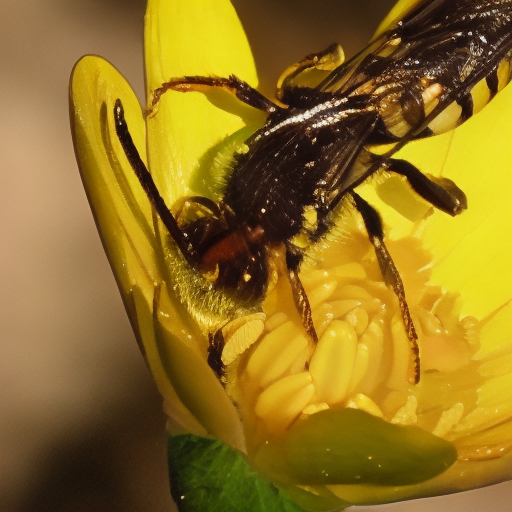} 
        };
        
        \spy on (-0.3, -0.2) in node at (0.8, -0.8); 
    \end{tikzpicture}
    \caption{Ours (w/ HF-VSD)}
    \end{subfigure} \\
  \end{tabular}
  }
  \endgroup 
  \caption{Qualitative comparison of training the model using different loss functions.}
  \label{fig:vsd_comparison}
\end{figure}
We trained three varieties of the proposed models with different regularizations. i) Do not use any regularizer, denoted as \textit{Ours (w/o VSD)}. ii) We use the standard VSD loss, denoted as \textit{Ours (w/ VSD)}. iii) We use the proposed HF-VSD loss, denoted as \textit{Ours (w/ HF-VSD). } Fig.~\ref{fig:vsd_comparison} qualitatively shows that the proposed HF-VSD produces the sharpest image details while preserving color fidelity compared to the other two.

\subsection{Guidance of the structure image and color cue for reconstruction.}
\begin{figure}
    \centering
    \includegraphics[width=\linewidth]{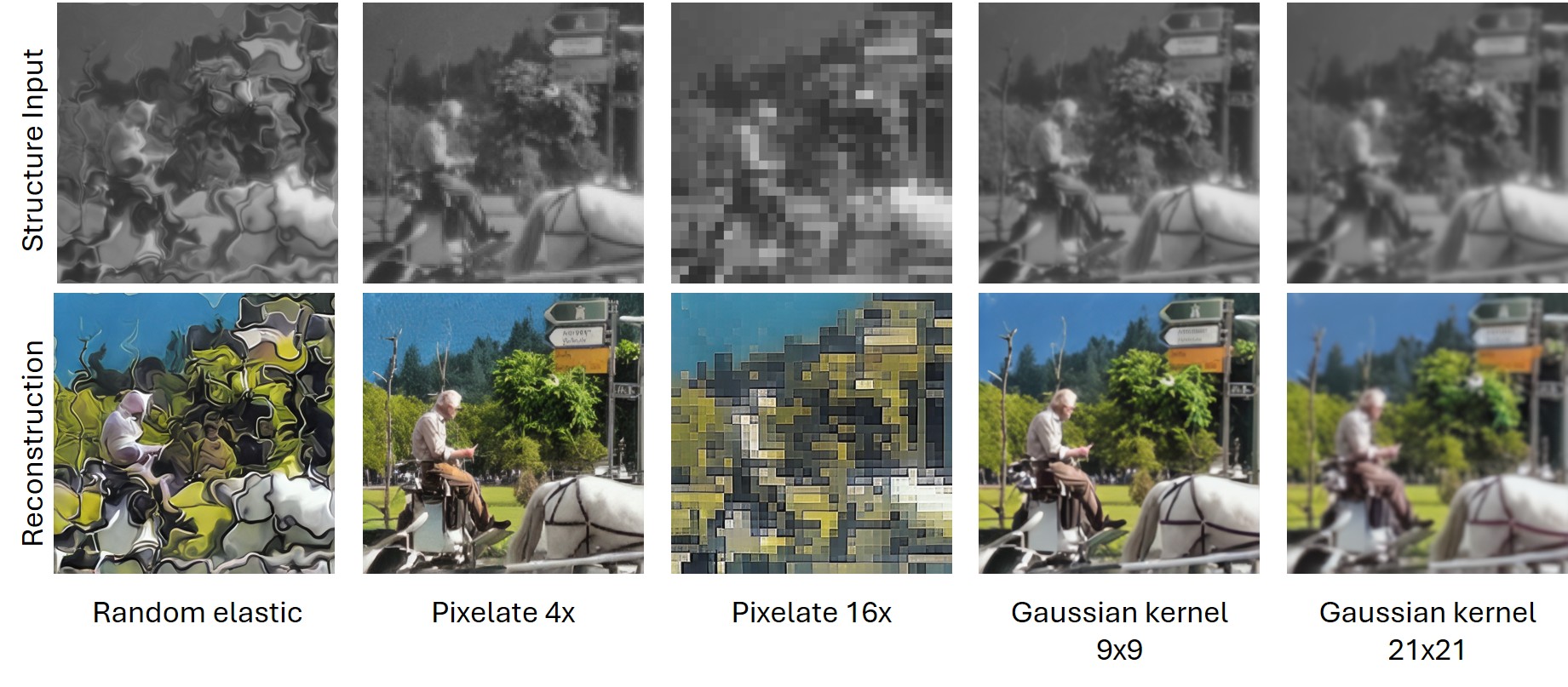}
    \caption{Effect on reconstructed image by using a deformed structural image. The first row shows the deformed structural image input, and the second row shows the reconstructed image. Each column represents a different deformation applied.}
    \label{fig:structure_ablation}
\end{figure}

\begin{figure}
    \centering
    \includegraphics[width=\linewidth]{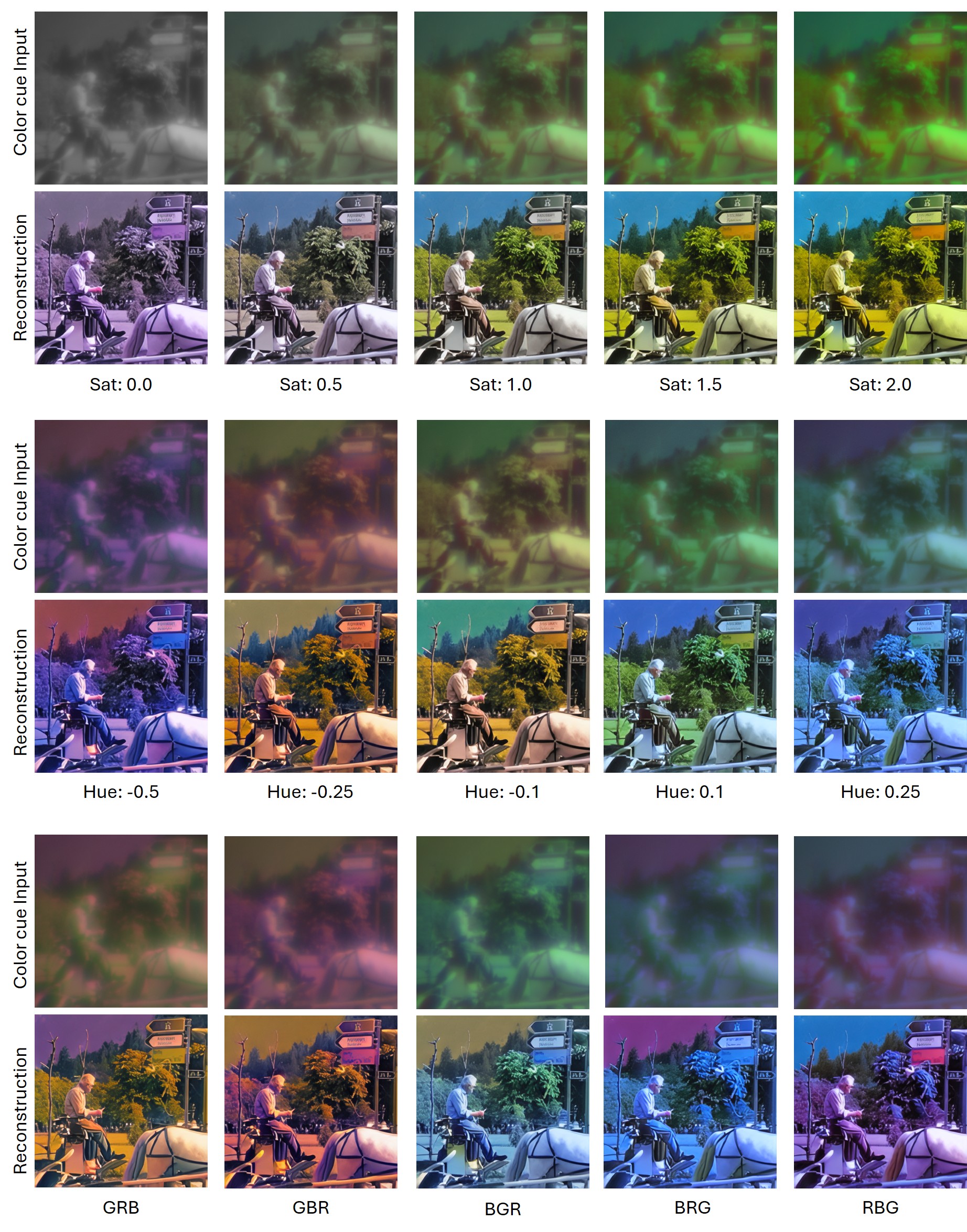}
    \caption{Effect on reconstructed image by intentionally degraded color cue. The 6 rows are grouped into 3 sections, each with a saturation adjustment, a hue adjustment, and a channel permutation. Each row pair shows the altered color cue image input and the corresponding reconstructed image.}
    \label{fig:colorcue_ablation}
\end{figure}

We investigate whether the proposed computational model is truly guided by the information provided in the color cue and structure image, rather than producing outputs dominated by generative priors. 
As illustrated in Fig.~\ref{fig:colorcue_ablation} and Fig.~\ref{fig:structure_ablation}, we intentionally degrade either the color cue or the structure image using several perturbation strategies and examine the resulting reconstructions. The outputs consistently reflect the degradations introduced in the corresponding inputs, without restoring them to visually plausible natural images. This indicates that the reconstruction prioritizes fidelity to the measured color and structural information, demonstrating that the model operates under strong measurement guidance rather than hallucinating content based on natural image statistics.

\subsection{Frequency-domain analysis.}
\begin{figure}
    \centering
    \includegraphics[width=\linewidth]{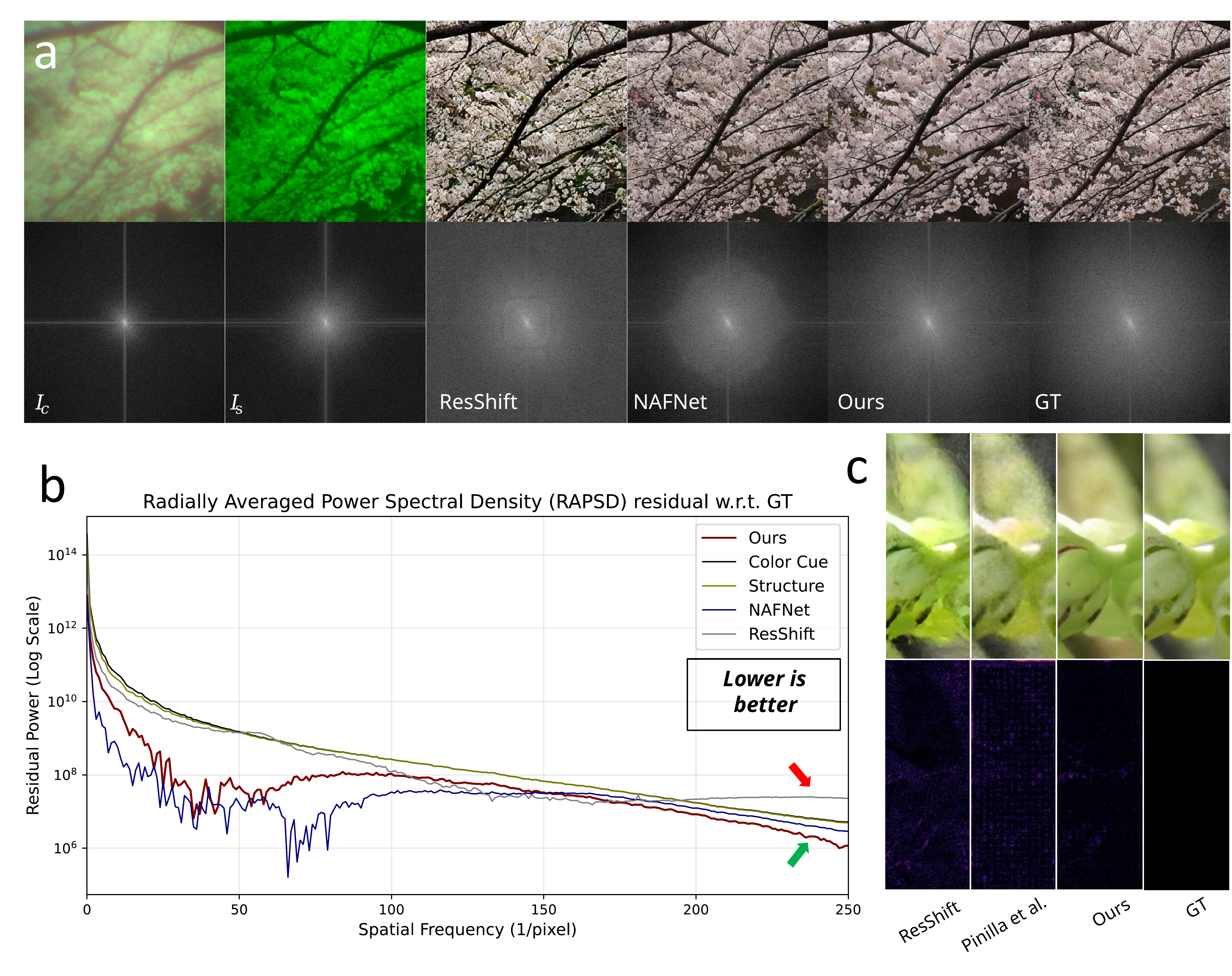}
    \caption{Frequency analysis of sample reconstructed images. (a) Spatial- and frequency-domain comparison of the color cue, structure image, reconstruction from different methods, and the ground truth.  (b) Radially averaged power spectral density (RAPSD) residuals computed by subtracting each competing method’s RAPSD from that of the ground truth. (c) Representative reconstructions from selected baselines (top row) and their corresponding high-frequency residual maps with respect to the ground truth (bottom row).}
    \label{fig:freq_comparison}
\end{figure}

Fig.~\ref{fig:freq_comparison}(a) compares the frequency spectra of sample raw measurements and the corresponding reconstructions produced by different post-processing algorithms. Our method yields a spectral distribution that most closely matches the ground truth. Fig.~\ref{fig:freq_comparison}(b) further quantifies this by visualizing the radially averaged power spectral density (RAPSD) residual with respect to the ground truth. Our approach exhibits the lowest residual energy in the high spatial frequency range, indicating its ability to recover realistic fine-scale details. This advantage is also qualitatively evident in Fig.~\ref{fig:freq_comparison}(c), which presents a zoomed-in region of the reconstructed images along with their residual maps relative to the ground truth. Our method achieves the highest visual fidelity and the lowest reconstruction error.

\end{document}